\documentclass[aps,prd,longbibliography,preprintnumbers,nofootinbib,
superscriptaddress,amsmath,floatfix,10pt]{revtex4}

\usepackage{amssymb}  
\usepackage{amsmath}
\usepackage{hyperref}
\usepackage{graphicx
}
\usepackage{exscale}
\usepackage{bbold}
\usepackage{textcomp}
\usepackage{enumerate}
\usepackage [english]{babel}
\usepackage[utf8]{inputenc}
\usepackage [autostyle, english = american]{csquotes}
\MakeOuterQuote{"}

\usepackage{color}

\usepackage[normalem]{ulem}  

\newcommand{\non}{\nonumber\\}

\newcommand{\be}{\begin{equation}}
\newcommand{\ee}{\end{equation}}
\newcommand{\bea}{\begin{eqnarray}}
\newcommand{\eea}{\end{eqnarray}}
\newcommand{\ba}[1]{\begin{array}{#1}}
\newcommand{\ea}{\end{array}}

\newcommand{\uq}{\hat{\vec{q}}} 
\newcommand{\up}{\hat{\vec{p}}}
\newcommand{\hk}{\hat{k}}
\newcommand{\hq}{\hat{q}}

\newcommand{\Tr}{{\rm Tr}}

\newcommand{\uk}{\hat{\vec{k}}}
\newcommand{\bm}[1]{\mbox{\boldmath${#1}$}}

\renewcommand{\vec}[1]{\bm{#1}}

\begin{document}

\title{Phases and properties of color superconductors}

\author{Andreas Schmitt}
\email{a.schmitt@soton.ac.uk}
\affiliation{Mathematical Sciences and STAG Research Centre, University of Southampton, Southampton SO17 1BJ, United Kingdom}

\date{10 November 2025}


\begin{abstract} 
Cold and dense matter is expected to be in a color-superconducting state. Here we review two calculations, relevant 
for fundamental properties and applications of color superconductivity, respectively: the weak-coupling QCD calculation of the fermionic energy gap together with the magnetic screening masses of the gauge bosons, and the calculation of bulk viscosity from a non-leptonic electroweak process. These calculations are supplemented by a discussion of color superconductors with mismatched Fermi momenta, and they are embedded in the context of the state of the art by giving an overview of previous and ongoing work and future directions. 
\end{abstract}

\maketitle

\tableofcontents

\section{Introduction} 
\label{sec:CFL}

QCD matter at large baryon densities and small temperatures exhibits many phenomena known from ordinary, low-energy condensed matter physics. The most important one 
is Cooper pairing. Recall Cooper pairing in an ordinary superconductor: electrons immersed in a lattice of ions effectively attract each other, which leads to the formation of a Cooper
pair condensate, which in turn induces an energy gap in the quasi-particle spectrum of the electrons. As a consequence, single electrons cannot be excited at low energies, which prevents them from undergoing scattering processes with the ion lattice while they are confined in a Cooper pair. Since electrons carry electric charge, this leads to the 
phenomenon of superconductivity. The underlying mechanism of Cooper pairing is very general. What we need is a sharp Fermi surface, i.e., a system of fermions in which the 
temperature is small compared to the Fermi energy. Then, if there is any attractive interaction between the fermions, even if it is arbitrarily small, Cooper pairing will occur. This
is Cooper's Theorem, and the underlying microscopic theory is  BCS theory, after Bardeen, Cooper and Schrieffer \cite{PhysRev.108.1175}. In an 
intuitive picture, Cooper pairing can be viewed as a way of fermions undergoing Bose-Einstein condensation. Since they are fermions, they cannot condense directly. So they pair, to form 
bosons, and then they condense. This picture has to be taken with some care, especially at weak coupling, where the size of a Cooper pair is much larger than the mean particle distance. However,  experiments with ultra-cold atoms have shown that the transition to a strongly coupled system where fermion pairs {\it are} well-separated bosons is  
continuous. Therefore, the picture of Bose-Einstein condensation of Cooper pairs is, at least qualitatively, also applicable to weak coupling.  Although best known in the 
context of electronic superconductivity, Cooper pairs do not necessarily carry electric charge. They are neutral for instance in ultra-cold fermionic gases or in $^3$He. In this case, the system 
is superfluid rather than superconducting. From a theoretical point of view, superconductors and superfluids are thus very similar. The difference lies in the charge of the Cooper pairs. Or, more formally speaking, superfluidity is associated with spontaneous symmetry breaking of a global symmetry (particle number conservation), while in a superconductor the broken 
group is local (the gauge group of electromagnetism or, as we shall see now, the color gauge group). 

Due to the generality of Cooper's Theorem, we may ask whether Cooper pairing also occurs in QCD. At high densities and low temperatures, there is a sharp Fermi sphere of quarks,
which interact by the exchange of gluons. We know -- for instance because of the existence of baryons -- that there must be an attractive channel for this interaction. 
Therefore, we may apply Cooper's Theorem and can expect the formation of Cooper pairs \cite{Barrois:1977xd,Frautschi:1978rz,Bailin:1983bm}. Here the relevant interaction is the fundamental force of QCD. This is 
in contrast to electronic systems, where the attractive force is provided in a less direct way by the interaction with the ion lattice. Despite the common and very general mechanism, there 
are, of course, important differences between quark Cooper pairing and Cooper pairing of electrons or $^3$He atoms. For instance, in three-flavor quark matter, there are $N_cN_f =9$ 
fermion species, where $N_c$ is the number of colors and $N_f$ the number of flavors. As a consequence, there is a multitude of possible pairing patterns, i.e., if we want to describe a phase of dense quark matter we have to specify who pairs with who. Moreover, in principle all nine fermion species can have different Fermi momenta. Since weak-coupling Cooper pairing occurs in a small vicinity 
of a common Fermi surface, a splitting of Fermi momenta may break some of the Cooper pairs. Indeed, due to the strange quark mass and if we impose electric charge neutrality and equilibrium with respect to the weak interactions, (unpaired) quark matter at densities relevant for neutron stars does not have a single common Fermi surface. This complication can be neglected at ultra-high densities, where the quark chemical potential $\mu$ is much larger than all three quark masses, $\mu\gg m_u\simeq m_d\simeq m_s $, which can therefore be neglected.  
In this case, (unpaired) three-flavor quark matter has equal densities of up, down and strange quarks, $n_u=n_d=n_s$. It is thus electrically neutral and in electroweak equilibrium without the presence of any electrons or positrons, and Cooper pairing is undisrupted.

Quarks carry color charge, electric charge, and baryon number, and thus we may ask whether dense quark matter is a color superconductor, an electronic superconductor, 
and/or a superfluid. The answer depends on the pairing pattern and is best discussed in terms of the broken symmetries. At ultra-high densities, the color-flavor locked (CFL) 
phase \cite{Alford:1997zt,Alford:1998mk} is energetically favored. The reason is that it is the only phase in which all nine quark species participate in pairing and thus it has the largest condensation energy. The CFL symmetry breaking pattern is
\be \label{CFLpattern}
[SU(3)_c]\times \underbrace{SU(3)_L\times SU(3)_R}_{\displaystyle{\supset[ U(1)_Q]}}\times U(1)_B \to \underbrace{SU(3)_{c+L+R}}_{\displaystyle{\supset [U(1)_{{\tilde Q} }]}}\times \mathbb{Z}_2 \, .
\ee        
Here, local symmetries are distinguished from global ones by square brackets. The left-hand side shows the symmetries of QCD: the color gauge group, the three-flavor chiral symmetry group (which is only approximate once quark masses are taken into account) and 
baryon number conservation, together with the QED gauge group, which is a subgroup of the flavor symmetry group. We have omitted the axial $U(1)_A$, 
which is broken on a quantum level at non-asymptotic densities.  The right-hand side is the symmetry group under which the CFL state is invariant. This residual group locks transformations in color and flavor space. CFL
is a color superconductor in the sense that the QCD gauge group is spontaneously broken, which manifests itself in nonzero Meissner masses of the gluons \cite{Rischke:2000ra,Schmitt:2003aa}. CFL is a superfluid because baryon number conservation 
is broken. This manifests itself for instance in the formation of vortices if CFL is rotated (possibly in the interior of neutron stars)
\cite{Forbes:2001gj,Balachandran:2005ev} and in the existence of a Goldstone mode. This Goldstone mode is exactly massless because baryon number conservation is an exact symmetry of QCD.   
The electromagnetic gauge group $U(1)_Q$ survives in a "rotated" form, where the generator 
$\tilde{Q}$ is given by a linear combination of the original generator $Q$ and one of the generators of the color gauge group $SU(3)_c$. As a consequence, one gauge boson, a mixture of a gluon and the photon, remains massless. 

The symmetry breaking pattern (\ref{CFLpattern}) also 
shows that the chiral symmetry group $SU(3)_L\times SU(3)_R$ is 
spontaneously broken. The Cooper pair condensate breaks it down to the same residual group as a chiral condensate does in low-density 
QCD. In the asymptotic limit $\mu\to\infty$, where the chiral group is an exact symmetry of QCD, the symmetry breaking pattern indicates the
existence of eight Goldstone modes. At moderate densities, these Goldstone modes acquire a (small) mass since 
non-negligible quark masses break chiral symmetry explicitly. Due to the gap in the spectrum of all quasi-fermions, the pseudo-Goldstone modes from chiral symmetry breaking and the exact superfluid Goldstone mode dominate the transport properties of CFL, and one can write down an effective low-energy theory \cite{Son:1999cm,Bedaque:2001je}, very similar to chiral perturbation theory in the QCD vacuum. 

This review article is structured as follows. In Sec.\ \ref{sec:calc}, we shall go through the weak-coupling calculation of the 
color-superconducting gap, the resulting condensation energy, and the Meissner masses of the gauge bosons. These calculations provide rigorous QCD results which, due to asymptotic freedom, apply to asymptotically large densities. We shall mostly restrict ourselves to CFL, but we will mention other possible phases and how they differ from CFL. In Sec.\ \ref{sec:cscstar} we discuss -- in a less detailed way -- the fate of the CFL phase as we go down in density. We present another detailed and exemplary calculation in Sec.\ \ref{sec:bulk}, where we calculate the bulk viscosity in the 2SC phase, making use of the quark propagators in the presence of Cooper pairing discussed in Sec.\ \ref{sec:calc}. The bulk viscosity is one example of a transport coefficient, and Sec.\ \ref{sec:bulk} points out the typical features encountered in transport of (partially) paired fermions. This section also contains a comparison to the bulk viscosity of CFL and a brief general discussion of transport in neutron stars, focusing on the effect of color superconductivity. By concentrating on the pedagogical presentation of two detailed calculations many aspects of color superconductivity can only be touched upon very briefly. More details can be found in the references and in the already existing reviews, which are recommended for further reading \cite{Rajagopal:2000wf,Buballa:2003qv,Shovkovy:2004me,Shovkovy:2005fy,Alford:2007xm,Wang:2009xf,Schmitt:2010pn,Fukushima:2010bq,Pang:2011qh,Anglani:2013gfu,Mannarelli:2014jsa}.

\section{Cooper pairing in QCD} 
\label{sec:calc}

\subsection{Quark propagator}
\label{sec:cooper}

\begin{figure}[h]
   \begin{center}
     \includegraphics[width=0.45\textwidth]{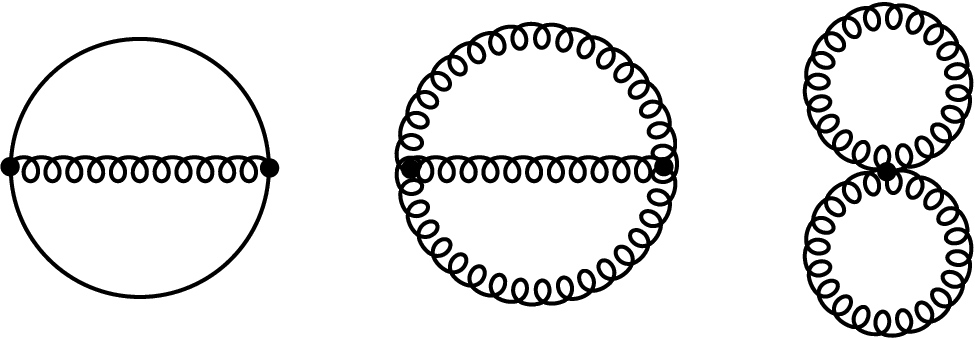}
     \end{center}
  \caption{\label{fig:2pi} Diagrams for the 2-loop truncation of the 2PI effective action. Solid (curly) lines correspond to the quark (gluon) propagator $S$ ($D$). The pressure is dominated by the left diagram because the gluon contributions are suppressed at small temperatures. Also for the other quantities calculated here the middle and right diagrams are not relevant.} 
\end{figure}

A very useful formalism for many-body systems with spontaneous symmetry breaking is the 2-particle-irreducible (2PI) formalism, also called Cornwall-Jackiw-Tomboulis (CJT) formalism \cite{Cornwall:1974vz,Baym:1962sx,Luttinger:1960ua}. The starting point is an effective action $\Gamma$, which is a functional of the quark and gluon propagators $S$ and $D$,
\be
\Gamma[S,D]=\frac{1}{2}\Tr\ln S^{-1} -\frac{1}{2}\Tr(1-S_0^{-1}S)
-\frac{1}{2}\Tr\ln D^{-1} + \frac{1}{2}\Tr(1-D_0^{-1}D) +\Gamma_2[S,D] \, , 
\ee
where $S_0$ and $D_0$ are the free (non-interacting) propagators. The factor 1/2 in front of the fermionic terms is due to the use of 
Nambu-Gorkov propagators, which is necessary in the presence of Cooper pairing, see below. The contribution $\Gamma_2[S,D]$ contains all 2-particle-irreducible diagrams, i.e., diagrams that remain connected after cutting one or two lines. Truncating at two-loop order, the relevant diagrams are shown in Fig.\ \ref{fig:2pi}.   

\begin{figure}[h]
   \begin{center}
     \includegraphics[width=0.5\textwidth]{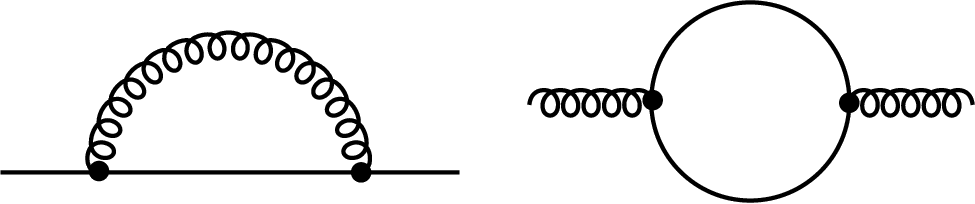}
     \end{center}
  \caption{\label{fig:SigmaPi} The quark self-energy $\Sigma$ (left diagram) is obtained by cutting a fermion line in the left diagram of 
  Fig.\ \ref{fig:2pi}. The gluon self-energy $\Pi$ is obtained by cutting a gluon line in the diagrams of Fig.\ \ref{fig:2pi}. We will only be interested in the contribution to the gluon self-energy from the quark loop (right diagram). } 
\end{figure}

The full propagators are obtained by the stationarity equations, which are obtained by 
taking the functional derivatives of the effective action,
\be
0 = \frac{\delta\Gamma}{\delta S} = \frac{\delta\Gamma}{\delta D} \, , 
\ee
which yields
\begin{subequations}
\bea
S^{-1} &=& S_0^{-1}+\Sigma \, , \label{DS} \\[2ex]
D^{-1} &=& D_0^{-1} + \Pi \, ,  \label{DS1}
\eea
\end{subequations}
where 
\be
\Sigma = 2\frac{\delta\Gamma_2}{\delta S} \, , \qquad 
\Pi = -2\frac{\delta\Gamma_2}{\delta D}
\ee
are the quark and gluon self-energies, respectively, see Fig.\ \ref{fig:SigmaPi}. 

The quark self-energy $\Sigma$ will 
be relevant for the QCD gap equation, see Sec.\ \ref{sec:gap}. The gluon self-energy $\Pi$ -- more precisely, the quark loop contribution to it -- will be needed in two instances: for the gap equation it is sufficient to 
use the so-called Hard-Dense-Loop (HDL) approximation for $\Pi$, where the quark loop contains ungapped quarks, while for the calculation of the 
screening masses from $\Pi$ the gapped quark propagator is needed, see Sec.\ \ref{sec:Meissner}. 

The grand-canonical potential (or free energy density) $\Omega$  is obtained by evaluating the effective action at the stationary point. With $\Gamma_2=\frac{1}{4}\Tr(\Sigma S)$ we obtain 
\be \label{Omega}
\Omega 
\simeq -\frac{1}{2}\Tr\ln S^{-1} +\frac{1}{4}\Tr(1-S_0^{-1}S)
\, .
\ee
Here we have neglected the contribution of the gluons because we are only interested in small temperatures $T\ll\mu$, where their 
contribution is negligible.

With the formalism at hand, we now include Cooper pairing. Cooper pairing gives rise to a nonzero expectation value of two fermion fields, so 
naively we expect a nonzero $\langle\psi\psi\rangle$, where $\psi$ is the 
quark spinor.
However, the product $\psi\psi$ is not defined (think of $\psi$ as a column vector in Dirac space). This is in contrast to the product $\bar{\psi}\psi$, which appears in the chiral condensate $\langle\bar{\psi}\psi\rangle$ and which can be understood as the product of a row vector with a column vector, yielding a scalar. Instead, Cooper pairing is described by a condensate of the form $\langle \psi\bar{\psi}_C\rangle$, where $\psi_C = C\bar{\psi}^T$ with $C=i\gamma^2\gamma^0$ is the charge-conjugate spinor. (In $\bar{\psi}_C$, first charge-conjugate, then take the Hermitian conjugate and then multiply by $\gamma^0$.) It is therefore convenient to work in an extended 
spinor space, where fermions and charge-conjugate fermions are subsumed in the so-called Nambu-Gorkov spinor
\be
\Psi = \left(\begin{array}{c}\psi\\ \psi_C \end{array}\right) \, .
\ee
As a consequence, the fermion propagators $S_0^{-1}$ and $S$ acquire an additional $2\times2$
structure. Together with Dirac, color, and flavor structure, they are therefore 
$72\times 72$ matrices ($2\times 4N_cN_f=72$). The free inverse 
propagator in Nambu-Gorkov space is
\be \label{S0inv}
S_0^{-1} = \left(\begin{array}{cc} [G_0^+]^{-1} & 0 \\ 0 & [G_0^-]^{-1} \end{array}\right)
\, ,  
\ee
where the free inverse fermion propagator and the free inverse charge-conjugate fermion propagator 
in momentum space are  
\be \label{G0pm}
[G_0^\pm]^{-1} = \gamma^\mu K_\mu \pm \mu\gamma^0 = \sum_{e=\pm}[k_0\pm(\mu-ek)]\gamma^0\Lambda_k^{\pm e} \, ,
\ee
with the four-momentum $K^\mu=(k_0,\vec{k})$, and $k=|\vec{k}|$. The temporal component is given by the fermionic Matsubara frequencies, $k_0=-i\omega_n$, with $\omega_n = (2n+1)\pi T$, where $n\in\mathbb{Z}$.  
We have restricted ourselves for simplicity to massless quarks (having in mind asymptotically large densities), and we have introduced the energy projectors
\be \label{Lambdake}
\Lambda_k^{e} = \frac{1}{2}(1+ e\gamma^0\vec{\gamma}\cdot\hat{\vec{k}}) \, , 
\ee
which are orthogonal, $\Lambda_k^+\Lambda_k^-=0$, and complete, $\Lambda_k^++\Lambda_k^-=1$. 
For the following, it is useful to note that $\gamma^0\Lambda_k^e = \Lambda_k^{-e}\gamma^0$.
Writing the propagator in terms of projectors is very useful for  inversion. For instance, from Eq.\ (\ref{G0pm}) we immediately obtain 
\be \label{G0}
G_0^\pm = \sum_e\frac{\Lambda_k^{\pm e}\gamma^0}{k_0\pm(\mu-ek)} \, .
\ee
For now, by defining the inverse Nambu-Gorkov propagator (\ref{S0inv}) we have simply doubled the degrees of freedom since fermions and charge-conjugate fermions remain uncoupled. They couple 
in the self-energy, which we write as 
\be \label{Sigma}
\Sigma = \left(\begin{array}{cc} \Sigma^+ & \Phi^- \\ \Phi^+ & \Sigma^- \end{array}\right) \, .
\ee
The diagonal elements can be approximated by \cite{Gerhold:2005uu}
\be \label{Sigdiag}
\Sigma^\pm \simeq \gamma^0\Lambda_k^\pm\frac{g^2}{18\pi^2} k_0\ln\frac{48e^2m_g^2}{\pi^2k_0^2} \, , 
\ee
where $e$ is Euler's constant, $g$ is the strong coupling constant [$\alpha_s = g^2/(4\pi)$], and 
\be \label{mgsq}
m_g^2 \equiv N_f\frac{g^2\mu^2}{6\pi^2}
\ee
is the effective gluon mass squared in a dense QCD plasma with $N_f$ quark flavors and 
$\mu \gg T$. 
The off-diagonal 
elements of the self-energy $\Sigma$ describe Cooper pairing. 
We will refer to $\Phi^+$ as the gap matrix; $\Phi^-$ is related to $\Phi^+$ via $\Phi^-=\gamma^0(\Phi^+)^\dag\gamma^0$. The gap matrix is a matrix 
in color, flavor, and Dirac space and determines the pairing pattern. We shall continue the calculation with the ansatz
\be \label{Phiplus}
\Phi^+ = \Delta {\cal M} \gamma^5 \, .
\ee
The gap function $\Delta$ depends on the four-momentum $K$ and 
will be determined 
by solving the gap equation. The Dirac structure $\gamma^5$ ensures that Cooper pairs are 
made of quarks with the same chirality and have total spin zero. In general, the Dirac structure can be more complicated, for instance 
if pairing occurs in the spin-1 channel. For a discussion of a more general Dirac structure see Refs.\ \cite{Bailin:1983bm,Pisarski:1999av}. In our ansatz Dirac and color/flavor structures factorize, such 
that ${\cal M}$ is a $9\times 9$ matrix in color/flavor space. In order to discuss the structure of ${\cal M}$ we consider the irreducible representations 
of diquarks regarding color and flavor symmetries, 
\begin{subequations} \label{SU3cf}
\bea
SU(3)_c: \qquad [3]_c\otimes[3]_c &=& [\bar{3}]_c^a\oplus[6]^s_c \, , \\[2ex]
SU(3)_f: \qquad [3]_f\otimes[3]_f &=& [\bar{3}]_f^a\oplus[6]^s_f \, .
\eea
\end{subequations}
Here, the flavor sector `$f$' represents both left-handed and right-handed sectors. 
The anti-symmetric anti-triplet channel in color space $[\bar{3}]_c^a$ is attractive. Since the overall wave function of a quark Cooper pair has to be anti-symmetric with respect to exchange of the two fermions, this determines the flavor channel: if pairing occurs in the anti-symmetric spin-0 channel, the flavor channel has to be anti-symmetric as well, and thus 
${\cal M}\in  [\bar{3}]_c^a\otimes [\bar{3}]_f^a$. The general form of the color/flavor 
structure can thus be written as 
\be \label{Mphi}
{\cal M} = \phi_{AB}J_A\otimes I_B \, , \qquad {\cal M}^{ij}_{\alpha\beta} = - \phi_{AB}\epsilon_{A\alpha\beta}\epsilon_{Bij} \, , 
\ee
where $\{J_1,J_2,J_3\}$ with $(J_A)_{\alpha\beta}=-i\epsilon_{A\alpha\beta}$ and $\{I_1,I_2,I_3\}$ with $(I_B)_{ij}=-i\epsilon_{Bij}$ are bases of $[\bar{3}]_c^a$ and 
$[\bar{3}]_f^a$, respectively. The $3\times 3$ matrix $\phi$ determines the pairing pattern
and thus the particular color-superconducting phase.  
We shall mostly be concerned with CFL, but also mention the so-called 
2SC phase,
\be \label{CFL2SC}
{\rm CFL:} \quad \phi_{AB}=\delta_{AB} \, , \qquad {\rm 2SC:} \quad \phi_{AB}=\delta_{A3}\delta_{B3} \, .
\ee
In the 2SC phase, quarks of one color, say blue, remain unpaired, as well as the quarks of one flavor. If all quark masses can be neglected (and if we consider pure QCD, i.e., if the electric charges of the quarks play no role) there is no difference between the flavors and we can randomly pick one quark flavor that remains unpaired. However, the 2SC phase becomes particularly interesting if the strange quark mass is taken into account. In that case, the 2SC phase 
usually refers to the phase where strange quarks remain unpaired. 
As we shall see below, at asymptotically large densities the CFL phase is favored. (In the presence of a large magnetic field the 2SC phase may become favored even in the massless limit \cite{Haber:2017oqb}.) One can check that the CFL order parameter in Eq.\ (\ref{CFL2SC}) gives rise to
the symmetry breaking pattern (\ref{CFLpattern}). 

\begin{figure}[h]
   \begin{center}
     \includegraphics[width=0.45\textwidth]{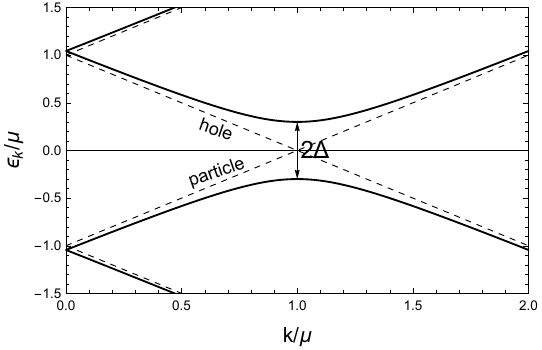}
     \end{center}
  \caption{\label{fig:dispersion} Dispersion relations with (solid) and without (dashed) Cooper pairing. In the absence of Cooper pairing, quasi-fermions can be excited with infinitesimally small energy at the Fermi surface. If Cooper pairs have formed, an energy $2\Delta$ is necessary to excite quasi-fermions. These quasi-fermions are (momentum-dependent) mixtures of particles and holes. Also anti-particle (uppermost branch) and  anti-hole (lowermost branch) excitations are shown for completeness.} 
\end{figure}

We may now compute the full quark propagator. We shall do so for the CFL phase and neglect the diagonal elements of the self-energy, $\Sigma^\pm\simeq 0$. Their effect is not very crucial for the understanding of the main physics, and neglecting them renders the calculation simpler. 
At the end, for completeness we will reinstate their (subleading) 
effect on the superconducting gap. With this simplification, putting together equations (\ref{DS}), (\ref{S0inv}), and (\ref{Sigma}), we obtain the inverse propagator
\be \label{Sinv}
S^{-1} = \left(\begin{array}{cc} [G_0^+]^{-1} & \Phi^- \\ \Phi^+ & [G_0^-]^{-1} \end{array}\right) \, .
\ee
Inversion of this matrix yields 
\be \label{Sfull}
S = \left(\begin{array}{cc} G^+ & F^- \\ F^+ & G^- \end{array}\right) \, , 
\ee
with 
\begin{subequations} \label{GpmFpm}
\bea
G^\pm &=& \left([G_0^\pm]^{-1} - \Phi^{\mp}G_0^\mp \Phi^\pm\right)^{-1} \, , \label{Gpm}\\[2ex]
F^\pm &=& -G_0^\mp \Phi^\pm G^\pm \, .
\eea
\end{subequations}
The off-diagonal elements in Nambu-Gorkov space $F^\pm$ are called anomalous propagators. 
They are only nonzero in the presence of Cooper pairing and describe (upper sign) the propagation of a charge-conjugate fermion which turns into a propagating fermion through
the Cooper pair condensate. This is a manifestation of the spontaneous breaking of baryon number conservation: fermions can be created and annihilated by the condensate. 
In order to perform the inversion in Eq.\ (\ref{Gpm}) it is useful to write the matrix 
${\cal M}^\dag{\cal M}$ (for CFL, ${\cal M}^\dag={\cal M}$) in its spectral representation,
\be
{\cal M}^2 = \lambda_1{\cal P}_1 + \lambda_2{\cal P}_2 \, , 
\ee
where $\lambda_1=1$ and $\lambda_2=4$ are the eigenvalues of ${\cal M}^2$, and ${\cal P}_1$ 
and ${\cal P}_2$ are the projectors onto the corresponding eigenspaces,
\be
{\cal P}_1=-\frac{{\cal M}^2-4}{3} \,, \qquad {\cal P}_2 = \frac{{\cal M}^2-1}{3} \, .
\ee
The degeneracies of the eigenvalues are $\Tr({\cal P}_1)=8$ and $\Tr({\cal P}_2)=1$. 
In terms of color and flavor indices ($\alpha,\beta$ and $i,j$, respectively) we have
\be \label{Mij}
{\cal M}_{\alpha\beta}^{ij} = \delta_\alpha^j\delta_\beta^i-\delta_\alpha^i\delta_\beta^j \quad \Rightarrow\qquad ({\cal M}^2)_{\alpha\beta}^{ij} = \delta_\alpha^i\delta_\beta^j+\delta_{\alpha\beta}\delta^{ij} \, , 
\ee
which yields
\be \label{P1P2ij}
{\cal P}_1 = \delta_{\alpha\beta}\delta^{ij}-\frac{1}{3}\delta_\alpha^i\delta_\beta^j \, ,\qquad 
{\cal P}_2 = \frac{1}{3}\delta_\alpha^i\delta_\beta^j \, .
\ee
From these expressions one finds ${\cal M}^4=5{\cal M}^2-4$,  which can be used to show that the projectors are orthogonal, ${\cal P}_1{\cal P}_2=0$. They are, obviously, also complete, ${\cal P}_1+{\cal P}_2=1$. With the help of these projectors we compute from Eq.\ (\ref{GpmFpm})
\begin{subequations} \label{GF}
\bea
G^\pm &=& \sum_{e=\pm}\sum_{r=1,2}\frac{[k_0\mp (\mu-ek)]{\cal P}_r\gamma^0\Lambda_k^{\mp e}}{k_0^2-(\epsilon_{k,r}^e)^2} \, , \label{GFone}\\[2ex]
F^\pm &=& \sum_{e=\pm}\sum_{r=1,2}\frac{\Phi^\pm{\cal P}_r\Lambda_k^{\mp e}}{k_0^2-(\epsilon_{k,r}^e)^2} \, , \label{anomalous}
\eea
\end{subequations}
where 
\be \label{dispersion}
\epsilon_{k,r}^e=\sqrt{(\mu-ek)^2+\lambda_r \Delta^2} \, .
\ee
This result shows that the propagator has poles at $k_0=\pm \epsilon_{k,r}^e$.
These are the quasi-particle and quasi-antiparticle excitations (upper sign) and 
quasi-hole and quasi-antihole excitations (lower sign), see Fig.\ \ref{fig:dispersion}. 

The structure of these excitations is 
solely determined by the pairing pattern, i.e., 
the spectrum of the matrix ${\cal M}^2$: in the CFL
phase, 8 quasi-particles are gapped with gap $\Delta$, while one quasi-particle is gapped 
with gap $2\Delta$. From the matrix ${\cal M}$ one can also read off the 
pairing pattern: if we label the color directions by $r$, $g$, $b$ and the flavor directions with $u$, $d$, $s$, then we have pairing of $rd$ with $gu$, of $bu$ with $rs$, of $bd$ with $gs$, and pairing among the three quark species $ru$, $gd$, $bs$. 
The calculation for the 2SC phase is completely analogous and yields 4 gapped quasi-particles and 5 ungapped quasi-particles. Here, $ru$ pairs with $gd$, $rd$ with $gu$, while $bu$, $bd$, $bs$, $rs$, $gs$ remain unpaired.

\subsection{Solving the gap equation}
\label{sec:gap}

\begin{figure}[h]
   \begin{center}
     \includegraphics[width=0.5\textwidth]{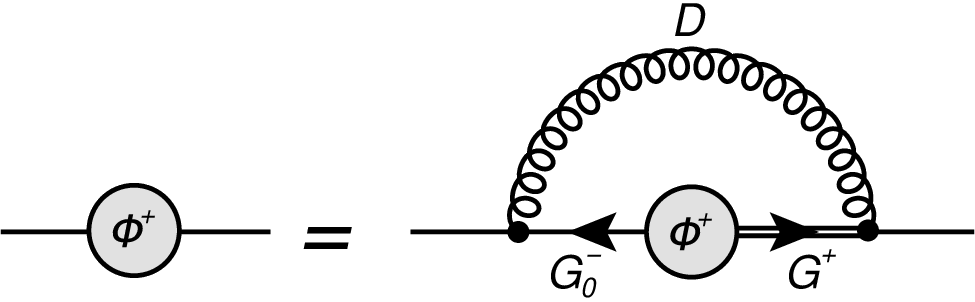}
     \end{center}
  \caption{\label{fig:gapeq} Diagrammatic representation of the QCD gap equation.
  The right-hand side is the (lower) off-diagonal element in Nambu-Gorkov space of the left diagram in Fig.\ \ref{fig:SigmaPi}, containing the 
  anomalous propagator $F^+=-G_0^-\Phi^+G^+$. The vertices are given by $g\gamma^\mu T_a^T$ and 
  $g\gamma^\nu T_b$, whose indices are contracted by the gluon propagator $D_{\mu\nu}^{ab}$. } 
\end{figure}

To compute the gap function $\Delta(K)$ we go back to the quark-self energy $\Sigma$, which is 
given by the left diagram in Fig.\ \ref{fig:SigmaPi}. This diagram contains the full propagator in 
Nambu-Gorkov space, i.e., it has to be understood as a $2\times 2$ matrix, whose four elements are given by the quark propagators $G^\pm$ and $F^\pm$, see Eq.\ (\ref{Sfull}). On the other hand, the self-energy is given by the gap matrix according to Eq.\ (\ref{Sigma}). Therefore, 
the lower off-diagonal element in Nambu-Gorkov space relates the gap matrix to the self-energy diagram. This is the gap equation, which is shown diagrammatically in Fig.\ \ref{fig:gapeq} (equivalently, we could have considered the upper off-diagonal element). 

The algebraic version of the gap equation is
\be \label{gapeq1}
\Phi^+(K) = g^2\frac{T}{V}\sum_Q\gamma^\mu T_a^TF^+(Q)\gamma^\nu T_bD_{\mu\nu}^{ab}(P) \, ,
\ee
where $V$ is the volume of the system, where $T_a=\lambda_a/2$ with the Gell-Mann matrices $\lambda_a$ ($a=1,\ldots,8$), and where  
we have denoted the gluon four-momentum by $P\equiv K-Q$. For our purposes we may assume the gluon propagator to be diagonal 
in color space, $D_{\mu\nu}^{ab} = \delta^{ab} D_{\mu\nu}$, and 
employ the HDL 
approximation. In this approximation, the gluon-self 
energy is computed from a quark loop where the quark momentum 
is `hard', i.e., of the order of the chemical potential, and much larger than the 
`soft' gluon energy and momenta. In particular, in this approximation, the effect of Cooper pairing is not taken into account. In other words, in computing the color-superconducting gap, the backreaction of the gap on the gluon propagator, which in turn affects the gap, can be neglected
up to the order we will be working \cite{Rischke:2001py}. 

In this 
approximation the gluon self-energy $\Pi$ is (4-)transverse to the gluon momentum $P$, as in abelian gauge theories. This allows us to decompose the self-energy into (3-)longitudinal and (3-)transverse components with respect to $\vec{q}$,
\be \label{PiFG}
\Pi_{\mu\nu} = {\cal F}P_{L,\mu\nu}+{\cal G}P_{T,\mu\nu} \, , 
\ee
where transverse and longitudinal projectors are defined as
\begin{subequations}
\bea
P_{T,00}&=&P_{T,0i}=P_{T,i0}=0 \, , \qquad P_{T,ij}=\delta_{ij}-\hat{p}_i\hat{p}_j \, , \\[2ex]
P_{L,\mu\nu} &=& \frac{P_\mu P_\nu}{P^2}-g_{\mu\nu}-P_{T,\mu\nu} \, .
\eea
\end{subequations}
The components ${\cal F}$ and ${\cal G}$ can be computed from the self-energy by appropriate projections, 
\be \label{FPi00}
{\cal F} = \frac{P^2}{p^2}\Pi_{00} \, , \qquad {\cal G} = \frac{1}{2}(\delta_{ij}-\hat{p}_i\hat{p}_j)\Pi_{ji} \, .
\ee
In the HDL approximation \cite{Bellac:2011kqa},
\begin{subequations} \label{calFG}
\bea
{\cal F}(P) &=& -3m_g^2\frac{P^2}{p^2}\left(1-\frac{p_0}{2p}\ln\frac{p_0+p}{p_0-p}\right) \,  , \\[2ex]
{\cal G}(P) &=& \frac{3m_g^2p_0}{2p}\left(\frac{p_0}{p}-\frac{P^2}{2p^2}\ln\frac{p_0+p}{p_0-p}\right) \,  , 
\eea
\end{subequations}
with the effective gluon mass $m_g$ from Eq.\ (\ref{mgsq}).
In Coulomb gauge, the gluon propagator in the HDL approximation is 
\be \label{DLDT}
D_{00}=D_L \, \qquad D_{0i}=0 \, , \qquad D_{ij} = (\delta_{ij}-\hat{p}_i\hat{p}_j)D_T \, ,
\ee
where the longitudinal and transverse components are 
\be \label{DLT}
D_L(P) = \frac{P^2}{p^2}\frac{1}{{\cal F}(P)-P^2} \, , \qquad 
D_T(P) = \frac{1}{{\cal G}(P)-P^2} \, .
\ee
The physical meaning of this form of the propagator is best understood by computing the 
corresponding spectral densities, which are given by the imaginary part of the retarded 
propagators,
\be
\rho_{L,T} = \frac{1}{\pi}\lim_{\epsilon\to0} {\rm Im}\,D_{L,T}(p_0+i\epsilon,\vec{p}) \, ,
\ee
where $\epsilon>0$. With Eqs.\ (\ref{calFG}) and (\ref{DLT}) one computes
\begin{subequations} \label{rhoLT}
\bea
\rho_L &=& \frac{\omega_L(\omega_L^2-p^2)\delta(p_0-\omega_L)}{p^2(3m_g^2+p^2-\omega_L^2)}
-\frac{1}{\pi}\frac{P^2}{p^2}\frac{\Theta(p-p_0){\rm Im}\,{\cal F}}{({\rm Re}\, {\cal F} -P^2)^2+({\rm Im}\,{\cal F})^2}  \, , \\[2ex]
\rho_T &=& \frac{\omega_T(\omega_T^2-p^2)\delta(p_0-\omega_T)}{3m_g^2\omega_T^2-(\omega_T^2-q^2)^2} -\frac{1}{\pi}\frac{\Theta(p-p_0){\rm Im}\,{\cal G}}{({\rm Re}\, {\cal G} -P^2)^2+({\rm Im}\,{\cal G})^2} \, ,
\eea
\end{subequations}
where $\omega_L$ and $\omega_T$ are the poles of the propagator, i.e., the 
numerical solutions for $p_0$ of $\Pi_{00}-p^2=0$ and ${\cal G}-P^2=0$, respectively.
We see that the gluon propagator exhibits two quasi-particle excitations in the timelike 
regime, $P^2>0$, given by the delta peak in the spectral density. These are the longitudinal and transverse plasmon modes. In addition, the spectral density is nonzero in the entire spacelike regime, $P^2<0$. These are Landau-damped 
gluons, unstable excitations due to the scattering with the quarks in the 
plasma, which is a well-known phenomenon also for photons in a 
QED plasma. In fact, in the given HDL approximation, the diagonal elements of the gluon propagator 
are identical to the photon propagator in an electromagnetic plasma.

Coming back to the gap equation (\ref{gapeq1}), we insert the anomalous propagator (\ref{anomalous}) and neglect the antiparticle contribution, i.e., we drop the term with $e=-$ and from now on denote $\epsilon_{k,r}\equiv\epsilon_{k,r}^+$. 
Moreover, we use the form of the gluon propagator from Eq.\ (\ref{DLDT}), multiply both sides of the gap equation by $\gamma^5{\cal M}\Lambda_k^+$ from the right and take the trace over color, flavor, and Dirac space. With the 
color/flavor traces 
\be
\Tr[T_a^T{\cal M}{\cal P}_1 T_a{\cal M}]=2\Tr[T_a^T{\cal M}{\cal P}_2T_a{\cal M}]=-\frac{16}{3}
\,,
\ee
and the Dirac traces
\begin{subequations}
\bea
\Tr[\gamma^0\gamma^5\Lambda_q^{-}\gamma^0\gamma^5\Lambda_k^+] &=& -(1+\hat{\vec{q}}\cdot\hat{\vec{k}}) \, , \\[2ex]
\Tr[\gamma^i\gamma^5\Lambda_q^-\gamma^j\gamma^5\Lambda_k^+] &=& \delta_{ij}(1-\hat{\vec{q}}\cdot\hat{\vec{k}})+\hat{q}_i\hat{k}_j+\hat{q}_j\hat{k}_i \, , 
\eea
\end{subequations}
we find
\bea \label{DelKg}
\Delta(K) &=& \frac{g^2}{3}\frac{T}{V}\sum_Q\left(\frac{2}{3}\frac{\Delta(Q)}{q_0^2-\epsilon_{q,1}^2}
+\frac{1}{3}\frac{\Delta(Q)}{q_0^2-\epsilon_{q,2}^2}\right) \left[(1+\hat{\vec{q}}\cdot\hat{\vec{k}})D_L(P)
-2(1-\hat{\vec{p}}\cdot\hat{\vec{q}}\,\hat{\vec{p}}\cdot\hat{\vec{k}}) D_T(P)\right] \, .
\eea
The sum over 4-momenta $Q$ contains the sum over Matsubara frequencies and the sum over 3-momenta, which turns into an integral in the thermodynamic limit, 
\bea
\frac{T}{V}\sum_Q \to 
T\sum_{q_0}\int\frac{d^3\vec{q}}{(2\pi)^3} \,. 
\eea
It is instructive to discuss the gap equation (\ref{DelKg}) first in the simple case of a pointlike interaction 
between fermions. So, let us (unrealistically) assume that quarks interact by the exchange of a heavy scalar field with mass $M$ such that we can approximate $D_L\simeq - 1/M^2$ and $D_T=0$. In this case, the sum over the fermionic Matsubara frequencies is easily performed with the help of 
\be \label{matsu1}
T\sum_{q_0}\frac{\Delta(Q)}{q_0^2-\epsilon_q^2} = -\frac{\Delta_q}{2\epsilon_q}\tanh\frac{\epsilon_q}{2T} \, ,
\ee
where $\Delta_q\equiv \Delta(\epsilon_q,\vec{q})$. 
Moreover, we can simply ignore the energy and momentum dependence of the gap function and thus 
divide the whole equation by $\Delta$. ($\Delta=0$ is obviously a solution to the gap equation for all temperatures. Eventually one has to check that the phase with a nonzero $\Delta$ has lower free energy than the unpaired phase.)  We perform the (trivial) angular integral and restrict the integral over momenta $q$ to a small vicinity of the Fermi surface,
$q\in[\mu-\delta,\mu+\delta]$, where $\delta$ is much larger than the gap but much smaller 
than the Fermi momentum, $\Delta\ll\delta\ll\mu$. This allows us to approximate the 
integration measure by $dq\, q^2 \simeq \mu^2 dq$. Finally, with the new integration variable $\xi=q-\mu$ we obtain
\be \label{gapeqT}
1\simeq \frac{g^2}{c}\int_0^\delta d\xi\left(\frac{2}{3\epsilon_{q,1}}\tanh\frac{\epsilon_{q,1}}{2T}+\frac{1}{3\epsilon_{q,2}}\tanh\frac{\epsilon_{q,2}}{2T}\right) \, ,
\ee
with the dimensionless constant $c\equiv 6M^2\pi^2/\mu^2$. At zero temperature, 
we have $\tanh\frac{\epsilon_{q,r}}{2T}=1$ and 
we can perform the resulting integral analytically. Denoting the 
zero-temperature gap by $\Delta_0$ and using $\delta\gg\Delta_0$, 
we find 
\be \label{Dpointlike}
\mbox{pointlike interaction:} \qquad \Delta_0 = 2\delta\cdot 2^{-1/3} \exp\left(-\frac{c}{g^2}\right) \, .
\ee
This detour has shown us, firstly, how to solve the gap equation in the BCS scenario, where there are no long-range interactions. Notice that the effective dimensional reduction of the momentum integral was crucial and can thus be considered as the formal reason for the instability towards Cooper pair condensation. Secondly, the result shows the BCS behavior
of the gap: at weak coupling -- where this calculation is valid --  the 
gap is exponentially suppressed. The result is non-perturbative in the sense that there is no 
Taylor expansion of $\Delta_0$ for small $g$. By solving the gap equation, we have implicitly resummed infinitely many diagrams since the quark self energy, from which the gap is computed, contains the full quark propagator, which in turn contains the gap etc. Thirdly, we see that the two-gap structure of the CFL phase creates a factor $2^{-1/3}$ in the gap. Below, we shall quote the general form of this factor, assuming a 
two-gap structure but keeping the degeneracies and prefactors of the gap 
in the dispersion relation general. 

One can use Eq.\ (\ref{gapeqT}) also to compute the temperature dependence of the 
gap. This has to be done numerically, and one finds a monotonically decreasing gap which goes to zero at a certain temperature, the critical temperature $T_c$. This temperature can be calculated analytically. To this end, we 
set $\Delta=0$ in the excitation energies, such that $\epsilon_{q,1}=\epsilon_{q,2}=\xi$. As a 
consequence, the two-gap structure disappears and we have
\be 
1=\frac{g^2}{c}\int_0^{\delta}\frac{d\xi}{\xi}\tanh\frac{\xi}{2T_c} \, .
\ee
With the new integration variable $z=\xi/(2T_c)$ we compute via partial integration 
\be\label{Tc1}
\frac{c}{g^2}=\ln z\,\tanh z\Big|_0^{\delta/(2T_c)}-\int_0^{\delta/(2T_c)}dz\frac{\ln z}{\cosh^2z}
\simeq \ln \frac{\delta}{2T_c}+\gamma-\ln\frac{\pi}{4} \, , 
\ee
where $\gamma\simeq 0.577$ is the Euler-Mascheroni constant, and where we have used $\delta\gg T_c$, such that the upper boundary of the integral can be approximated by infinity. This is a consistent assumption since, as we shall see now, 
the critical temperature is of the same order as the zero-temperature gap. Solving 
Eq.\ (\ref{Tc1}) for $T_c$ yields 
\be \label{TcBCS}
T_c = \frac{e^\gamma}{\pi}2\delta e^{-c/g^2} = \frac{e^\gamma}{\pi}2^{1/3}\Delta_0 \simeq 2^{1/3}\cdot 0.567\Delta_0 \, . 
\ee
We see that the nontrivial factor due to the two-gap structure, 
in this case $2^{1/3}$, appears in the relation between $T_c$ 
and $\Delta_0$, but not if $T_c$ is expressed directly in terms of the coupling constant.  

Let us return to QCD and realistic gluons. For brevity we ignore the two-gap 
structure and set $\epsilon_{1,q}=\epsilon_{2,q}\equiv \epsilon_q$. Its effect on the QCD gap is the same as just shown for the BCS case. We may thus simply reinstate the resulting factor in the final result. It is convenient to work with the spectral representation of the gluon propagator, based on the spectral densities (\ref{rhoLT}). Details of this procedure are worked out in Ref.\ \cite{Pisarski:1999tv}. It turns out that 
the gap equation is sensitive to gluon momenta with $p_0\ll p$. This allows us to simplify the gluon propagator from the beginning,
\begin{subequations}
\bea
D_L(p) &\simeq& -\frac{1}{p^2+3m_g^2} \,  , \label{DLsim}\\[2ex]
D_T(p_0,p) &\simeq& \frac{\Theta(p-M_g)}{p^2}+\frac{\Theta(M_g-p)p^4}{p^6+M_g^4p_0^2} \, ,
\label{DTsim}
\eea
\end{subequations}
where $M_g^2\equiv 3\pi m_g^2/4$. 
This yields, after performing the Matsubara sum and the (now nontrivial) angular integral,
\bea \label{gapeq2}
\Delta_k \simeq \frac{g^2}{24\pi^2}\int_{\mu-\delta}^{\mu+\delta} dq\frac{\Delta_q}{\epsilon_q}
\left(\ln\frac{4\mu^2}{3m_g^2}+\ln\frac{4\mu^2}{M_g^2}+\frac{1}{3}\ln\frac{M_g^2}{|\epsilon_q^2-\epsilon_k^2|} \right) \tanh\frac{\epsilon_q}{2T}\, .
\eea
The three terms in parentheses arise, respectively, from static electric gluons (\ref{DLsim}), 
non-static magnetic gluons [first term of (\ref{DTsim})] and almost static, Landau-damped
magnetic gluons [second term in (\ref{DTsim})]. Combining the 
three logarithms, we obtain
\bea \label{Deltak}
\Delta_k \simeq \bar{g}^2 \int_0^{\delta} d(q-\mu) \frac{\Delta_q}{\epsilon_q}
\frac{1}{2}\ln\frac{b^2\mu^2}{|\epsilon_q^2-\epsilon_k^2|}\tanh\frac{\epsilon_q}{2T} \, ,
\eea
where
\be
\bar{g} \equiv \frac{g}{3\sqrt{2}\pi} \, , \qquad b\equiv 256\pi^4\left(\frac{2}{N_fg^2}\right)^{5/2} \, .
\ee
We can now approximate the logarithm by 
\be
\frac{1}{2} \ln\frac{b^2\mu^2}{|\epsilon_q^2-\epsilon_k^2|} \simeq \Theta(k-q)\ln\frac{b\mu}{\epsilon_k} +\Theta(q-k)\ln\frac{b\mu}{\epsilon_q} \, , 
\ee
and introduce the logarithmic integration variable 
\be
y\equiv \bar{g}\ln\frac{2b\mu}{q-\mu+\epsilon_q} \, . 
\ee
With 
\be
x\equiv \bar{g}\ln\frac{2b\mu}{k-\mu+\epsilon_k} \, , \qquad x^*\equiv \bar{g}\ln\frac{2b\mu}{\Delta_0} \, , \qquad x_0\equiv 
\bar{g}\ln\frac{b\mu}{\delta} \, , 
\ee
where $\Delta_0$ is the zero-temperature value of the gap at the Fermi surface, the zero-temperature
gap equation can be written as 
\be \label{gapeqxy}
\Delta(x) \simeq x\int_x^{x^*}dy\, \Delta(y) + \int_{x_0}^x dy\, y\,\Delta(y) \, .
\ee
Differentiating twice with respect to $x$, this becomes a 
differential equation, $\Delta''(x) = -\Delta(x)$,
with boundary conditions $\Delta(x^*)=\Delta_0$ and 
$\Delta'(x^*)=0$ (the gap peaks at the Fermi surface). This yields $\Delta(x) = \Delta_0\cos(x^*-x)$ and by 
reinserting this solution into the integral version of the gap equation (\ref{gapeqxy})
one finds $\cos x^*\simeq 0$ and thus $\Delta(x) \simeq \Delta_0\sin x$. The gap at the 
Fermi surface $\Delta_0$ is then found from $x^*\simeq 1$, which yields
the final result
\be \label{QCDgap}
\Delta_0 = 2b\mu e^{b_0'}e^{-\zeta}e^{-d}\exp\left(-\frac{3\pi^2}{\sqrt{2} g}\right) \, .
\ee
Here we have included various effects that we have omitted in our derivation. Our derivation would yield the same result with $b_0'=d=\zeta=0$. The most important feature of the weak-coupling QCD gap is the leading behavior in $g$, which 
is different compared to a four-fermion interaction: comparing with Eq.\ (\ref{Dpointlike}) we see that the QCD gap is parametrically enhanced due to the different power of $g$ in the exponential \cite{Barrois:1979pv,Son:1998uk}. This behavior arises from the exchange of almost static magnetic gluons [third term in Eq.\ (\ref{gapeq1})]. Due to the (one-loop) running of the coupling,
$1/g^2$ increases logarithmically with $\mu$ at asymptotically large $\mu$. 
Therefore, $\exp(-{\rm const}/g)$ decreases more slowly than $1/\mu$ and as a consequence 
$\Delta_0$ {\it increases} with $\mu$ at asymptotically large $\mu$, while $\Delta_0/\mu$ 
decreases. 

The origin and values of the prefactors of the gap (\ref{QCDgap}) are as follows:
\begin{itemize}
\item The factor 
\be
e^{b_0'} = \exp\left(-\frac{\pi^2+4}{8}\right) \simeq 0.177 
\ee
arises if the quark self-energy (\ref{Sigdiag}) is carried through the calculation of the gap \cite{Wang:2001aq}.

\item  From Eq.\ (\ref{Dpointlike}) we know that the two-gap structure of CFL generates a prefactor $2^{-1/3}$. This factor can be written more generally as $e^{-\zeta}$ with
\be
\zeta = \frac{1}{2}\frac{\langle\Tr({\cal P}_1)\lambda_1\ln\lambda_1 +\Tr({\cal P}_2)\lambda_2\ln\lambda_2\rangle}{\langle\Tr({\cal P}_1)\lambda_1 +\Tr({\cal P}_2)\lambda_2\rangle} \, .
\ee
Using the same notation as above, $\lambda_1$ and $\lambda_2$ are the eigenvalues of ${\cal M}{\cal M}^\dag$, while ${\cal P}_1$ and ${\cal P}_2$ are the projectors onto the respective eigenspaces, such that $\Tr({\cal P}_1)$ and $\Tr({\cal P}_2)$ are the degeneracies. For CFL, we verify that with $\lambda_1=1$, $\lambda_2=4$, $\Tr({\cal P}_1)=8$ and $\Tr({\cal P}_2)=1$ we obtain $e^{-\zeta}=2^{-1/3}$. If there is only a single isotropic gap, as in the 2SC phase, we have $\lambda_1=1$ and $\lambda_2=0$, and thus the prefactor becomes trivial, $e^{-\zeta}=1$. The general formula also allows for anisotropic gaps, with the angular brackets denoting angular average. This becomes relevant for spin-1 Cooper pairing, where ${\cal M}$ includes the spin structure of the gap matrix, such that the eigenvalues $\lambda_r$ contain the angular dependence of the gap in momentum space. In this case, $e^{-\zeta}$ becomes nontrivial even for a single gap \cite{Schmitt:2004et}.

\item Finally, the factor $e^{-d}$ becomes nontrivial only for spin-1 
Cooper pairing. This factor, in contrast to the factor $e^{-\zeta}$, 
does depend on the details of the interaction. Depending on the 
particular spin-1 phase, one finds values of $d$ varying 
between $d=6$ for pairing of quarks with the same chirality and $d=4.5$ for opposite-chirality pairing \cite{Schafer:2000tw,Schmitt:2004et}. This suppresses the weak-coupling 
gap in the spin-1 channel by 2-3 orders of magnitude compared to the spin-0 channel.

\end{itemize}

The zero-temperature gap (\ref{QCDgap}) is a result from first principles, systematically collecting all 
contributions to $\ln(\Delta_0/\mu)$ of order 
$g^{-1}$, $\ln g$ , and $g^0$. (Higher order contributions arise for instance from terms of the HDL gluon propagator that we have neglected 
or terms beyond the HDL approximation, but also from going beyond the one-loop truncation of the quark self-energy that we have used to derive the gap equation, for instance from vertex corrections.) The result is valid for sufficiently large chemical potentials; it was estimated that it starts to lose its validity at around $\mu\lesssim 10^5\, {\rm GeV}$ \cite{Rajagopal:2000rs}. This corresponds to chemical potentials 
many orders of magnitude larger than in the center of neutron stars. To get an idea of the size of the gap at more moderate densities we can extrapolate the QCD result down in density. According to the two-loop running of the coupling constant, we estimate that 
for neutron star interiors with $\mu\sim 400\, {\rm MeV}$ we have 
$\alpha_s\sim 1$ and thus $g\sim 3.5$. This yields a color-superconducting gap of $\Delta_0\sim 10\, {\rm MeV}$. Given that this is a bold extrapolation over many orders of magnitude, it is reassuring that phenomenological models such as the Nambu-Jona-Lasinio (NJL) model \cite{Nambu:1961tp,Nambu:1961fr} give similar
-- although somewhat larger -- results, $\Delta_0\sim (10-100)\, {\rm MeV}$
\cite{Buballa:2001gj,Steiner:2002gx,Abuki:2004zk,Blaschke:2005uj,Ruster:2005jc,Warringa:2006dk,Gholami:2024diy}. These numbers suggest that the energy gap from quark Cooper pairing is significantly larger than 
the corresponding quantity from Cooper pairing of nucleons, which is calculated to be of the order of $1\, {\rm MeV}$ at most 
\cite{Wambach:1992ik,Schulze:1996zz,Fabrocini:2005lvb,Cao:2006gq}. 

It is of course desirable to understand the color-superconducting gap better at moderate densities, beyond comparing the weak-coupling extrapolation with phenomenological models. Possible approaches to capture strong-coupling effects are based on Dyson-Schwinger equations    \cite{Nickel:2006vf,Marhauser:2006hy,Muller:2013pya,Muller:2016fdr,Murgana:2025dfa} and the functional renormalization group \cite{Braun:2021uua,Gholami:2025afm}. It has also been attempted to treat color superconductivity (or at least a simplified variant thereof) within the gauge-gravity correspondence, which makes the strong-coupling limit accessible; currently, however, only in theories (very) different from QCD \cite{Chen:2009kx,Basu:2011yg,BitaghsirFadafan:2018iqr,Faedo:2018fjw,Preau:2025ubr,CruzRojas:2025fzs}. It was also pointed out that it is possible to obtain an upper limit for $\Delta_0$ from a completely different approach, independent of any model description and perturbative QCD calculation: by asking for the maximal change to the equation of state due to quark Cooper pairing that is thermodynamically consistent with the known low-density nuclear matter equation of state and with known astrophysical constraints from neutron star data, it was estimated within a Bayesian analysis that $\Delta_0\lesssim 450\, {\rm MeV}$ at $\mu = 2.6\, {\rm GeV}$ \cite{Kurkela:2024xfh}.

One can also use the gap equation (\ref{Deltak}) to compute the critical temperature for color superconductivity. Details of this calculation can be found for instance in Ref.\ \cite{Schmitt:2002sc}. Expressed in terms of the zero-temperature gap, the result is the same as 
derived above for the pointlike interaction, see right-hand side of Eq.\ (\ref{TcBCS}). Therefore, the critical temperature of CFL is $T_c = e^\gamma/\pi\,2^{1/3} \Delta_0 \simeq 0.72\,  \Delta_0$. With the above estimate 
of $\Delta_0\sim 10\, {\rm MeV}$ or larger, this suggests that the temperature in the 
interior of neutron stars is -- except for the very early stages of the life of the star \cite{Prakash:2000jr} -- much smaller than the critical temperature of color superconductors. Even in neutron star mergers, where the temperature can be a few tens of MeV \cite{Hammond:2021vtv}, it is conceivable that quark Cooper pairing survives. 

At asymptotically large densities, color superconductors are type-I superconductors since the coherence length
$\xi\sim 1/\Delta$ is much larger than the penetration depth $\lambda \simeq 1/(g\mu)$ and thus the Ginzburg-Landau parameter $\kappa =\lambda/\xi \ll 1/\sqrt{2}$. This is relevant 
for the critical temperature because it is known from electronic superconductors that gauge field fluctuations become important in the type-I regime. As a consequence, the 
finite-temperature phase transition becomes first order and the critical temperature receives a correction of order $g$. The corrected critical temperature $T_c^*$ for color superconductivity at weak coupling is \cite{Giannakis:2004xt},
\be
T_c^*=T_c\left(1+\frac{\pi^2}{12\sqrt{2}}g\right) \, .
\ee

\subsection{Condensation energy}
\label{sec:condE}

We have seen that the gap equation has two solutions: the trivial one,
corresponding to the unpaired phase, and the nontrivial one, corresponding to the formation of a Cooper pair condensate. To establish that the color-superconducting phase 
is indeed the ground state of QCD at large densities, we have to compute its free energy and check whether it is lower than the free energy of the unpaired phase. We shall go through 
this calculation in this section. In principle, there could also be other phases whose free energy competes with the unpaired and Cooper-paired phases. For instance, for QCD with 
a large number of colors, we know that the so-called chiral density wave \cite{Deryagin:1992rw}, which is based on 
fermion-hole pairing (as opposed to fermion-fermion pairing) becomes a strong contender. It has been estimated, however, that at $\mu \sim 1\, {\rm GeV}$ this phase is preferred over color superconductivity 
only for $N_c\gtrsim 1000$ \cite{Shuster:1999tn}. To find the ground state, we also must compare different color-superconducting phases, which we will be able to do with the result derived in this section.

The free energy density is calculated from Eq.\ (\ref{Omega}). Let us calculate the two terms in this 
equation separately. As for the gap equation, we shall continue working in the limit of massless quarks and neglect the diagonal components of the 
quark self-energy for simplicity, $\Sigma^\pm\simeq 0$. With $\Tr\ln=\ln{\rm det}$ and 
\be
{\rm det} = \left(\begin{array}{cc} A&B\\ C&D\end{array}\right) = {\rm det}(AD-BD^{-1}CD) 
\ee
for matrices $A,B,C$ and an invertible matrix $D$, we find with Eq.\ (\ref{Sinv})
\be
\frac{1}{2}{\rm Tr}\ln \frac{S^{-1}}{T^2} = \frac{1}{2}\Tr\ln\frac{[G_0^+]^{-1}[G_0^-]^{-1}-\Phi^-G_0^-\Phi^+[G_0^-]^{-1}}{T^2} \, .
\ee
We have thus performed the trace over Nambu-Gorkov space, i.e., the trace on the right-hand side is over color, flavor, and Dirac space. We have also included the scale $T^2$ in the logarithm, which was not needed in the derivation of the gap equation.   
Now, with Eqs.\ (\ref{G0pm}), (\ref{G0}), and (\ref{Phiplus}) we have
\be
[G_0^+]^{-1}[G_0^-]^{-1} = \sum_e[k_0^2-(\mu-ek)^2]\Lambda_k^{-e} \, , 
\ee
and
\be
\Phi^-G_0^-\Phi^+[G_0^-]^{-1} = \Delta^2{\cal M}^2 = \Delta^2\sum_r\lambda_r{\cal P}_r \, .
\ee
Consequently, we can write
\be
\frac{1}{2}{\rm Tr}\ln \frac{S^{-1}}{T^2} = \frac{1}{2}\Tr\ln\sum_{e,r}\Lambda_k^{-e}{\cal P}_r\frac{k_0^2-(\epsilon_{k,r}^e)^2}{T^2} \, .
\ee
The trace of the logarithm of a matrix is the sum over the logarithms of the eigenvalues of the matrix, weighted with their
degeneracy. Having written the matrix in terms of projectors, this
trace can thus be evaluated easily. With the Dirac trace $\Tr[\Lambda_k^{-e}]=2$ we obtain 
\bea
\frac{1}{2}{\rm Tr}\ln \frac{S^{-1}}{T^2} &=& \frac{T}{V}\sum_K\sum_{e,r}\Tr({\cal P}_r)\ln\frac{k_0^2-(\epsilon_{k,r}^e)^2}{T^2} \nonumber \\
&=& \sum_{e,r}\int\frac{d^3\vec{k}}{(2\pi)^3}\Tr({\cal P}_r)\left[\epsilon_{k,r}^e+2T\ln\left(1+e^{-\epsilon_{k,r}^e/T}\right)\right] \, , \label{TrlnS}
\eea
where we have taken the thermodynamic limit and employed the Matsubara sum 
\be
T\sum_{k_0}\ln\frac{k_0^2-\epsilon_k^2}{T^2} = \epsilon_k+2T\ln\left(1+e^{-\epsilon_k/T}\right) +{\rm const.} \, , 
\ee
where the (infinite) terms constant in $k$, $\mu$, and $T$ can be ignored. 

For the second term of the free energy density (\ref{Omega}) we use Eqs.\ (\ref{S0inv}) and (\ref{Sfull}) to perform the Nambu-Gorkov trace,
\be
\Tr(1-S_0^{-1}S) = \Tr(2-[G_0^+]^{-1}G^+-[G_0^-]^{-1}G^-) \, .
\ee
With Eqs.\ (\ref{G0pm}) and (\ref{GFone}) we have 
\be
[G_0^\pm]^{-1}G^\pm=\sum_{e,r}\left(1+\frac{\lambda_r\Delta^2}{k_0^2-(\epsilon_{k,r}^e)^2}
\right)\Lambda_k^{\mp e}{\cal P}_r \, .
\ee
Therefore, 
\bea
\frac{1}{4}\Tr(1-S_0^{-1}S) &=& -\frac{T}{V}\sum_K\sum_{e,r}\Tr({\cal P}_r)\frac{\lambda_r\Delta^2}{k_0^2-(\epsilon_{k,r}^e)^2} = \sum_{e,r}\int\frac{d^3\vec{k}}{(2\pi)^3}\Tr({\cal P}_r)\frac{\lambda_r\Delta^2}{2\epsilon_{k,r}^e}\tanh\frac{\epsilon_{k,r}^e}{2T} \, , \label{S0S}
\eea
where we used the Matsubara sum from Eq.\ (\ref{matsu1}).
Putting both terms (\ref{TrlnS}) and (\ref{S0S}) together, we have
\bea
\Omega &=& -\sum_{e,r}\int\frac{d^3\vec{k}}{(2\pi)^3}\Tr({\cal P}_r)\left[\epsilon_{k,r}^e+2T\ln\left(1+e^{-\epsilon_{k,r}^e/T}\right)-\frac{\lambda_r\Delta^2}{2\epsilon_{k,r}^e}\tanh\frac{\epsilon_{k,r}^e}{2T}\right]\nonumber\\[2ex]
&=&-\sum_{e,r}\int\frac{d^3\vec{k}}{(2\pi)^3}\Tr({\cal P}_r)\left(\epsilon_{k,r}^e-\frac{\lambda_r\Delta_0^2}{2\epsilon_{k,r}^e}\right) \, , 
\eea
where, in the second step, we have taken the zero-temperature limit
(since $\epsilon_{k,r}^e>0$, the logarithm vanishes in this limit). 
In the absence of a gap, $\Delta_0=0$, we find
\be
\Omega_0 = -\frac{N_cN_f\mu^4}{12\pi^2} \, ,
\ee
where we have subtracted the (infinite) vacuum contribution $\Omega(\mu=T=0)$. The result 
$\Omega_0$ is the free energy density of a non-interacting gas of massless quarks because in the
present approximation the only effect of interactions is to create a nonzero gap.
To compute the contribution of the Cooper pair condensate to the free energy we assume the gap 
to be nonzero only in the small vicinity of the Fermi surface $k\in [\mu-\delta,\mu+\delta]$
and in this vicinity to assume the constant value $\Delta_0$. Then, setting the anti-particle 
gap to zero, and using 
\be
\int_{\mu-\delta}^{\mu+\delta} dk\, k^2\left(\epsilon_{k,r}^+-\frac{\lambda_r\Delta_0^2}{2\epsilon_{k,r}} -|k-\mu|\right) = \frac{\mu^2}{2}\left[\lambda_r\Delta_0^2+{\cal O}(\Delta_0^2/\delta^2)\right] \, ,
\ee
we compute 
\be
\Omega\simeq \Omega_0 - \frac{\mu^2}{4\pi^2}\sum_r\Tr({\cal P}_r)\lambda_r \Delta_0^2 \, . \label{OmOm0}
\ee
The second term, which is negative, is called condensation energy (density) and shows that 
the free energy density is indeed reduced by the formation 
of a Cooper pair condensate. At weak coupling, where $\Delta\ll \mu$, the 
condensation energy is well approximated by the term of the form $\mu^2\Delta_0^2$. We may use the result to determine the preferred pairing pattern. For CFL, $\sum_r\Tr({\cal P}_r)\lambda_r = 8\times 1 + 1\times 4 = 12$, while for 2SC $\sum_r\Tr({\cal P}_r)\lambda_r = 4\times 1 + 5\times 0 = 4$. This number measures how many quasi-particles are gapped and 
what their gap is in multiples of $\Delta_0$. In addition, we need to take into account that $\Delta_0$ may vary from phase to phase. We have seen that due to the two-gap structure of CFL, $\Delta_0$ is smaller by a factor of $2^{-1/3}\simeq 0.79$ compared to the 2SC phase. This gives a ratio of condensation energies between CFL and 2SC of $3\times 2^{-2/3} \simeq 1.89$. As a consequence, CFL is the favored phase of three-flavor QCD at asymptotically large   $\mu$. How far down in density CFL persists and if it (or a variant of it) is still preferred
at neutron star densities is unknown. We shall discuss this question on a qualitative level 
in Sec.\ \ref{sec:mismatch}. 

Without Cooper pairing, the free energy density (or pressure $P=-\Omega$) is of course known beyond the non-interacting limit. More precisely, the pressure is known up to ${\cal O}(\alpha_s^2)$ without \cite{Freedman:1976ub} and with strange quark mass \cite{Kurkela:2009gj}, and it is an ongoing effort to compute the complete ${\cal O}(\alpha_s^3)$ contributions
\cite{Gorda:2021znl,Fernandez:2021jfr,Gorda:2023mkk}. It is a challenge to combine these perturbative calculations with the non-perturbative nature of Cooper pairing. Studies in this direction have been made to compute the high-density behavior of the speed of sound \cite{Geissel:2024nmx,Geissel:2025vnp}.

\subsection{Meissner masses and rotated electromagnetism}
\label{sec:Meissner}

\begin{figure}[h]
   \begin{center}
     \includegraphics[width=0.7\textwidth]{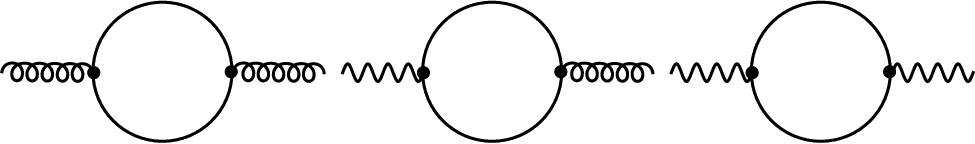}
     \end{center}
  \caption{\label{fig:meissnerloops} One-loop diagrams for the gluon and photon self-energy in 
  a color superconductor, from which the screening masses are calculated.
  Solid lines are quark propagators in Nambu-Gorkov space, such that each diagram receives contributions from normal and anomalous propagators, see Fig.\ \ref{fig:anomloop}. Curly external legs are gluons while wavy external legs correspond to photons. In particular, the loop with mixed photon and gluon legs  can be nonzero in a color superconductor.} 
\end{figure}

One of the defining properties of a superconductor is the Meissner effect, i.e., the expulsion of magnetic fields. 
In field-theoretical terms, the Meissner effect is described by a non-zero magnetic screening mass (the `Meissner mass'), whose inverse is the penetration depth, up to which magnetic fields penetrate the superconductor. We can generalize this 
concept to a color superconductor and compute the magnetic screening masses for the gauge bosons, which, in this case, include the gluons. The screening masses are calculated from the
gluon and photon self-energies, the polarization tensors. (For a pedagogical presentation of the calculation in the case of pure electromagnetism, i.e., without gluons, in the same formalism as used here see Ref.\ \cite{Schmitt:2014eka}.) At one-loop order the self energies are given by the diagrams in Fig.\ \ref{fig:meissnerloops}, whose algebraic form is 
\be \label{Pimunu}
\Pi_{\mu\nu}^{ab}(P) = \frac{1}{2}\frac{T}{V}\sum_K\Tr[\hat{\Gamma}_a^\mu S(K) \hat{\Gamma}_b^\nu S(K-P)] \, .
\ee
Here, we have introduced the vertices in Nambu-Gorkov space $\hat{\Gamma}_a^\mu = {\rm diag}(\Gamma^\mu_a,\bar{\Gamma}_a^\mu)$, where $\Gamma_a^\mu = g\gamma^\mu T_a$ for $a=1,\ldots , 8$ and $\Gamma_a^\mu=e\gamma^\mu Q$ for $a=9$, while for the charge-conjugate components $\bar{\Gamma}_a^\mu = -g\gamma^\mu T_a^T$ for $a=1,\ldots , 8$ and $\bar{\Gamma}_a^\mu=-e\gamma^\mu Q$ for $a=9$, where $Q$ is the generator of the 
electromagnetic gauge group $U(1)_Q$ and $e$ is the elementary charge. For the following calculation it is convenient to put the quark flavors in the order $d$, $s$, $u$, such that $Q={\rm diag} (-1/3,-1/3,2/3)=-2/\sqrt{3}\, T_8$. 

In the combined gluon/photon space, the polarization tensor is a $9\times 9$ matrix
and potentially has off-diagonal components that mix gluons with the photon, see Fig.\ \ref{fig:meissnerloops}. If the full momentum dependence is kept, the polarization tensor can be used to compute the spectral density of the gauge bosons \cite{Rischke:2001py,Rischke:2002rz,Malekzadeh:2006ud}, as briefly discussed for the HDL approximation in Sec.\ \ref{sec:gap}. While this approximation was sufficient for the calculation of the gap up to a certain order, now we are interested in
the effect of Cooper pairing on the gauge bosons themselves. We shall  not discuss the full spectral density but are rather interested in the limit $p_0=0, p\to 0$. This limit yields the electric (`Debye')
and magnetic (`Meissner') screening masses, which are defined through the longitudinal and transverse components of the self-energy [see Eqs.\ (\ref{PiFG}) and (\ref{FPi00})],
\begin{subequations}
\bea
m_{D,ab}^2 &=& -\lim_{p\to 0} \Pi^{00}_{ab}(0,\vec{p}) \, , \\[2ex]
m_{M,ab}^2 &=&\frac{1}{2}\lim_{p\to 0}(\delta_{ij}-\hat{p}_i\hat{p}_{j})\Pi^{ij}_{ab}(0,\vec{p}) \label{Meissner}
\, .
\eea
\end{subequations}
In the following we will focus on the calculation of the Meissner masses in the CFL phase. We will comment on the Debye masses and other color-superconductors briefly at the end of this section.

\begin{figure}[h]
   \begin{center}
     \includegraphics[width=0.5\textwidth]{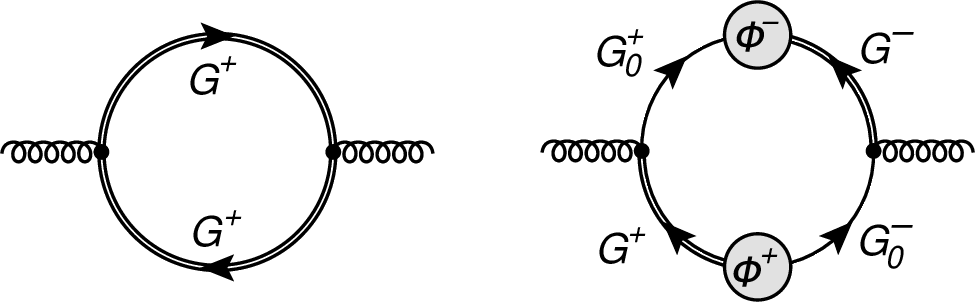}
     \end{center}
  \caption{\label{fig:anomloop} The gluon (shown here), photon, and mixed polarization tensors receive contributions from normal propagators (left) and anomalous propagators (right), see Eq.\ (\ref{Pitrace}). The anomalous propagators are represented according to $F^\pm = -G_0^\mp\Phi^\pm G^\pm$, as in the diagram for the gap equation, Fig.\ \ref{fig:gapeq}. Each diagram also appears with $+$ and $-$ exchanged. 
  } 
\end{figure}

Inserting the quark propagator (\ref{Sfull}) into Eq.\ (\ref{Pimunu}) and performing the trace
over Nambu-Gorkov space yields 
\bea \label{Pitrace}
\Pi^{\mu\nu}_{ab}(P) &=& \frac{1}{2}\frac{T}{V}\sum_K\Tr[\Gamma_a^\mu G^+(K)\Gamma_b^\nu G^+(Q) 
+\bar{\Gamma}_a^\mu G^-(K)\bar{\Gamma}_b^\nu G^-(Q) \nonumber \\[2ex]
&& +\Gamma_a^\mu F^-(K)\bar{\Gamma}_b^\nu F^+(Q)+\bar{\Gamma}_a^\mu F^+(K)\Gamma_b^\nu F^-(Q)] \, ,
\eea
where we have abbreviated $Q\equiv K-P$. We see that there are loops formed of normal propagators $G^\pm$ and loops formed of anomalous propagators $F^\pm$, see Fig.\ \ref{fig:anomloop}. 

Mixed terms, containing one normal and one anomalous propagator, do not appear. 
Next, we insert the propagators $G^\pm$ and $F^\pm$ from Eq.\ (\ref{GF}). For the spatial components, $\mu=i, \nu=j$, which are needed to compute the Meissner mass, we need the 
Dirac traces
\bea
{\cal T}^{ij}_{e_1e_2}(\uk,\uq)&\equiv& \Tr[\gamma^i\gamma^0\Lambda_k^{\mp e_1}\gamma^j\gamma^0\Lambda_q^{\mp e_2}] 
= \Tr[\gamma^i\gamma^5\Lambda_k^{\pm e_1}\gamma^j\gamma^5\Lambda_q^{\mp e_2}] \nonumber\\[2ex]
&=& \delta^{ij}(1-e_1e_2\uk\cdot\uq)+e_1e_2(\hk^i\hq^j+\hk^j\hq^i) \, . 
\eea
The color and flavor traces are performed with the help of Eqs.\ (\ref{Mij}) and (\ref{P1P2ij}). With $\Tr[T_aT_b]=\delta_{ab}/2$ and $\Tr[T_a]=0$ we find 
for $a,b\le 8$ 
\bea \label{colortr1}
{\cal U}_{ab}^{rs}&\equiv&g^2\Tr[T_a{\cal P}_rT_b{\cal P}_s] = g^2\Tr[T_a^T{\cal P}_rT_b^T{\cal P}_s]
=g^2\left\{\begin{array}{cc} \displaystyle{\frac{7\delta_{ab}}{6}} & \mbox{for} \;\; r=s=1 \\[2ex] 0 & \mbox{for}\;\; r=s=2 \\[2ex] \displaystyle{\frac{\delta_{ab}}{6}} & \mbox{for} \;\; r\neq s \end{array}\right. \, ,
\eea
and
\bea \label{colortr2}
{\cal V}_{ab}^{rs}&\equiv&g^2\Tr[T_a{\cal M}{\cal P}_rT_b^T{\cal M}{\cal P}_s] =g^2\Tr[T_a^T{\cal M}{\cal P}_rT_b{\cal M}{\cal P}_s] = g^2\left\{\begin{array}{cc} 0 & \mbox{for} \;\; r=s=2 \\[2ex] \displaystyle{-\frac{\delta_{ab}}{3}} & \mbox{otherwise}   \end{array}\right. \, .
\eea
The results for $a=9$ and/or $b=9$ are easily obtained from these expressions: Since $Q=-2/\sqrt{3}\,T_8$ we simply multiply 
these results by $-\frac{2e}{\sqrt{3}g}$ for the traces where one $SU(3)$ generator is replaced by $Q$ and by $\frac{4e^2}{9g^2}$ if both $SU(3)$ generators are replaced by $Q$.
Hence we can write the spatial components of the polarization tensor as
\bea
\Pi^{ij}_{ab} = \frac{T}{V}\sum_{e_1,e_2}\sum_{r,s}\sum_K 
{\cal T}^{ij}_{e_1e_2}(\uk,\uq)
 \frac{{\cal U}_{ab}^{rs}(k_0q_0+e_1e_2\xi_k^{e_1}\xi_q^{e_2})+{\cal V}_{ab}^{rs}\Delta^2}{[k_0^2-(\epsilon_{k,r}^{e_1})^2]
 [q_0^2-(\epsilon_{q,s}^{e_2})^2]} \, ,
\eea
where we have defined
\be
\xi_k^e \equiv k-e\mu \, .
\ee
Next we perform the sum over $k_0=-i\omega_n$, where $\omega_n$ are 
fermionic Matsubara frequencies. To this end, we make use of the 
general result 
\bea
T\sum_{k_0}\frac{k_0q_0+\xi_1\xi_2+\Delta^2}{(k_0^2-\epsilon_1^2)(k_0^2-\epsilon_2^2)} &=& \frac{\epsilon_1\epsilon_2+\xi_1\xi_2+\Delta^2}{2\epsilon_1\epsilon_2}\left[\frac{f(\epsilon_1)-f(\epsilon_2)}{p_0-\epsilon_1-\epsilon_2}-\frac{f(\epsilon_1)-f(\epsilon_2)}{p_0+\epsilon_1+\epsilon_2}\right]\nonumber\\[2ex]
&&+\frac{\epsilon_1\epsilon_2-\xi_1\xi_2-\Delta^2}{2\epsilon_1\epsilon_2}\left[\frac{1-f(\epsilon_1)-f(\epsilon_2)}{p_0-\epsilon_1-\epsilon_2}-\frac{1-f(\epsilon_1)-f(\epsilon_2)}{p_0+\epsilon_1+\epsilon_2}\right] \nonumber \\[2ex]
&&\to -\frac{\epsilon_1\epsilon_2-\xi_1\xi_2-\Delta_0^2}{2\epsilon_1\epsilon_2(\epsilon_1+\epsilon_2)}
\eea 
where $q_0=k_0-p_0$ and $p_0=-i\omega_m$ with the bosonic Matsubara frequencies $\omega_m=2\pi mT$, where $f$ is the Fermi-Dirac distribution, and where, in the second step, we have set $p_0=0$ and taken the zero-temperature limit, using 
that $\epsilon_1,\epsilon_2>0$. Moreover, after taking the thermodynamic limit, we need the angular integral
\be
\int\frac{d\Omega}{4\pi}{\cal T}^{ij}_{e_1e_2}(\uk,\uk) 
= \delta^{ij}\left(1-\frac{e_1e_2}{3}\right) \, .
\ee
Putting everything together, we obtain 
\bea\label{Pi0ij}
\Pi^{ij}_{ab}(0,\vec{p}\to0) &=& -\frac{\delta^{ij}}{3\pi^2} \sum_{e}\sum_{r,s}\int_0^\infty dk\, k^2  
\left\{ \frac{{\cal U}_{ab}^{rs}(\epsilon_{k,r}^e\epsilon_{k,s}^e-\xi_k^e\xi_k^e)-{\cal V}_{ab}^{rs}\Delta_0^2}{2\epsilon_{k,r}^e\epsilon_{k,s}^e(\epsilon_{k,r}^e+\epsilon_{k,s}^e)} 
 +2\frac{{\cal U}_{ab}^{rs}(\epsilon_{k,r}^e\epsilon_{k,s}^{-e}-\xi_k^e\xi^{-e})-{\cal V}_{ab}^{rs}\Delta_0^2}{2\epsilon_{k,r}^e\epsilon_{k,s}^{-e}(\epsilon_{k,r}^e+\epsilon_{k,s}^{-e})} \right\} \, . \hspace{0.5cm}
\eea
It remains to perform the momentum integral. For now we replace the upper boundary with a finite 
cutoff $\Lambda$. We first consider the integral over the terms proportional to 
${\cal U}_{ab}^{rs}$. We have to distinguish the cases where $r=s$, i.e., where the 
excitation energies belong to the same quasi-particle branch, and where $r\neq s$, i.e., the 
terms that mix the two quasi-particle branches with different gaps. For $r=s$ we have
\bea \label{MomInt1}
\int_0^\Lambda dk\, k^2 \left[\frac{(\epsilon_{k,r}^+)^2-(\xi_k^+)^2}{2(\epsilon_{k,r}^+)^3}
+\frac{(\epsilon_{k,r}^-)^2-(\xi_k^-)^2}{2(\epsilon_{k,r}^-)^3}+4\frac{\epsilon_{k,r}^+\epsilon_{k,r}^-+\xi_k^+\xi_k^-}{\epsilon_{k,r}^+\epsilon_{k,r}^-(\epsilon_{k,r}^++\epsilon_{k,r}^-)}
\right] = 2\Lambda^2-\mu^2+2\Delta_0^2-3\Delta_0^2\ln\frac{2\Lambda}{\Delta_0} +{\cal O}\left(\frac{1}{\Lambda^2}\right) . 
\eea
Although the integrand looks complicated one can perform this integral analytically without approximations. After doing so we have dropped all terms that 
vanish for $\Lambda\to\infty$. 
We have produced a divergent term proportional to $\Lambda^2$. This is expected and can be removed by subtracting the vacuum contribution. We have also produced a logarithmic divergence 
that depends on the gap $\Delta_0$. If we choose the cutoff to be of the order of $\mu$, say
$\Lambda=2\mu$, then we find that this term is exponentially suppressed due to the 
exponential suppression of the gap. We will thus omit this term in the following. Its appearance is an artifact of the assumption of a momentum-independent gap when we performed the momentum integral. Had we used the full momentum dependence, we would expect a natural cutoff since the gap is peaked at the Fermi surface. As a result, we can approximate the integral (\ref{MomInt1}) by $-\mu^2$ since $\mu\gg \Delta_0$. With the same arguments we compute the other integrals: for 
$r\neq s$ we find the same leading order result, such that 
\bea
\sum_e \int_0^\infty dk\, k^2  
\left[\frac{\epsilon_{k,r}^e\epsilon_{k,s}^e-\xi_k^e\xi_k^e}{2\epsilon_{k,r}^e\epsilon_{k,s}^e(\epsilon_{k,r}^e+\epsilon_{k,s}^e)} 
+2\frac{\epsilon_{k,r}^e\epsilon_{k,s}^{-e}-\xi_k^e\xi^{-e}}{2\epsilon_{k,r}^e\epsilon_{k,s}^{-e}(\epsilon_{k,r}^e+\epsilon_{k,s}^{-e})} \right] \simeq -\mu^2\, ,
\eea
independent of $\lambda_r$ and $\lambda_s$. For the integrals proportional to ${\cal V}_{ab}^{rs}$ (originating from the anomalous propagators) we find
\bea
 \sum_e\int_0^\infty dk\, k^2  \left[ \frac{\Delta_0^2}{2\epsilon_{k,r}^e\epsilon_{k,s}^e(\epsilon_{k,r}^e+\epsilon_{k,s}^e)} 
 +2\frac{\Delta_0^2}{2\epsilon_{k,r}^e\epsilon_{k,s}^{-e}(\epsilon_{k,r}^e+\epsilon_{k,s}^{-e})} \right] \simeq \left\{\begin{array}{cc} \mu^2 & \mbox{for} \;\; r=s \\[2ex]
\displaystyle{\frac{\mu^2}{\lambda_r-\lambda_s}\ln\frac{\lambda_r}{\lambda_s}} & \mbox{for} \;\; r\neq s 
\end{array}\right. \, .
\eea
Inserting these results into Eq.\ (\ref{Pi0ij}), using the traces (\ref{colortr1}) and 
(\ref{colortr2}) and the definition (\ref{Meissner}) we obtain
for the Meissner masses of gluons and the photon in the CFL phase
\be \label{MeissnerCFL}
m_{M,ab}^2=\frac{\mu^2}{54\pi^2}(21-8\ln 2)\left(\begin{array}{cccc} g^2&&&\\&\ddots&&\\&&g^2&\displaystyle{-2/\sqrt{3}\,eg}\\&&\displaystyle{-2/\sqrt{3}\,eg}&\displaystyle{4/3\,e^2}
\end{array}\right) \, ,
\ee
where we have omitted all zeros, and the dots indicate that the diagonal entries for $a=b\le 8$ are identical. Several observations can be made from this result. We see that Cooper pairing of quarks leads to magnetic screening of gauge bosons, as expected from the analogous effect in 
ordinary superconductors. In the CFL phase, all 8 gluons acquire a Meissner mass, and the 
Meissner mass of all gluons is the same. One might wonder why there is no dependence on the gap $\Delta_0$. After all, the Meissner mass should vanish in the unpaired phase, where $\Delta_0=0$. The reason is that we have taken the limit of vanishing gluon momentum 
$\vec{p}\to 0$ at a fixed nonzero $\Delta_0$. Had we first set $\Delta_0 =0$ and then computed the screening mass we would have found $m_{M,ab}=0$, as expected. Maybe the most striking 
feature of the result (\ref{MeissnerCFL}) is the presence of off-diagonal elements. This indicates mixing of the eighth gluon with the photon. By an orthogonal transformation ${\cal O}$ 
of the gauge fields we can diagonalize the gauge boson propagator, such that $A_\mu^a\Pi_{ab}^{\mu\nu}A_\nu^b = \tilde{A}_\mu^a\tilde{\Pi}_{ab}^{\mu\nu}\tilde{A}_\nu^b$ with rotated 
gauge fields $\tilde{A}_\mu^a={\cal O}_{ab}A_\mu^b$ and $\tilde{\Pi}_{ab}^{\mu\nu}={\cal O}_{ac}{\Pi}_{cd}^{\mu\nu}{\cal O}_{db}^{T}\propto \delta_{ab}$. In 
general, this transformation 
depends on the gauge boson momentum. In the zero-energy, low-momentum case considered here this 
transformation leaves the gauge fields $A_\mu^1, \ldots, A_\mu^7$ invariant while the eighth gluon $A_\mu^8$ and the photon $A_\mu$
are rotated,
\begin{subequations}\label{mixA}
\bea
\tilde{A}_\mu^8 &=& \cos\theta \, A_\mu^8 + \sin\theta \, A_\mu \, , \\[2ex]
\tilde{A}_\mu &=& -\sin\theta \, A_\mu^8 + \cos\theta \, A_\mu  \, .
\eea
\end{subequations}
The mixing angle, which diagonalizes the nontrivial $2\times 2$ sector of
the Meissner mass matrix (\ref{MeissnerCFL}) is 
\be \label{mixCFL}
\cos\theta = \frac{3g^2}{3g^2+4e^2} \, .
\ee
After diagonalization, we have $\tilde{m}_{M,88}^2=(4e^2+3g^2)/6\,m_{M,88}^2$ and $\tilde{m}_{M,99}^2=0$, i.e., one combination of the gluon and the photon is massless. This means that there exists a `rotated gauge boson' whose magnetic field {\it can} 
penetrate the CFL phase\footnote{This is analogous to the Higgs mechanism of the standard model. There, the original symmetry group is $SU(2)\times U(1)$
with gauge fields $W_1,W_2,W_3$ and $W_0$. Symmetry breaking leaves a residual group $U(1)$, which is the electromagnetic group. The new fields are the massive $W^\pm$ and $Z$ and the massless photon, where the $Z$ and the photon are given by a rotation of the original fields $W_3$ and $W_0$ --  analogous to Eq.\ (\ref{mixA}) -- and the mixing angle is the Weinberg angle.}. In other words, there is a certain charge with respect to which the CFL Cooper pairs are neutral. The generator of the corresponding symmetry is given by a linear combination of the 
generators of the color and electromagnetic gauge groups, $\tilde{Q} = {\bf 1}\otimes Q+\frac{2}{\sqrt{3}}T_8\otimes {\bf 1}$, see also the CFL symmetry breaking pattern (\ref{CFLpattern}). 
At strong coupling, $g\gg e$, which is relevant for the physics of neutron stars, the mixing angle (\ref{mixCFL}) becomes very small, $\cos\theta\simeq 1$. This means that the magnetic gauge boson that penetrates the CFL phase 
is almost identical to the original photon. 

The above calculation can also be performed for the electric screening masses and for other 
color superconductors, see Ref.\ \cite{Schmitt:2003aa} for a general discussion and a collection of the results for various phases. In the CFL phase, one finds that the electric screening mass squared is 3 times the magnetic screening mass squared for all $a,b$. In particular, the 
mixing angle is the same in electric and magnetic sectors. This is different in the 2SC phase, where different mixing angles are found. Moreover, due to the the presence of unpaired quarks, the 2SC screening masses are less symmetric than in the CFL phase: the Debye and Meissner masses of the gluons 1--3 vanish, while the gluons 4--7 have the same nonzero screening, which is different from that of the 8th gluon. As in the CFL phase, there is a residual rotated 
electromagnetic group, such that a certain rotation of the photon and the 8th gluon has a vanishing Meissner mass. 

Rotated electromagnetism plays a role for the calculation of the gap in an external magnetic field. For the CFL phase, the gap equation was solved within an NJL model in Refs.\ \cite{Ferrer:2005vd,Noronha:2007wg}, and the QCD gap equation for the 2SC phase with external magnetic field was studied in Ref.\ \cite{Yu:2012jn}. It is also interesting to ask if and how the rotated electromagnetism could affect observables of a neutron star with a quark matter core. Within Ginzburg-Landau theory with parameters computed from perturbative QCD, one finds that at strong coupling 2SC and CFL become a type-II superconductor (at $g=3.5$ if at the same time $T_c\gtrsim 0.05\mu$ \cite{Haber:2017oqb}). The corresponding critical magnetic fields for the entrance into a phase of magnetic flux tubes is likely too large to be reached in neutron stars. However, a flux tube lattice is conceivable taking into account the evolution of the star: the magnetic field penetrates the core when the core cools into a color superconductor, and it is unknown on which timescale the magnetic field (more precisely, its rotated version according to the mixing with the gluons) gets expelled or if it stays trapped inside flux tubes. Magnetic flux tubes in CFL can come in different variants, further complicated by the broken $U(1)_B$ in the CFL phase, which allows for defects that have magnetic flux and baryon circulation at the same time, i.e., they are superfluid vortices and magnetic flux tubes \cite{Iida:2002ev,Iida:2004if,Balachandran:2005ev,Eto:2009kg,Vinci:2012mc,Eto:2013hoa,Alford:2016dco,Haber:2017oqb,Haber:2018tqw}. In contrast, 2SC is not a superfluid, its magnetic defects were discussed for instance in Ref.\ \cite{Haber:2017oqb,Alford:2010qf}, including magnetic domain walls that exist in the massless limit and turn into multi-winding flux tubes in the presence of a strange quark mass \cite{Evans:2020uui}. If CFL or 2SC flux tubes are present in a compact star, they may stabilize `mountains' of the star, resulting in  detectable continuous gravitational waves \cite{Glampedakis:2012qp}.
Rotated electromagnetism in CFL and in particular the resulting flux tubes also play an important role in the question of a potential quark-hadron continuity. This continuity was originally pointed out based on symmetries and the fermionic spectrum \cite{Schafer:1998ef,Alford:1999pa}. New aspects arise when flux tubes are taken into account, which is relevant for instance for a potential CFL-hadron interface in a (rotating) neutron star \cite{Alford:2018mqj,Chatterjee:2018nxe,Cherman:2018jir,Cherman:2020hbe}.

\section{Color superconductors under neutron star conditions}
\label{sec:cscstar}

\subsection{Splitting of Fermi momenta}

The calculations of the previous subsections were all performed under the premise of weak coupling, treating the coupling constant $g$ as a small parameter. We have briefly discussed extrapolations to large values of $g$ and have mentioned approaches to understand this regime. 
Besides the problem of strong coupling there is another assumption we have made so far in our discussion of color superconductors. We have assumed all quark masses to vanish. Just like the assumption of the coupling strength to be weak, this assumption is also certain to fail as we move towards lower densities. The reason is that as we go to smaller densities, the relative magnitude of the (strange) quark mass, compared to the chemical potential, increases. This is, firstly, 
because the chemical potential is smaller at lower densities and, secondly, also because we can expect the (medium-dependent) mass to increase due to strong coupling effects. At moderate densities we expect the strange quark mass to lie somewhere 
between its current and constituent values, $m_s\sim (100 -500)\, {\rm MeV}$. This is not
negligible compared to the quark chemical potential in the interior of neutron stars $\mu\lesssim 500\, {\rm MeV}$.  A difference in quark masses generates a difference in Fermi momenta and thus the usual mechanism of Cooper pairing is challenged. Let us discuss this problem in some more detail, but without going through any full-fledged calculations. 

Let us assume that neutron star matter is electrically neutral at every point in space and is in $\beta$-equilibrium, i.e., in equilibrium with respect to the quark analogues of $\beta$-decay and electron capture.  
(Ignoring mixed phases, where local neutrality can be violated \cite{Glendenning:1992vb,Heiselberg:1992dx,Schmitt:2020tac}, and ignoring fast dynamics as in neutron star mergers, where matter may be out of chemical equilibrium \cite{Hammond:2021vtv,Chabanov:2023abq}.) This imposes a constraint on quark and electron densities, 
\be \label{neutralityQ}
\frac{2}{3}n_u - \frac{1}{3} n_d - \frac{1}{3} n_s - n_e = 0 \, , 
\ee
and on the chemical potentials 
\be \label{betaQ}
\mu_u+\mu_e=\mu_d \, , \qquad \mu_u+\mu_e=\mu_s \, , 
\ee
where we have neglected the neutrino chemical potential, assuming that the neutrino mean free path is of the order of or larger than the size of the star. We also have to impose {\it color} neutrality, although we will largely ignore this point in the following, mostly qualitative, discussion. Massless three-flavor quark matter is particularly symmetric: equal quark densities and chemical potentials fulfill these conditions, and no electrons are needed. The reason is that the electric charges of the three lightest quarks happen to add up to zero. Therefore, in the previous subsections it was justified to consider a single, common quark chemical potential $\mu$ for all quark species. 
If we now `switch on' a mass for the strange quark, its Fermi momentum is reduced to $k_{F,s}=\sqrt{\mu_s^2-m_s^2}$; it has become energetically more costly to populate the
system with strange quarks. Therefore, since $n_s\propto k_{F,s}^3$ (at $T=0$), the number density of strange quarks is reduced as well. As a consequence, and because the up and down quarks remain approximately massless, we have disrupted the symmetric situation and the quark 
and electron densities -- under the conditions (\ref{neutralityQ}) and (\ref{betaQ}) -- 
will adjust themselves in a less symmetric way. If all interactions are neglected, one 
computes that, to lowest order in the strange quark mass, 
\be
\mu_e \simeq \frac{m_s^2}{4\mu} \, , 
\ee
and
\be
k_{F,u}\simeq \mu-\frac{m_s^2}{6\mu} \, , \qquad 
k_{F,d}\simeq \mu+\frac{m_s^2}{12\mu} \, , \qquad 
k_{F,s}\simeq \mu-\frac{5m_s^2}{12\mu} \, .
\ee
We see that the Fermi momenta of the quarks are split by an equal amount of 
\be \label{deltakF}
\delta k_F \simeq \frac{m_s^2}{4\mu} \, ,
\ee
such that $k_{F,s}<k_{F,u}<k_{F,d}$. The lack of negative charge due to the smaller number of strange quarks is compensated by an increased Fermi momentum of the down quark and by the presence of electrons. (Note that
`the down goes up and the up goes down'.) The split $\delta k_F$ becomes larger as we go down 
in density since $m_s$ becomes larger and $\mu$ becomes smaller. 

\subsection{Cooper pairing beyond CFL}

How does a split of Fermi momenta affect Cooper pairing? This is a very general question, which, in its simplest form, is: consider a (zero-temperature) system of two fermion species with attractive interaction and chemical potentials $\mu_1$ and $\mu_2$, such that the mismatch in chemical potentials is $\delta\mu=(\mu_1-\mu_2)/2$ and the average chemical potential is $\bar{\mu}=(\mu_1+\mu_2)/2$. Is `usual' Cooper pairing possible at a nonzero $\delta\mu$ and if yes at what value of $\delta\mu$ does it break down? The answer is that it is still energetically favorable to form Cooper pairs if the mismatch $\delta\mu$ is sufficiently small compared to the gap $\Delta$. In this case, the pairing process can be viewed as follows. Starting from the two different Fermi spheres, first create a fictitious state in which both fermion species are filled up to a common Fermi momentum. This costs free energy. Then pair. 
If the energy gain from pairing is larger than the cost from creating a common Fermi momentum, pairing will happen. In this case, the quasi-fermion dispersion relations are modified by the 
mismatch to $\epsilon_k^e\pm\delta\mu$ with $\epsilon_k^e$ from Eq.\ (\ref{dispersion}). In particular, if $\delta\mu>\Delta$, one excitation branch  becomes gapless despite the presence of a nonzero $\Delta$. However, one finds that Cooper pairing breaks down before gapless modes appear. The critical value for the mismatch beyond which the usual pairing is no longer favored over the unpaired phase turns out to be $\delta\mu =\Delta/\sqrt{2}$. 
This is called the Chandrasekhar-Clogston limit \cite{Chandrasekhar:1962,Clogston:1962}. For a pedagogical discussion of the calculation leading to this result see for instance Ref.\ \cite{Schmitt:2014eka}. 

In quark matter, the situation is more complicated because we have 9 fermion species, not just two, and we have the conditions of neutrality and $\beta$-equilibrium. In QCD, the split between the Fermi surfaces as well as the gap $\Delta$ are all functions of the single thermodynamic parameter $\mu$ (at $T=0$). Nevertheless, the general principle is the same: a mismatch in Fermi momenta tends to break Cooper pairs. And, it is not possible to avoid the mismatch and maintain `conventional pairing' by choosing a different pairing pattern \cite{Rajagopal:2005dg}. It turns out that CFL 
becomes energetically disfavored compared to the unpaired phase at $m_s^2/\mu = 4 \Delta$ \cite{Alford:2002kj}. However, one finds that gapless excitations already set in at the smaller mismatch $m_s^2/\mu = 2 \Delta$. (Gaplessness first occurs in the sector where $bd$ quarks pair with $gs$ quarks. Due to the presence of color chemical potentials the mismatch for this sector is $\delta\mu=m_s^2/(2\mu)$, not $m_s^2/(4\mu)$, as the result of unpaired quark matter (\ref{deltakF}) would suggest.) Although, by 
comparing the free energy with the unpaired phase, `gapless CFL' appears to be stable, it suffers an instability with respect to the 
formation of anisotropic or even inhomogeneous phases. This instability manifests itself in a negative Meissner mass squared, i.e., if the calculation of Sec.\ \ref{sec:Meissner} is repeated in the presence of a strange quark mass, one finds 
unphysical Meissner masses as soon as at least one excitation branch becomes gapless \cite{Casalbuoni:2004tb,Fukushima:2005cm}. This was first pointed out in the context of the 2SC phase \cite{Huang:2004bg}.

These results give us some idea about whether we can expect CFL in the interior of neutron stars, but due to the lack of knowledge of the medium-dependent $m_s$ and the color-superconducting gap $\Delta$ at moderate densities we do not have a conclusive answer to this question. 
What if CFL breaks down before the transition to nuclear matter is reached? In this case, rather than being completely unpaired, quark matter is expected to be in a more exotic paired phase. Almost all of these phases are `partially paired' 
phases, where some, but not all, fermion species remain unpaired and/or where certain fermion 
species only form pairs in certain directions in momentum space. In either case, these phases 
are interesting for the phenomenology of quark matter in neutron stars because they 
differ in their transport properties from the completely gapped CFL phase. Let us discuss these phases a bit more systematically. 

If the parameter $m_s^2/(\mu\Delta)$ is small, one can discuss the fate of CFL in a controlled 
manner within weak coupling. It turns out that the next phase down in density then is the 
so-called CFL-$K^0$ phase, where a condensate of neutral kaons forms on top of the 
fermionic Cooper pair condensate \cite{Bedaque:2001je,Buballa:2004sx,Forbes:2004ww,Alford:2007qa}. To understand this phase we recall that CFL breaks the 
approximate chiral symmetry of QCD, such that there is an octet of pseudo-Goldstone modes. 
The quantum numbers of these modes are the same as the ones of the usual meson
octet in low-density QCD. This is due to the identical symmetry breaking pattern regarding the global chiral symmetry. Therefore, there are CFL versions of the 
 pions, kaons, and the eta, which can be described by the chiral field 
\be \label{Schiral}
\Sigma = \phi_L^\dag\phi_R \in SU(3) \, , 
\ee
where $\phi_L$ and $\phi_R$ are $3\times 3$ matrices for the left-handed and right-handed sectors. This is to be compared to Eq.\ (\ref{Mphi}), where we assumed $\phi_L=\phi_R\equiv \phi$.
In the `pure' CFL phase, therefore, $\Sigma=1$. Each $\phi$ in Eq.\ (\ref{Schiral}) represents diquarks, and thus the CFL mesons are four-fermion objects, made of two 
fermions and two holes. Therefore, even though their quantum numbers are the same as for the usual mesons, their properties, in particular their masses, are different. For instance, the 
CFL neutral kaon can be viewed as an object $K^0 \sim \bar{u}\bar{s}du$. This can be 
obtained from the `usual' kaon $K^0\sim \bar{s} d$ by the replacements $s\to \bar{u}\bar{d}$
and $d\to \bar{u}\bar{s}$. These replacements reflect the {\it anti}-triplet, which the 
flavor index of $\phi_R$ and $\phi_L$ represents (the second index is a color index). As a consequence, the meson masses are ordered differently: while the quark flavors $(u,d,s)$ 
have mass ordering $m_u<m_d<m_s$, the corresponding anti-triplet $(\bar{d}\bar{s},\bar{u}\bar{s},\bar{u}\bar{d})$ has the opposite ordering 
$m_dm_s>m_um_s>m_um_d$. Therefore, in the QCD vacuum, $m_{\pi^0}<m_{K^0}$ since  $m_{\pi^0}\propto m_u+m_d$ and $m_{K^0}\propto m_s+m_d$, while in CFL $m_{K^0}<m_{\pi^0}$ since $m_{K^0}\propto m_um_d+m_um_s$ while $m_{\pi^0}\propto m_dm_s+m_um_s$. Indeed, the neutral kaon is the lightest CFL meson. 
Masses and other properties of CFL mesons can be studied within a chiral effective theory with Lagrangian \cite{Son:1999cm,Bedaque:2001je}
\be \label{calLSigma}
{\cal L} = \frac{f_\pi^2}{4}\Tr[\nabla_0\Sigma\nabla_0\Sigma^\dag -v_\pi^2\partial_i\Sigma\partial_i\Sigma^\dag]+\frac{af_\pi^2}{2}{\rm det}\,M\,
\Tr[M^{-1}(\Sigma+\Sigma^\dag)] \, , 
\ee
with $v_\pi=1/\sqrt{3}$, the covariant derivative $\nabla_0\Sigma \equiv \partial_0\Sigma + i[A,\Sigma]$, which includes effective chemical potentials of the mesons through $A\equiv -\frac{M^2}{2\mu}$, and the quark mass matrix $M={\rm diag}(m_u,m_d,m_s)$. The constants 
$f_\pi$ and $a$ can be computed at weak coupling, 
\be \label{fpia}
f_\pi^2 = \frac{21-8\ln 2}{18}\frac{\mu^2}{2\pi^2}  \, , \qquad a=\frac{3\Delta^2}{\pi^2 f_\pi^2}  \, .
\ee
Since the structure of this effective theory is determined entirely by symmetries, it gives us a rigorous low-energy description of the CFL phase. Notice that all quarks are gapped in `pure' CFL and thus its transport properties are entirely determined by the (pseudo-)Goldstone modes. Thus, if fully gapped CFL survives in neutron star conditions, the chiral Lagrangian (\ref{calLSigma}) provides us with a rigorous tool to study quark matter at moderate densities, up to the uncertainties in the parameters $f_\pi$ and $a$. The Lagrangian (\ref{calLSigma}) can be used to demonstrate that a condensate of neutral kaons becomes preferred if the effective 
kaon chemical potential $\mu_{K^0}$ becomes larger than the kaon mass $m_{K^0}$. With 
\be
\mu_{K^0} = \frac{m_s^2-m_d^2}{2\mu} \, , \qquad m_{K^0}^2 = am_u(m_s+m_d) \, , 
\ee
and the weak-coupling results (\ref{fpia}) for $f_\pi$ and $a$ we find that the 
critical strange mass for kaon condensation scales as $m_u^{1/3}\Delta^{2/3}$.  In this kaon-condensed CFL phase, all fermion
excitations remain fully gapped. As we go further down in density, however, ungapped fermions appear. They do so first in certain regions in momentum space in the so-called curCFL-$K^0$ phase \cite{Schafer:2005ym,Kryjevski:2008zz}. In this phase, a meson supercurrent appears, which is balanced by a counter-propagating current of ungapped fermions. This phase thus spontaneously breaks rotational invariance. More counter-propagating currents in more directions 
may appear as we go further down in density. The resulting phases break translational invariance and exhibit periodically varying gaps \cite{Alford:2000ze,Casalbuoni:2001gt,Giannakis:2002jh,Casalbuoni:2005zp,Carignano:2017meb}. 
These crystalline phases are also called Larkin-Ovchinnikov-Fulde-Ferrell (LOFF) phases \cite{FuldeFerrell,LarkinOvchinnikov}, and different lattice structures have been studied to find the preferred phase \cite{Rajagopal:2006ig}. These calculations are based on Ginzburg-Landau studies or phenomenological models, and thus we are starting to enter theoretically uncontrolled territory in our journey down in density. Nevertheless, the possible existence of crystalline quark phases is of interest for neutron star physics, for example in the context of pulsar glitches \cite{Mannarelli:2007bs}, continuous gravitational wave emission \cite{Knippel:2009st}, or the tidal deformability \cite{Lau:2017qtz,Lau:2018mae}.   

Besides breaking CFL Cooper pairs in some directions in momentum space, the system 
might also react by leaving strange quarks completely unpaired, which is the 2SC phase. We have seen that at asymptotically large densities the 2SC phase has larger free energy than CFL and thus is not preferred. This remains true if a small strange quark mass is switched on. However, as the coupling and/or the mismatch in chemical potentials is increased, a transition from 
CFL to 2SC becomes possible. For instance in NJL calculations with strong diquark coupling the 2SC phase tends to occupy a large region of the phase diagram between CFL and nuclear matter \cite{Ruester:2005jc,Gholami:2024diy}. Since 2SC is also based on cross-flavor pairing, a split between $u$ and $d$ Fermi surfaces tends to break 2SC for the same reasons as discussed for CFL, and a 2SC version of the LOFF phase becomes a candidate phase. 

If the mismatch between Fermi surfaces is too large, be it in CFL or 2SC, the system may resort to single-flavor pairing \cite{Schafer:2000tw,Schmitt:2004hg,Schmitt:2004et,Aguilera:2005tg,Schmitt:2005wg,Alford:2005yy,Feng:2007bg,Brauner:2008ma,Pang:2010wk,Fujimoto:2019sxg,Fujimoto:2021bes,Sogabe:2024yfl}. Due to the anti-symmetry of the 
Cooper pair wave function [see discussion below Eq.\ (\ref{SU3cf})], this requires pairing in the spin-1 channel.  As we have seen above, the weak-coupling gap in the spin-1 channel is about 2-3 orders of magnitude smaller than the spin-0 gap. Therefore, the gain in condensation energy is much smaller as well, and spin-1 pairing is only expected to occur if cross-flavor pairing becomes impossible due to a large mismatch in Fermi momenta or as a "residual" pairing of fermionic modes that are otherwise left unpaired, for instance strange quarks  \cite{Alford:2005yy} or blue down quarks \cite{Fujimoto:2019sxg} in the 2SC phase.
Spin-one color superconductivity 
allows for a multitude of possible pairing patterns in itself. The symmetries involved are not unlike the ones in superfluid $^3$He, and thus many candidate phases have their analogues in $^3$He. The energetically preferred phase at weak coupling 
is the color-spin locked phase \cite{Schafer:2000tw,Schmitt:2004et}, analogous to the  B-phase in $^3$He \cite{Vollhardt1990}. This is the only 
spin-1 color superconductor in which the gap function is isotropic. Moreover, all Meissner masses, including that of the photon are nonzero \cite{Schmitt:2003xq}. Other spin-1 phases have anisotropic gaps and may have nodal points or lines at the Fermi surface, such as the polar and A phases \cite{Schmitt:2004et}.

\label{sec:mismatch}

\section{Bulk viscosity of color superconductors}
\label{sec:bulk}

\subsection{Transport in neutron stars}

Transport plays a crucial role for the understanding of color-superconducting phases. While the equation of state is only mildly sensitive to Cooper pairing, transport properties can change dramatically if matter becomes  superconducting or superfluid. All particles in the system contribute to thermodynamic properties such as the pressure, even if the particles are excluded from any scattering processes due to Pauli blocking. Therefore, if a  small fraction of particles in a small  vicinity of the Fermi surface undergoes Cooper pairing (having in mind a weakly coupled and sufficiently cold system) this does not significantly change the bulk properties such as the pressure. Transport, however, {\it is} sensitive to a change in properties of the particles at the Fermi surface, because these are the particles that dominate transport to begin with, through  scattering processes or other reactions such as flavor-changing processes. (The stronger the coupling, the less accurate this distinction between bulk physics and the physics at the Fermi surface becomes, and at very strong coupling the whole picture of transport through quasi-particles becomes questionable.) 

Here, `transport' describes  the transfer of conserved quantities from one spatial region to another. This occurs if the system is brought out of equilibrium and attempts to re-equilibrate. For instance, a temperature gradient will induce a transfer of energy from the hotter to the colder region, or a gradient in chemical potentials will induce processes that change the chemical composition of matter. Calculating `transport properties' therefore amounts to determine how efficiently, i.e., on which time scale, the system is able to re-equilibrate. 

There are many observables of neutron stars that are sensitive to transport properties and thus are able to discriminate between paired and unpaired phases or between different (color-)superconducting phases. Examples are  neutrino emissivity and the resulting cooling behavior \cite{Potekhin:2015qsa,Marino:2024gpm}, damping of oscillations and consequences for the rotational frequency of the star \cite{Andersson:1997xt,Kraav:2024cus}, dissipation in neutron star mergers \cite{Alford:2017rxf}, pulsar glitches \cite{Antonopoulou:2022rpq}, and the evolution of the magnetic field \cite{Pons:2019zyc}.  

In a neutron star, there are many particle species and many possible scattering and reaction processes that can potentially contribute to transport, see Ref.\ \cite{Schmitt:2017efp} for a review that takes into account all layers of the star, from the crust to a potential quark matter core. Typically, if present, fermionic contributions (electrons, nucleons, quarks, $\ldots$) 
dominate over bosonic ones (kaons, pions, superfluid phonons, $\ldots$), at least at sufficiently small temperatures, where, due to the presence of the Fermi surface, fermionic contributions are usually enhanced by powers of $\mu/T$ compared to bosonic ones. Fermions, however, may  become `unavailable' for low-energy transport through Cooper pairing if the temperature is much smaller than the energy gap induced in the fermionic spectrum due to pairing.  In most forms of neutron star matter, not all fermionic degrees of freedom are gapped. For instance in nuclear matter, even if protons and neutrons are paired, unpaired electrons are present and will dominate most transport properties. In quark matter, CFL is special because all quarks are gapped and no electrons are expected at zero temperature. In this case,  transport is dominated by mesons and the superfluid phonon \cite{Jaikumar:2002vg,Shovkovy:2002sg,Alford:2007rw,Manuel:2007pz,Alford:2008pb,Mannarelli:2009ia,Alford:2009jm,Anglani:2011cw,Bierkandt:2011zp}. In contrast, in the 2SC phase, there are ungapped quarks and a sizable fraction of electrons, both of which are relevant for the calculation of transport properties \cite{Jaikumar:2005hy,Alford:2006gy,Schmitt:2010pn,Alford:2014doa,Alford:2025tbp,Alford:2025jtm}. This is similar in all color-superconducting phases other than CFL, for instance in spin-one color superconductors \cite{Schmitt:2005wg,Sad:2006egl,Wang:2009if,Wang:2010ydb,Berdermann:2016mwt} or in the LOFF phase \cite{Anglani:2006br,Sedrakian:2013pva,Sarkar:2016gib}. 

We shall use bulk viscosity of the 2SC phase from the processes $u+d\leftrightarrow u+s$ as an example for a transport coefficient of a color-superconducting phase. Going through the details of the calculation will show the complications of calculating transport coefficients in a superconducting phase, and also will allow us to derive a simple analytical result, at least in certain limits, for unpaired quark matter. 
Bulk viscosity is different from shear viscosity, thermal conductivity, and electric conductivity in the sense that its dominant contribution originates from flavor-changing electroweak processes. These processes are most efficient in creating dissipation because they can occur on the same time scales as given by the oscillations of the star, or by the merger of two neutron stars. These time scales are of the order of milliseconds or larger and thus always much larger than those of scattering processes of the strong interaction. Therefore (and in contrast to heavy-ion collisions) bulk viscosity of neutron star matter is dominated by processes of the electroweak interaction, not by  QCD scattering. Strong interaction effects may nevertheless provide corrections to the electroweak rate,  for instance through so-called non-Fermi liquid effects, which alter the dispersion relations of (unpaired) quarks \cite{Schafer:2004zf,Schwenzer:2012ga}. In the following calculation we shall ignore these corrections for simplicity. To make our calculation self-contained, we start with the derivation of the bulk viscosity.

\subsection{Derivation of bulk viscosity}
\label{sec:derivebulk}

Bulk viscosity is responsible for dissipation due to compression and expansion of a fluid. We denote the  angular frequency of the external volume oscillation by $\omega$, such that we can write the volume as $V(t) = V_0[1+\delta v(t)]$, 
with a dimensionless volume perturbation 
\be
\delta v(t) = \delta v_0\cos\omega t \, ,
\ee
which we assume to be small, $\delta v_0\ll 1$.
Denoting averaging over one oscillation period $\tau=2\pi/\omega$ by angular brackets, we can write the dissipated energy density, on the one hand, as
\be \label{Epsdot}
\langle\dot{{\cal E}}\rangle_\tau = -\frac{\zeta}{\tau}\int_0^\tau dt \, (\nabla\cdot{\bm v})^2 \simeq  -\frac{\zeta \omega^2\delta v_0^2}{2} \, ,
\ee
where $\zeta$ is the bulk viscosity. 
Here we have used the continuity equation at zero three-velocity ${\bm v}=0$ (but nonzero derivative of $\vec{v}$) to relate the divergence of the velocity field to the change in the total 
particle number density, $n\nabla\cdot \vec{v} = -\partial_t n$. With $n=N/V$ and holding the particle number $N$ fixed, this is then rewritten in terms of the change in volume. On the other hand, the dissipated energy density
can be expressed in terms of the mechanical work done by the induced pressure oscillations,  
\be \label{Epsdot1}
\langle\dot{{\cal E}}\rangle_\tau = \frac{1}{\tau}\int_0^\tau dt \, P(t)\frac{d\delta v}{d t} \, .
\ee
We write the pressure of three-flavor quark matter as 
\be \label{pt}
P(t) = P_0 + \frac{\partial P}{\partial V}V_0\delta v(t) + \sum_{x=u,d,s} \frac{\partial P}{\partial n_x}\delta n_x(t) \, ,
\ee
where the derivatives on the right-hand side are evaluated in equilibrium. 
The oscillation in the pressure is in general out of phase compared to the volume oscillation because of the microscopic re-equilibration processes which induce changes 
in the number densities of the particle species $\delta n_x$. 
From Eqs.\ (\ref{Epsdot}) and (\ref{Epsdot1}) we obtain the bulk viscosity in the form 
\be \label{zeta1}
\zeta = -\frac{2}{\omega^2\delta v_0^2}\frac{1}{\tau} \int_0^\tau dt \, P(t)\frac{d\delta v}{d t} \, . 
\ee
To calculate this expression, we assume that the dominant electroweak re-equilibration processes are the non-leptonic reactions  
\bea
u+d &\leftrightarrow& u + s    \label{udus}  \, . 
\eea
There are other processes that potentially contribute to the bulk viscosity, such as the
leptonic processes (`direct Urca' processes)
\begin{subequations} \label{leptonic}
\bea
u+e &\to& d + \nu_e \, , \qquad d \to u+e+ \bar{\nu}_e\label{uuu2} \\[2ex]
u+e &\to& s + \nu_e \, , \qquad s \to u+e+ \bar{\nu}_e\, . \label{uuu3}
\eea
\end{subequations}
Their contribution becomes relevant for temperatures larger than typically found in isolated neutron stars, see Ref.\ \cite{Alford:2025tbp} and below discussion of Fig.\ \ref{fig:bulkvisc}. (See Refs.\ \cite{Iwamoto:1980eb,Iwamoto:1982zz} for the calculation of the rates of these processes in unpaired quark matter.)

In chemical equilibrium, the two reactions (\ref{udus}) do not change the various densities because they occur with the same rate, and according to the principle of detailed balance in this situation the sum of the chemical potentials of the 
ingoing particles equals the sum of the chemical potentials of the outgoing particles. Therefore, defining 
\be \label{deltamu}
\delta\mu\equiv \mu_s-\mu_d \, , 
\ee
chemical equilibrium implies $\delta \mu=0$. 
A deviation from chemical equilibrium occurs if the equality of 
chemical potentials is disrupted, $\delta\mu\neq 0$. 
The situation considered here is particularly simple because there is a single process (\ref{udus}) and a single $\delta\mu$. In general, there can be multiple processes related to the 
same $\delta\mu$, or there can be multiple processes related to multiple $\delta\mu$'s. For instance, if the leptonic processes (\ref{leptonic}) were taken into account there would be two relevant differences in chemical potentials, $\delta\mu_1 = \mu_s-\mu_d$ and $\delta\mu_2 = \mu_d-\mu_u-\mu_e$
(neglecting the neutrino chemical potential since neutrinos and anti-neutrinos leave the system once they are created, at least if the  temperature is sufficiently low). The resulting expression for the bulk viscosity then becomes significantly more cumbersome, see for instance Ref.\ \cite{Sad:2007afd} or appendix A of Ref.\ \cite{Alford:2006gy}. Here we proceed with a single process and a single $\delta\mu$, in which case we can write the change in quark number densities due to the electroweak processes (\ref{udus}) as
\be\label{dnd}
\delta n_d (t)  =-\delta n_s(t)  =\int_0^t dt'\,\Gamma[\delta\mu(t')]  \simeq \lambda \int_0^t dt'\delta\mu(t') \, , \qquad \delta n_u (t) = 0\, .
\ee
Here, $\Gamma[\delta\mu(t)]$ is the net number of $d$ quarks produced per unit time and volume in the processes  $u+s \leftrightarrow u+d$. Calculating $\Gamma$ is the main task in calculating the bulk viscosity and will be explained in the next subsection. We have linearized the result for small $\delta\mu(t)$, such that now all microscopic input is encoded in $\lambda$.  According to our definition of $\delta\mu$, a net production 
of $d$ quarks sets in for $\delta\mu>0$, from which we conclude $\lambda >0$.  

The difference in chemical potentials $\delta\mu$ oscillates due to the volume oscillation and due to the electroweak reactions,
\bea \label{dmu0}
\frac{d\delta\mu}{dt} &=& \frac{\partial\delta\mu}{\partial V}\frac{dV}{dt} + \sum_{x=d,s} \frac{\partial \delta\mu }{\partial n_x}\frac{dn_x}{dt} \, . 
\eea
Using Eqs.\ (\ref{deltamu}), (\ref{dnd}), the thermodynamic equilibrium relation 
\be
\frac{\partial P}{\partial n_x} = -V_0\frac{\partial \mu_x}{\partial V} \, , 
\ee
and abbreviating 
\be \label{BCabbr}
B \equiv \frac{\partial P}{\partial n_s}  - \frac{\partial P}{\partial n_d} \, , \qquad C\equiv \frac{\partial \mu_s }{\partial n_s}+\frac{\partial \mu_d }{\partial n_d}-\frac{\partial \mu_d }{\partial n_s}-\frac{\partial \mu_s }{\partial n_d} \, ,
\ee
we obtain 
\bea  \label{ddmudt}
\frac{d\delta\mu}{dt} 
&=& - B \frac{d\delta v}{dt}  - \lambda C  \delta\mu(t) \, . 
\eea
Note that $B$ and $C$ are thermodynamic functions evaluated in equilibrium, i.e., they do not depend on the electroweak reaction rate. The pressure (\ref{pt}) can now be expressed with the help of $B$,
\be \label{ptB}
P(t) = P_0 + \frac{\partial P}{\partial V}V_0\delta v(t) - B \delta n_d(t) \, ,
\ee
which, with Eq.\ (\ref{Epsdot1}), yields the   dissipated energy density in the form
\be \label{Edot2}
\langle\dot{{\cal E}}\rangle_\tau = \left\langle P(t)\frac{d\delta v}{dt}\right\rangle = \lambda B \langle\delta\mu(t)\delta v(t) \rangle  \,, 
\ee
where partial integration and Eq.\ (\ref{dnd}) have been employed\footnote{It is instructive to note that the differential equation (\ref{ddmudt})  and the dissipated energy (\ref{Edot2}) are in exact analogy to the equations for an electric circuit with an external alternating voltage $U(t)$ and an induced current $I(t)$,
\bea
\langle\dot{{\cal E}}\rangle_\tau = \langle I(t) U(t) \rangle \, , \qquad \dot{U}(t) = \dot{I}(t)R+L\ddot{I}(t)+\frac{I(t)}{C} \, , \nonumber
\eea
with resistance $R$, inductance $I$, and capacitance $C$. Identifying our external oscillation $\delta v$ with the voltage and the response $\delta \mu$ with the current, we can identify `resistance' and `capacitance' of our system (the `inductance' turns out to be zero).}. Notice that there is only dissipation if $B$ is nonzero. The reason is that $B$ tells us to what extent the system is brought out of equilibrium by a volume change. For instance, for non-interacting quark matter at zero temperature and neglecting the light quark masses, 
\be \label{Bms}
B = \left(n_s\frac{\partial \mu_s}{\partial n_s} - n_d\frac{\partial \mu_d}{\partial n_d}\right)_{\delta\mu=0} = -\frac{m_s^2}{3\mu_d} \, .
\ee
In other words, if strangeness does not introduce an asymmetry in the thermodynamic relations in equilibrium, there will be no dissipation and thus no bulk viscosity due to a strangeness changing process.  

In general, $\delta\mu(t)$ oscillates out of phase with the volume $\delta v(t)$, and thus we make the ansatz $\delta\mu(t) = {\rm Re}[\delta\mu_{0}e^{i\omega t}]$, with 
a complex amplitude $\delta\mu_{0}$. The differential equation (\ref{dmu0}) then yields algebraic equations for real and imaginary parts of $\delta\mu_{0}$, which can easily be solved,
\be \label{ReImdmu}
{\rm Re}\,\delta\mu_0 = -\frac{B\omega^2\delta v_0}{\omega^2+(C\lambda)^2} \, , \qquad 
{\rm Im}\,\delta\mu_0 = -\frac{BC\lambda\omega\delta v_0}{\omega^2+(C\lambda)^2} \, .
\ee
In order to obtain this simple analytical result it was crucial to use the linear approximation of the rate (\ref{dnd}). As we shall see,  higher powers in $\delta\mu$ do appear in the rate in general. The linear approximation is valid in the so-called `sub-thermal' regime $\delta\mu\ll T$. In the `supra-thermal' regime, however, the deviation from chemical equilibrium becomes of the order of the temperature and larger \cite{PhysRevD.46.3290,Alford:2010gw}.

With the help of Eq.\ (\ref{ReImdmu}) we can compute $\delta n_d$ from the integral in Eq.\ (\ref{dnd}), insert the result into the pressure (\ref{ptB}), and the result into the expression for the bulk viscosity (\ref{zeta1}). This yields 
\be\label{zeta2}
\zeta(\omega) = \frac{\lambda B^2}{(\lambda C)^2+\omega^2} \, . 
\ee
For given values of the external frequency $\omega$ and the equilibrium functions $B$ and $C$, the bulk viscosity has a maximum at $\lambda=\omega/C$. In this sense, bulk viscosity is a resonance phenomenon. In particular, the bulk viscosity relevant for the physics of neutron stars is frequency dependent. The zero-frequency limit is not sufficient because there are re-equilibration processes that occur on the same time scale as the external oscillation.  As we shall see, the linearized rate $\lambda$ increases monotonically with temperature (while for $B$ and $C$ we can, to a good approximation, use the zero-temperature limit). This implies that the viscosity for a given external frequency $\omega$ has a maximum at a certain temperature. It turns out that this temperature is of the order of 1 MeV or even larger. Thus, bulk viscosity, in contrast to shear viscosity, is particularly important for young neutron stars and potentially for neutron star mergers. The purpose of the following subsection is to calculate $\lambda$.

\subsection{Calculating the rate of $u+d\leftrightarrow u+s$ in 2SC quark matter}

\subsubsection{Setup}

Our starting point is the kinetic equation, which arises from the more general Kadanoff-Baym equation \cite{Kadanoff} via a gradient expansion,
\be \label{kinetic}
i\frac{\partial}{\partial t}{\rm Tr}[\gamma_0S^<(P_1)] = -{\rm Tr}[S^>(P_1)\Sigma^<(P_1) - 
\Sigma^>(P_1)S^<(P_1)] \, ,
\ee
where $S^<$ and $S^>$ are the lesser and greater quark propagators in momentum space, $\Sigma^<$ and $\Sigma^>$ are the lesser and greater quark self-energies, $P_1$ is the $d$ quark four-momentum, and the trace is taken over color, flavor, Dirac, and Nambu-Gorkov space.
The greater and lesser propagators are an important ingredient of the real-time formalism, see for instance Ref.\ \cite{lebellac}.  By inserting the equilibrium propagators, the kinetic equation will give us the net number of $d$ quarks produced per unit time and volume $\Gamma$  through the process $u+d\leftrightarrow u+s$. This process enters the equation through the appropriate choice of the self energy.  The left-hand side of the kinetic equation essentially is $\Gamma$ (after integrating over momentum and up to a numerical factor), and our task is to compute the right-hand side, which gives the gain and loss terms due to the processes $u+s\to u+d$ and $u+d\to u+s$. Starting from the kinetic equation rather than writing down the rate more directly from Fermi's Golden Rule is particularly useful for superconducting phases since it naturally yields all sub-processes that are specific to Cooper-paired systems and whose form and even existence are not obvious. 

\begin{figure}[h]
   \begin{center}
     \includegraphics[width=0.5\textwidth]{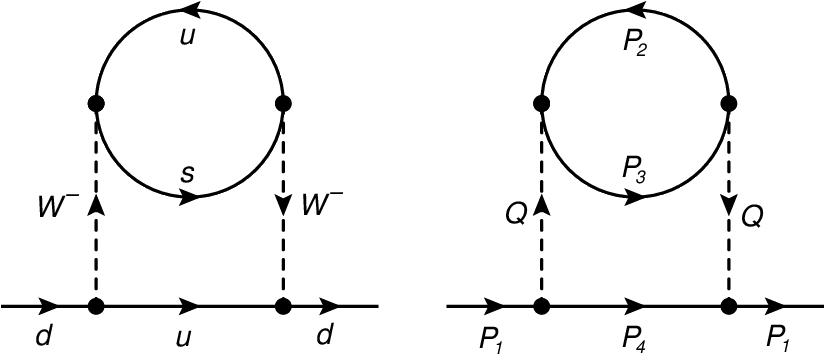}
     \end{center}
  \caption{\label{fig:Sigmaudus} {\it Left:} Quark self-energy needed for the calculation of the rate $u+d\leftrightarrow u+s$. {\it Right:} The diagram is duplicated to show our notation for the four-momenta of the various propagators. Due to energy-momentum conservation, $Q=P_1-P_4=P_3-P_2$.} 
\end{figure}

The self-energies $\Sigma^{<,>}$ are given by the diagram shown in Fig.\ \ref{fig:Sigmaudus}. 

Cutting this diagram in half yields the processes we are interested in. Since the $W$-boson mass $M_W$ is much larger than any other energy involved, the $W$-boson lines can be  reduced to pointlike interactions, and the self-energies are 
\be \label{dquarkself}
\Sigma^{<,>}(P_1) = \frac{i}{M_W^4}
\int_{P_4}\,\Gamma_{ud,-}^\mu \,S^{<,>}(P_4)\,\Gamma_{ud,+}^\nu\,
\Pi_{\mu\nu}^{>,<}(Q) \, ,
\ee
where we have abbreviated the four-momentum integration by 
\be
\int_{P}\equiv  \int\frac{d^4P}{(2\pi)^4} \, ,
\ee
and where $\Gamma_{ud,\pm}^\mu$ are  the vertices of the subprocesses
$d\leftrightarrow u + W^-$.  
The $W$-boson polarization tensors
are 
\be \label{polardef}
\Pi_{\mu\nu}^{<,>}(Q) = - i\int_{P_2}
{\rm Tr}[\Gamma_{us,+}^\mu \,S^{>,<}(P_3)\,\Gamma_{us,-}^\nu\,
S^{<,>}(P_2)] \, ,
\ee
where $\Gamma_{us,\pm}^\mu$ are the vertices of the subprocesses $u + W^-\leftrightarrow s$. In the diagram, the $W$-boson polarization tensor corresponds to the quark loop of $u$ and $s$ quarks. This is analogous to the photon and gluon polarization tensors used in the calculation of the Meissner masses in Sec.\ \ref{sec:Meissner}. 
As the quark-gluon vertices in the calculation of the Meissner masses, the electroweak vertices have a nontrivial structure in Nambu-Gorkov space,
\be \label{weakvertex}
\Gamma_{ud/us,\pm}^\mu = \frac{e\,V_{ud/us}}{2\sqrt{2}\sin\theta_W}\left(\begin{array}{cc} 
\gamma^\mu(1-\gamma^5)\,\tau_{ud/us,\pm} & 0 \\ 
0 &  -\gamma^\mu(1+\gamma^5)\,\tau_{ud/us,\mp}\end{array}\right) \, ,
\ee
where $V_{ud}\simeq 0.97$ and $V_{us}\simeq 0.22$ are elements of the Cabibbo-Kobayashi-Maskawa (CKM)  matrix, $\theta_W$ is the Weinberg angle, and 
\bea
\tau_{ud,+} &\equiv& \left(\begin{array}{ccc}   0 & 1 &0 \\ 0 &0&0 \\ 0&0&0 \end{array}\right) \, , 
\qquad \tau_{ud,-} \equiv \left(\begin{array}{ccc}   0 & 0 &0 \\ 1 &0&0 \\ 0&0&0 \end{array}\right) \, , \non[2ex]
\tau_{us,+} &\equiv& \left(\begin{array}{ccc}   0 & 0 &1 \\ 0 &0&0 \\ 0&0&0 \end{array}\right) \, , 
\qquad \tau_{us,-} \equiv \left(\begin{array}{ccc}   0 & 0 &0 \\ 0 &0&0 \\ 1&0&0 \end{array}\right) 
\eea
are matrices in flavor space, with rows and columns labeled in the order $u,d,s$.

Finally, we have to specify the 
quark propagators. The structure  in Nambu-Gorkov space from Eq.\ (\ref{Sfull}) carries over to the greater and lesser propagators,
\be \label{propnambu}
S^{<,>} = \left(\begin{array}{cc} G_+^{<,>} & F_-^{<,>} \\[2ex] F_+^{<,>} & G_-^{<,>}
\end{array}\right)\, . 
\ee
The normal propagators $G_\pm^{<,>}$ and anomalous propagators $F_\pm^{<,>}$  contain the color-flavor structure of the 2SC phase. From Eqs.\ (\ref{Mphi}) and (\ref{CFL2SC}) we read off the  2SC order parameter ${\cal M}=J_3I_3$, with $J_A$ and $I_B$ defined below Eq.\ (\ref{Mphi}). Therefore, ${\cal M}^2=J_3^2I_3^2$ with $J_3^2={\rm diag}(1,1,0)$ in color space and $I_3^2={\rm diag}(1,1,0)$ in flavor space. This shows that there are 4 gapped quasiparticles with identical gap and 5 ungapped quasiparticles. The corresponding projection operator for the gapped branches is simply ${\cal P}_1={\cal M}^2$. This structure shows that $ru$, $rd$, $gu$, and $gd$ quarks participate in pairing. For simplicity, we shall assume all quarks to be  massless and the chemical potentials for $u$ and $d$ quarks to be identical, 
\be
\mu\equiv \mu_u=\mu_d \, .
\ee
Since we are interested in a deviation from chemical equilibrium, the chemical potential for strange quarks $\mu_s$ of course must be allowed to be different. Because of this distinction between unpaired strange quarks and unpaired up and down quarks, we introduce two projectors onto the unpaired sector, namely ${\cal P}_2 = I_3^2(1-J_3^2)$ (for $bu$ and $bd$ quarks) and ${\cal P}_3=1-I_3^2$ (for $rs$, $gs$, and $bs$ quarks). Obviously, ${\cal P}_1+{\cal P}_2+{\cal P}_3=1$. Then, we can write the normal and anomalous propagators as 
\begin{subequations}\label{2SCpropagators}
\bea
G_\pm^{<,>}(K) &=& \sum_{r=1}^3 {\cal P}_r G_{\pm,r}^{<,>}(K)\,\gamma^0\Lambda_{k}^\mp \, , \\[2ex]
F_\pm^{<,>}(K) &=& J_3 I_3 F_{\pm,1}^{<,>}(K)\,\gamma^5\Lambda_{k}^\mp \, , 
\eea
\end{subequations}
with the energy projectors $\Lambda_{k}^\mp$ from Eq.\ (\ref{Lambdake}). We have omitted the anti-particle contribution (the $\pm$ signs correspond to charge conjugation). The anomalous propagators are proportional to the order parameter matrix ${\cal M} = J_3I_3$ and only the gapped branch, labeled by 1, appears.   Written in this way, we have made all color, flavor, and Dirac structure explicit, and introduced  greater and lesser propagators and their charge-conjugate counterparts for the three different branches $r=1,2,3$,
\begin{subequations}\allowdisplaybreaks
\label{greaterlesser}
\bea
G_{\pm,r}^>(K) &=& -2\pi i\, \left\{B_{k,r}^\pm \, f(\epsilon_{k,r})\, \delta(k_0\pm\mu_r-\epsilon_{k,r}) +B_{k,r}^\mp \, [1-f(\epsilon_{k,r})]\, \delta(k_0\pm\mu_r+\epsilon_{k,r})\right\} \, , \\[2ex]
G_{\pm,r}^<(K) &=& -2\pi i\, \left\{B_{k,r}^\pm \, [1-f(\epsilon_{k,r})]\, \delta(k_0\pm\mu_r-\epsilon_{k,r})+B_{k,r}^\mp\, f(\epsilon_{k,r})\,\delta(k_0\pm\mu_r+\epsilon_{k,r})\right\} \, ,\\[2ex]
F_{\pm,r}^>(K) &=& 2\pi i\,\frac{\Delta}{2\epsilon_{k,r}}\Big\{f(\epsilon_{k,r})\,\delta(k_0\mp\mu_r-\epsilon_{k,r}) -[1-f(\epsilon_{k,r})]\,\delta(k_0\mp\mu_r+\epsilon_{k,r})\Big\} \, ,\\[2ex]
F_{\pm,r}^<(K) &=& 2\pi i\,\frac{\Delta}{2\epsilon_{k,r}}\Big\{[1-f(\epsilon_{k,r})]\,\delta(k_0\pm\mu_r-\epsilon_{k,r})-f(\epsilon_{k,r})\,\delta(k_0\pm\mu_r+\epsilon_{k,r})\Big\} \, ,
\eea
\end{subequations}
with $\mu_1=\mu_2=\mu$, $\mu_3=\mu_s$, the Fermi-Dirac distribution function  $f(x)=1/(e^{x/T}+1)$, the dispersion relations 
\begin{subequations}
\bea
\epsilon_{k,1} &=& \sqrt{(k-\mu)^2+\Delta^2}\,  , \\[2ex]
\epsilon_{k,2} &=& |k-\mu|\,  , \\[2ex]
\epsilon_{k,3} &=& |k-\mu_s|\,  ,
\eea
\end{subequations}
and the Bogoliubov coefficients
\be \label{bogoliubov}
B_{k,r}^\pm\equiv \frac{\epsilon_{k,r}\pm (\mu_r-k)}{2\epsilon_{k,r}} \, .
\ee
These coefficients, and the existence of two terms in each of the propagators (\ref{greaterlesser}) reflects the fact that due to pairing the quasiparticles are mixtures of particles and holes, as shown in Fig.\ \ref{fig:dispersion}.

\subsubsection{$W$-boson polarization tensor}

Having collected all vertices and propagators, we can now compute the $W$-boson polarization tensor. Inserting Eqs.\ (\ref{weakvertex}) and (\ref{propnambu}) into Eq.\ (\ref{polardef}) and performing the trace over Nambu-Gorkov space yields 
\bea
\Pi_{\mu\nu}^{<,>}(Q) &=& -i \frac{e^2V_{us}^2}{8\sin^2\theta_W}\int_{P_2}\Big\{
{\rm Tr}[\gamma^\mu(1-\gamma^5)\,\tau_{us,+}\,G_+^{>,<}(P_3)\,\gamma^\nu(1-\gamma^5)\,\tau_{us,-}\,G_+^{<,>}(P_2)] \non[2ex]
&&+\;{\rm Tr}[\gamma^\mu(1+\gamma^5)\,\tau_{us,-}\,G_-^{>,<}(P_3)\,\gamma^\nu(1+\gamma^5)\,\tau_{us,+}\,G_-^{<,>}(P_2)] \non[2ex]
&&-\;{\rm Tr}[\gamma^\mu(1-\gamma^5)\,\tau_{us,+}\,F_-^{>,<}(P_3)\,\gamma^\nu(1+\gamma^5)\,\tau_{us,+}\,F_+^{<,>}(P_2)] \non[2ex]
&&-\;{\rm Tr}[\gamma^\mu(1+\gamma^5)\,\tau_{us,-}\,F_+^{>,<}(P_3)\,\gamma^\nu(1-\gamma^5)\,\tau_{us,-}\,F_-^{<,>}(P_2)] 
\Big\}
\, .\label{afternambu}
\eea
For the color-flavor traces we need
\begin{subequations} \label{tauP1}
\bea
 \tau_{us,-}{\cal P}_1&=&\tau_{us,-}J_3^2 \, ,\\[2ex]
 \tau_{us,-}{\cal P}_2&=&\tau_{us,-}(1-J_3^2) \, ,\\[2ex]
\tau_{us,+}{\cal P}_3 &=& \tau_{us,+} \, ,\\[2ex]
\tau_{us,+}{\cal P}_1&=&\tau_{us,+}{\cal P}_2=\tau_{us,-}{\cal P}_3=\tau_{us,\pm}I_3\tau_{us,\pm}I_3=0
\, .
\eea
\end{subequations}
In particular, this renders the contribution from the anomalous propagators zero. One can show that both terms in Eq.\ (\ref{afternambu}) with normal propagators yield the same contribution, and thus we can continue with twice the first term. 
Then, denoting  
\be \label{diracT}
{\cal T}^{\mu\nu}_{kq}\equiv {\rm Tr}[\gamma^\mu(1-\gamma^5)\gamma^0\Lambda_{k}^-\gamma^\nu(1-\gamma^5)\gamma^0\Lambda_{q}^-]\, ,
\ee
we obtain 
\bea
\Pi_{\mu\nu}^{<,>}(Q) &=& -i \frac{e^2V_{us}^2}{4\sin^2\theta_W}\int_{P_2}
{\cal T}^{\mu\nu}_{p_3p_2}  \left[2G_{+,3}^{>,<}(P_3)G_{+,1}^{<,>}(P_2) + 
G_{+,3}^{>,<}(P_3)G_{+,2}^{<,>}(P_2)\right]\,
 \, .
\eea
Here, the factor 2 in the first term is due to the color degeneracy and arises from the color trace over $J_3^2$. Inserting the propagators from Eqs.\ (\ref{greaterlesser}), using 
$1-f(\epsilon)=f(-\epsilon)$, and performing the integral over $u$ quark energies $p_{20}$  yields
\bea
\label{Pi}
\Pi_{\mu\nu}^{<,>}(Q)&=&  i \frac{\pi
e^2V_{us}^2}{2\sin^2\theta_W}\sum_{e_2e_3}\int_{\vec{p}_2}
{\cal T}^{\mu\nu}_{p_3p_2} \Big\{2B_{p_3,3}^{e_3}B_{p_2,1}^{e_2}
\, f(\pm e_3\epsilon_{p_3,3})f(\mp e_2\epsilon_{p_2,1})\,\delta(q_0-e_3\epsilon_{p_3,3}+e_2\epsilon_{p_2,1}+\delta\mu) \non[2ex]
&&+\; 
B_{p_3,3}^{e_3}B_{p_2,2}^{e_2}
\, f(\pm e_3\epsilon_{p_3,3})f(\mp e_2\epsilon_{p_2,2})\,\delta(q_0-e_3\epsilon_{p_3,3}+e_2
\epsilon_{p_2,2}+\delta\mu)\Big\} \, , \hspace{1cm}
\eea
where the sums are over $e_2,e_3=\pm 1$, where the upper (lower) sign corresponds to $\Pi^<$ ($\Pi^>$), and where we have abbreviated  
integration over  three-momentum by
\be
\int_{\vec{p}}\equiv \int\frac{d^3\vec{p}}{(2\pi)^3} \, .
\ee

\subsubsection{Collision integral}

We now turn to evaluating the right-hand side of the kinetic equation (\ref{kinetic}), the `collision integral'. With the help of Eq.\ (\ref{dquarkself}), we have
\bea \label{SSigma12}
{\rm Tr}[S^{>,<}(P_1)\Sigma^{<,>}(P_1)]= \frac{i}{M_W^4}\int_{P_4}
{\rm Tr}[S^{>,<}(P_1)\Gamma_{ud,-}^\mu S^{<,>}(P_4) \Gamma_{ud,+}^\nu] \Pi^{>,<}_{\mu\nu}(Q) \, .
\eea
Taking the trace over Nambu Gorkov space yields
\bea
{\rm Tr}[S^{>,<}(P_1)\Gamma_{ud,-}^\mu S^{<,>}(P_4)\Gamma_{ud,+}^\nu] &=& \frac{e^2V_{ud}^2}{8\sin^2\theta_W}\Big\{ {\rm Tr}[G_+^{>,<}(P_1)\gamma^\mu(1-\gamma^5)\tau_{ud,-}G_+^{<,>}
(P_4)\gamma^\nu(1-\gamma^5)\tau_{ud,+}] \non[2ex]
&&+\;{\rm Tr}[G_-^{>,<}(P_1)\gamma^\mu(1+\gamma^5)\tau_{ud,+}G_-^{<,>}(P_4)\gamma^\nu(1+\gamma^5)\tau_{ud,-}]\non[2ex]
&&-\;{\rm Tr}[F_-^{>,<}(P_1)\gamma^\mu(1+\gamma^5)\tau_{ud,+}F_+^{<,>}(P_4)\gamma^\nu(1-\gamma^5)\tau_{ud,+}]\non[2ex]
&&-\;{\rm Tr}[F_+^{>,<}(P_1)\gamma^\mu(1-\gamma^5)\tau_{ud,-}F_-^{<,>}(P_4)\gamma^\nu(1+\gamma^5)\tau_{ud,-}]
\Big\} \, .
\label{afternambu21}
\eea
For the traces over color and flavor space we 
need
\begin{subequations}
\bea
\tau_{ud,\pm}{\cal P}_1 &=& \tau_{ud,\pm} J_3^2 \, ,\\[2ex]
\tau_{ud,\pm}{\cal P}_2 &=& \tau_{ud,\pm} (1-J_3^2) \, ,\\[2ex]
\tau_{ud,\pm}{\cal P}_3 &=& 0 \, ,\\[2ex]
\tau_{ud,+}I_3\tau_{ud,+}I_3 &=& I_1^2-1\, , \\[2ex]
\tau_{ud,-}I_3\tau_{ud,-}I_3 &=& I_2^2-1 \, .
\eea
\end{subequations}
In contrast to the $W$-boson polarization tensor, now 
the traces containing anomalous propagators do not vanish. The four terms in Eq.\ (\ref{afternambu21}) yield pairwise identical contributions, such that we may keep one of the terms containing normal propagators and one of the terms containing anomalous propagators and multiply the result by 2.  We find
\bea \label{rhs1}
{\rm Tr}[S^{>,<}(P_1)\Sigma^{<,>}(P_1)]&=& i\frac{e^2V_{ud}^2}{4M^4_W\sin^2\theta_W}\int_{P_4}\Big\{\left[2G_{+,1}^{>,<}(P_1)G_{+,1}^{<,>}(P_4) + 
G_{+,2}^{>,<}(P_1)G_{+,2}^{<,>}(P_4)\right]{\cal T}^{\mu\nu}_{p_4p_1} \non[2ex]
&&+\; 2F_{+,1}^{>,<}(P_1)F_{-,1}^{<,>}(P_4)
\,{\cal U}^{\mu\nu}_{p_4p_1}\Big\}\,\Pi_{\mu\nu}^{>,<}(Q) \, , 
\eea
where we have abbreviated the Dirac trace 
\be \label{diracU}
{\cal U}^{\mu\nu}_{kq}\equiv {\rm Tr}[\gamma^\mu(1-\gamma^5)\gamma^5\Lambda_{k}^+\gamma^\nu
(1+\gamma^5)\gamma^5\Lambda_{q}^-] \, .
\ee
The different contributions in curly brackets in Eq.\ (\ref{rhs1}) describe the possible variants of the process $d\leftrightarrow W^- + u$ and can be understood as follows. The gapped $u$ and $d$ quarks, $r=1$, yield normal and anomalous contributions. Since there are two colors of gapped $u$ and $d$ quarks, these contributions come with a factor 2. The other contribution comes from the mode, $r=2$, the ungapped $u$ and $d$ quarks of a single color, which, of course, do not yield an 
anomalous contribution. Due to color conservation of the electroweak interaction, no mixing of the two 
modes is possible.

Eventually, we shall integrate both sides of the kinetic equation 
over the $d$-quark four-momentum $P_1=(p_{10},\vec{p}_1)$ because we are interested in the total rate, not in 
the rate as a function of momentum.
To this end, we need the result 
\bea 
\int_{-\infty}^{\infty}\frac{dp_{10}}{2\pi}{\rm Tr}[S^{>,<}(P_1)\Sigma^{<,>}(P_1)]
&=&-i\frac{e^2V_{ud}^2}{4M_W^4\sin^2\theta_W}\sum_{e_1e_4}\int_{\vec{p}_4}\Bigg\{2\left[B_{p_1,1}^{e_1}B_{p_4,1}^{e_4}\,
{\cal T}^{\mu\nu}_{p_4p_1} + \frac{e_1e_4\Delta^2}{4\epsilon_{p_1,1}\epsilon_{p_4,1}}{\cal U}^{\mu\nu}_{p_4p_1}
\right]\non[2ex]
&&\times f(\pm e_1\epsilon_{p_1,1})f(\mp e_4\epsilon_{p_4,1})\,
\Pi_{\mu\nu}^{>,<}(e_1\epsilon_{p_1,1}-e_4\epsilon_{p_4,1},\vec{q})\non[2ex]
&&+\,B_{p_1,2}^{e_1}B_{p_4,2}^{e_4}\,
{\cal T}^{\mu\nu}_{p_4p_1}\,f(\pm e_1\epsilon_{p_1,2})f(\mp e_4\epsilon_{p_4,2})\,
\Pi_{\mu\nu}^{>,<}(e_1\epsilon_{p_1,2}-e_4\epsilon_{p_4,2},\vec{q})\Bigg\} \,,
\label{trSSigma}
\eea
where we have performed the integration over the $u$ quark energy $p_{40}$ and the $d$ quark energy $p_{10}$, and where we have used the propagators (\ref{greaterlesser}). 
Now we insert the polarization tensors from Eq.\ (\ref{Pi}) into Eq.\ (\ref{trSSigma}).
Evaluating the result requires us to calculate the contractions
\begin{subequations}
\bea
{\cal T}^{\mu\nu}_{p_4p_1}{\cal T}_{\mu\nu}^{p_3p_2}&=& 16(1-x_{34})(1-x_{12}) \, , \\[2ex]
{\cal U}^{\mu\nu}_{p_4p_1}{\cal T}_{\mu\nu}^{p_3p_2} &=& 8\left[
(1-x_{34})(1-x_{12})+(1+x_{13})(1+x_{24})-(1+x_{14})(1+x_{23})
\right]\non[2ex]
&&+8i\left[(\up_2+\up_3)\cdot(\up_1\times\up_4)+
(\up_1+\up_4)\cdot(\up_2\times\up_3)\right] \, ,\hspace{0.3cm}\label{UT2}
\eea
\end{subequations}
where we have abbreviated  
\be \label{xmn}
x_{mn} \equiv \up_m\cdot\up_n = \cos\theta_{mn} \, ,
\ee
where $\theta_{mn}$ is the angle between the two 
three-momenta $\vec{p}_m$ and $\vec{p}_n$ with $m,n\in\{1,2,3,4\}$. 
Upon angular integration in the collision integral, the imaginary part in Eq.\ (\ref{UT2}) vanishes. Therefore, we drop this term from now on.

As mentioned above, the left-hand side of the kinetic equation (\ref{kinetic}) essentially becomes $\Gamma$, the 
net change of $d$ quarks per time and volume. In order to define $\Gamma$ properly and extract the correct numerical factor, we integrate the 
left-hand side over $P_1$ and project onto the $d$-flavor subspace. The latter is done by inserting 
the projector ${\cal Q}_d\equiv {\rm diag}(0,1,0)$. Moreover, to obtain the $d$ quark number density $n_d$ we also have to insert  
the matrix $\tau_3\equiv{\rm diag}(1,-1)$ in Nambu-Gorkov space (without pairing, the Nambu-Gorkov space would simply be a doubling of degrees of freedom, and the minus sign ensures that the charge-conjugate fermions yield the same contribution to the density as the original fermions). Then,    
\bea \label{leftside}
&&i\frac{\partial}{\partial t} \int_{P_1}{\rm Tr}[\gamma_0{\cal Q}_d \tau_3 S^<(P_1)] = -4\frac{\partial n_d}{\partial t} \equiv -4\Gamma \, ,
\eea
where the factor 4 originates from the Nambu-Gorkov doubling and spin degeneracy. 

\begin{figure}[h]
   \begin{center}
     \includegraphics[width=0.7\textwidth]{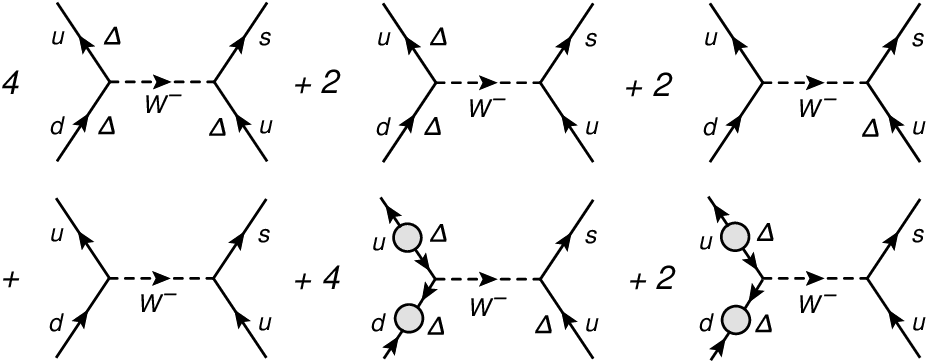}\end{center}
\caption{\label{fig:udus} Subprocesses contributing to the rate of the process $u+d\to u+s$ in the 2SC phase. Gapped branches are indicated by a $\Delta$ at the quark lines, with $s$ quarks being always ungapped. The two last diagrams contain anomalous propagators, indicated by the condensate insertion. }
\end{figure}

 Putting the results for the left-hand and right-hand sides of the kinetic equation
together and dividing both sides by 4 yields 
\be \label{collision}
\Gamma=
4\Gamma^{1131}  + 2 \Gamma^{1231}+2\Gamma^{2132} + \Gamma^{2232}+ 
4 \widetilde{\Gamma}^{1131}+ 2\widetilde{\Gamma}^{1231}  
 \, . 
\ee
There are contributions from normal propagators,
\bea \label{gammad}
&&\Gamma^{r_1r_2r_3r_4}\equiv 128\pi^4 G_F^2V^2_{ud}V^2_{us}\sum_{e_1e_2e_3e_4} 
\int_{\vec{p}_1}\int_{\vec{p}_2}\int_{\vec{p}_3}\int_{\vec{p}_4}\,\delta(\vec{p}_1+\vec{p}_2-\vec{p}_3-\vec{p}_4)\delta(e_1\epsilon_1+e_2\epsilon_2-e_3\epsilon_3-e_4\epsilon_4+\delta\mu) \non[2ex]
&&\times\,(1-x_{12})(1-x_{34})\,B^{e_1}_1B^{e_2}_2B^{e_3}_3B^{e_4}_4\left[f(e_1\epsilon_1)f(e_2\epsilon_2)f(-e_3\epsilon_3)f(-e_4\epsilon_4)-f(-e_1\epsilon_1)f(-e_2\epsilon_2)f(e_3\epsilon_3)f(e_4\epsilon_4)\right] \, , \hspace{0.5cm}
\eea
and from anomalous propagators, 
\bea
\widetilde{\Gamma}^{r_1r_2r_3r_4}&\equiv& 64\pi^4 G_F^2V^2_{ud}V^2_{us}\sum_{e_1e_2e_3e_4} 
\int_{\vec{p}_1}\int_{\vec{p}_2}\int_{\vec{p}_3}\int_{\vec{p}_4}\,\delta(\vec{p}_1+\vec{p}_2-\vec{p}_3-\vec{p}_4)\delta(e_1\epsilon_1+e_2\epsilon_2-e_3\epsilon_3-e_4\epsilon_4+\delta\mu)  \non[2ex]
&&\times\,\left[(1-x_{12})(1-x_{34})+(1+x_{13})(1+x_{24})
-(1+x_{14})(1+x_{23})\right]\frac{e_1\Delta}{2\epsilon_1}
B_2^{e_2}B_3^{e_3}\frac{e_4\Delta}{2\epsilon_4}
\non[2ex]
&&\times\,\left[f(e_1\epsilon_1)f(e_2\epsilon_2)f(-e_3\epsilon_3)f(-e_4\epsilon_4)
-f(-e_1\epsilon_1)f(-e_2\epsilon_2)f(e_3\epsilon_3)f(e_4\epsilon_4)\right] \, .\label{tildegammad}
\eea
We have expressed the factor from the electroweak vertices in terms of the  Fermi coupling constant 
\be \label{FermiCoupling}
G_F = \frac{\sqrt{2}e^2}{8M_W^2\sin^2\theta_W} = 1.16637\cdot 10^{-11} \, {\rm MeV}^{-2} \, .
\ee
Also, we have introduced an additional integration over $\vec{p}_3$ and the corresponding $\delta$-function for 
momentum conservation, which was  implicitly present in the previous expressions. Moreover, we have abbreviated
the quasiparticle energies and Bogoliubov coefficients by
\be
\epsilon_i\equiv \epsilon_{p_i,r_i} \, , \qquad 
B_i^{e_i}\equiv B_{p_i,r_i}^{e_i} \, .
\ee
Next we need to perform the angular integral. The 2SC phase is isotropic, i.e., none of the excitation energies and Bogoliubov coefficients depend on any angle, and thus we need to compute integrals of the form
\be
K\equiv \int d\Omega_1\int d\Omega_2\int d\Omega_3\int d\Omega_4 (1-x_{12})(1-x_{34})
\delta(\vec{p}_1+\vec{p}_2-\vec{p}_3-\vec{p}_4) \, .
\ee
We start with the $d\Omega_2$ integration and, to this end, abbreviate $\vec{P}\equiv \vec{p}_3 + \vec{p}_4$ and factorize the $\delta$-function in terms of modulus and angles,
\be
\delta[\vec{p}_2-(\vec{P}-\vec{p}_1)] = \frac{1}{p_2^2}\delta(p_2-|\vec{P}-\vec{p}_1|)
\delta(\Omega_2-\Omega) \, ,
\ee
where $\Omega$ stands for the angles of the vector $\vec{P}-\vec{p}_1$. Thus, 
the effect of the angular $\delta$-function is to replace the direction $\up_2$ with 
$(\vec{P}-\vec{p}_1)/|\vec{P}-\vec{p}_1|$. Consequently, recalling the definition (\ref{xmn}), we obtain
\be
K = \frac{1}{p_2^2}\int d\Omega_1\int d\Omega_3\int d\Omega_4\left(1-\frac{\up_1\cdot \vec{P}-p_1}{p_2}\right)(1-x_{34})\delta(p_2-|\vec{P}-\vec{p}_1|) \, .
\ee
Now, the $d\Omega_1$ integration can be performed by choosing the $z$-axis to 
be parallel to $\vec{P}$. We compute 
\be
\int d\Omega_1\left(1-\frac{\up_1\cdot \vec{P}-p_1}{p_2}\right)\delta(p_2-|\vec{P}-\vec{p}_1|) = \frac{\pi}{p_1^2}\frac{1}{P}\Theta(P-p_{12})\Theta(P_{12}-P)
(P_{12}^2-P^2) \, ,
\ee
where we have used  $\Theta(a-|b|)=\Theta(a-b)\Theta(a+b)$
and introduced the abbreviations 
\be
p_{ij}\equiv|p_i-p_j|  \, , \qquad P_{ij} \equiv p_i+p_j \, . 
\ee
The remaining angular dependence is reduced to the angle between $\vec{p}_3$ and $\vec{p}_4$, which is present in $P=|\vec{p}_3+\vec{p}_4|$.
Hence, one integration, say $d\Omega_4$, is trivial while we are left with the azimuthal integral from 
$d\Omega_3$,
\bea
K&=&\frac{8\pi^3}{p_1^2p_2^2}\int_{-1}^1 dx\,\frac{1-x}{P}\Theta(P-p_{12})\Theta(P_{12}-P)(P_{12}^2-P^2)
\non[2ex]
&=&\frac{4\pi^3}{p_1^2p_2^2p_3^2p_4^2}\int_{p_{34}}^{P_{34}}dP\,\Theta(P-p_{12})\Theta(P_{12}-P)(P_{12}^2-P^2)
(P_{34}^2-P^2) \, .\hspace{1cm}
\eea
Since $P_{ij}>p_{ij}$, there are four orders of $p_{12},p_{34},P_{12},P_{34}$ that yield non-vanishing
integrals,
\begin{subequations}\label{orders}
\bea
&&p_{12}<p_{34}<P_{12}<P_{34} \, ,\\[2ex]
&&p_{34}<p_{12}<P_{12}<P_{34} \, ,\\[2ex]
&&p_{34}<p_{12}<P_{34}<P_{12} \, , \\[2ex]
&&p_{12}<p_{34}<P_{34}<P_{12} \, .
\eea
\end{subequations}
The two other possibilities, $p_{12}<P_{12}<p_{34}<P_{34}$, and $p_{34}<P_{34}<p_{12}<P_{12}$ lead 
to vanishing integrals. Consequently, 
\be
K = \frac{4\pi^3}{p_1^2p_2^2p_3^2p_4^2}\,L(p_{12},P_{12},p_{34},P_{34}) \, ,
\ee
where we have defined
\bea \label{Labcd}
L(a,b,c,d)&\equiv&\Theta(c-a)\Theta(d-b)\Theta(b-c)
\,J(c,b,b,d) +\Theta(a-c)\Theta(d-b)
\,J(a,b,b,d) \non[2ex]
&&+\,\Theta(a-c)\Theta(b-d)\Theta(d-a)
\,J(a,d,b,d) + \Theta(c-a)\Theta(b-d)
\,J(c,d,b,d) \, , 
\eea     
with 
\bea \label{Jabcd}
J(a,b,c,d)&\equiv&\int_a^b dP\,(c^2-P^2)(d^2-P^2) 
=c^2d^2(b-a) - \frac{1}{3}(c^2+d^2)(b^3-a^3)+\frac{1}{5} (b^5-a^5) \, . 
\eea
This is the result for the angular integration in Eq.\ (\ref{gammad}) and for the first term in Eq.\ (\ref{tildegammad}). The two additional terms in Eq.\ (\ref{tildegammad}) are easily obtained by a relabeling of momenta. Then, with the 
definitions
\begin{subequations}\label{IItilde}
\bea
I(p_1,p_2,p_3,p_4)&\equiv& L(p_{12},P_{12},p_{34},P_{34}) \, , \label{I}\\[2ex]
\tilde{I}(p_1,p_2,p_3,p_4)&\equiv& L(p_{12},P_{12},p_{34},P_{34}) + L(p_{24},P_{24},p_{13},P_{13}) + L(p_{14},P_{14},p_{23},P_{23}) \, , 
\eea
\end{subequations} 
we arrive at 
\begin{subequations} \label{afterang}
\bea \label{afterang1}
\Gamma^{r_1r_2r_3r_4}&\equiv& \frac{G_F^2V^2_{ud}V^2_{us}}{8\pi^5}\sum_{e_1e_2e_3e_4} 
\int_{p_1p_2p_3p_4} \,I(p_1,p_2,p_3,p_4)
\,B^{e_1}_1B^{e_2}_2B^{e_3}_3B^{e_4}_4\delta(e_1\epsilon_1+e_2\epsilon_2-e_3\epsilon_3-e_4\epsilon_4+\delta\mu) \non[2ex] &&\times \left[f(e_1\epsilon_1)f(e_2\epsilon_2)f(-e_3\epsilon_3)f(-e_4\epsilon_4) -f(-e_1\epsilon_1)f(-e_2\epsilon_2)f(e_3\epsilon_3)f(e_4\epsilon_4)\right] \, , \\[2ex]
\label{afterang2}
\widetilde{\Gamma}_d^{r_1r_2r_3r_4}&\equiv& \frac{G_F^2V^2_{ud}V^2_{us}}{16\pi^5} \sum_{e_1e_2e_3e_4} 
\int_{p_1p_2p_3p_4} \,\tilde{I}(p_1,p_2,p_3,p_4)
\,\frac{e_1\Delta}{2\epsilon_1}
B_2^{e_2}B_3^{e_3}\frac{e_4\Delta}{2\epsilon_4}\delta(e_1\epsilon_1+e_2\epsilon_2-e_3\epsilon_3-e_4\epsilon_4+\delta\mu) \non[2ex]
&&\times\left[f(e_1\epsilon_1)f(e_2\epsilon_2)f(-e_3\epsilon_3)f(-e_4\epsilon_4) -f(-e_1\epsilon_1)f(-e_2\epsilon_2)f(e_3\epsilon_3)f(e_4\epsilon_4)\right] \, ,
\eea
\end{subequations}
where we have abbreviated
\be
\int_{p_1p_2p_3p_4} \equiv \int_0^\infty dp_1\int_0^\infty dp_2\int_0^\infty dp_3\int_0^\infty dp_4 \,.
\ee
Let us pause at this point in the calculation to explain the structure of the result we have obtained. We should be looking at Eqs.\ (\ref{collision}) and (\ref{afterang}). In Eq.\ (\ref{collision}) we see that the rate receives contributions from 6 different terms. They correspond to 4 contributions where only normal propagators appear and 2 contributions which contain anomalous propagators. The 4 normal contributions contain all possible combinations of quasiparticle branches that can appear in the process $u+d\leftrightarrow u+s$. Recall that we labeled the branches by 1 (gapped $u$ and $d$ quarks), 2 (ungapped $u$ and $d$ quarks), and 3 (strange quarks, which are always ungapped). Therefore, for instance the rate $\Gamma^{1131}$ comes from the ingoing gapped $d$ and $u$ quarks (first pair of indices) and outgoing $s$ and $u$ quarks, of which the $u$ quark is gapped (second pair of indices). (The inverse process is also included, i.e., with $s$ and $u$ ingoing and $d$ and $u$ ingoing.) This contribution comes with a degeneracy factor 4 because at each of the electroweak vertices there are two possible colors, namely red and green, corresponding to the two colors of gapped $u$ and $d$ quarks. In contrast, the second term $\Gamma^{1231}$ only has degeneracy 2 because now the incoming $u$ quark is ungapped (labeled by 2), which fixes the color of the $u+W^-\to s$ vertex to be blue. The terms where both ingoing and outgoing quarks of one of the vertices are gapped (this can only happen at one vertex since the other vertex always has an ungapped $s$ quark attached to it) have counterparts where both gapped quarks are represented by anomalous propagators, $\widetilde{\Gamma}^{1131}$ and  $\widetilde{\Gamma}^{1231}$ in Eq.\ (\ref{collision}). We show all 6 terms in diagrammatic form in Fig.\ \ref{fig:udus}. 

Notice that one diagram (with degeneracy 1) only contains ungapped quarks. At low temperatures, the total rate is well approximated by this diagram because a gap in at least one of the dispersions of the participating quasiparticles induces exponential suppression at  temperatures much smaller than the gap. This exponential suppression, however, can be alleviated by a sufficiently large $\delta\mu$. For a detailed discussion see Ref.\ \cite{Alford:2006gy}.

The calculation of the rate is further complicated by the fact that each term in Eq.\ (\ref{collision}) is in turn composed of various subprocesses, as Eq.\ (\ref{afterang}) shows. Due to the sum over $e_1, e_2, e_3, e_4 =\pm$ there are in fact 16 such subprocesses (not all of them actually giving a nonzero contribution). What is the meaning of these subprocesses? Naively, had we written down an `ordinary' collision integral for the process $u+d\to u+s$ we would have expected a structure of the Fermi distribution functions of the form $f_df_u(1-f_s)(1-f_u)$ because this corresponds to the probability of finding the ingoing states occupied and the outgoing states empty. However, with $f(-x)=1-f(x)$, Eq.\ (\ref{afterang}) shows that all possible combinations of $f$ and $1-f$ formally appear. Therefore, the quasiparticles, which actually are momentum-dependent mixtures of particles and holes, can appear on either side of the reaction. This is a typical property of fermionic superconductors and superfluids, where particles can be created from or deposited into the condensate since particle number is spontaneously broken. The same effect is thus also present in superfluid $^3$He \cite{Vollhardt1990}. 
For a more detailed study of the contribution of the various subprocesses, based on a numerical evaluation, see Ref.\ \cite{Alford:2006gy}.

\subsubsection{Unpaired limit}

It is instructive to take the limit of unpaired quark matter from our complicated result. To this end, it is useful to slightly rewrite the Bogoliubov coefficients and the dispersion relation. We define
\be \label{newbogol}
\tilde{B}_i^{e_i} \equiv \frac{1}{2}\left(1+e_i\frac{p_i-\mu_i}{\tilde{\epsilon}_i}\right) \, ,
\qquad \tilde{\epsilon}_i\equiv {\rm sgn}(p_i-\mu_i)\epsilon_i \, .
\ee
Then we note that for any function $F$ we have
\be
\sum_{e_i}\int_0^\infty dp_i\, B_i^{e_i} F(e_i\epsilon_i) =
\sum_{e_i}\int_0^\infty dp_i\,\tilde{B}_i^{e_i} F(-e_i\tilde{\epsilon}_i) \, .  
\ee
Consequently, in Eqs.\ (\ref{afterang})  we can simply replace $B_i^{e_i}$ by 
$\tilde{B}_i^{e_i}$ and $e_i\epsilon_i$ by $-e_i\tilde{\epsilon}_i$. The advantage of the new formulation is that now $\tilde{B}_i^{+} = 1$ and $\tilde{B}_i^{-}=0$ if $\Delta =0$. More physically speaking, as we let  $\Delta\to 0$, the dispersions $\tilde{\epsilon}_i$ and $-\tilde{\epsilon}_i$ reduce to pure particle and hole branches, respectively, while $\epsilon_i$  becomes a hole below the Fermi surface and a particle above the Fermi surface and vice versa for $-\epsilon_i$. Therefore, after the replacement with the new Bogoliubov coefficients and excitation energies, the rate of the unpaired phase is given by the term $e_1=e_2=e_3=e_4=+1$ in Eq.\ (\ref{afterang1}) [the anomalous contribution (\ref{afterang2}) obviously vanishes completely for $\Delta=0$]. We therefore obtain
\bea \label{normalphase}
&&\Gamma_0 = \frac{N_c^2G_F^2V^2_{ud}V^2_{us}}{8\pi^5}\int_{p_1p_2p_3p_4} \,I(p_1,p_2,p_3,p_4)
\,\delta(p_1+p_2-p_3-p_4)\non[2ex]
&&\times\,\Big\{[1-f(p_1-\mu)][1-f(p_2-\mu)]f(p_3-\mu_s)f(p_4-\mu)  -f(p_1-\mu)f(p_2-\mu)[1-f(p_3-\mu_s)][1-f(p_4-\mu)]\Big\} \, , \hspace{0.7cm}
\eea
where we have reinstated the degeneracy factor $N_c^2$ -- one factor of $N_c$ for each of the electroweak vertices -- to obtain the complete result for three-flavor unpaired quark matter. 

We can now make use of the $\delta$-function to perform the $p_4$ integral, which yields a factor $\Theta(p_1+p_2-p_3)$ in the integrand. From now on, we shall assume $\delta\mu>0$. This means that the chemical potential for strange quarks is larger than in equilibrium and thus the system will react by converting $s$ quarks into $d$ quarks. Therefore, the first term in the curly brackets will give a larger contribution 
than the second term. It is thus convenient to use the explicit form of the Fermi distribution functions to rewrite the second term,
\bea \label{normalphase1}
\Gamma_0 &=& \frac{2N_c^2G_F^2V^2_{ud}V^2_{us}}{15\pi^5}\left(1-e^{-\delta\mu/T}\right)\int_0^\infty dp_1\int_0^\infty dp_2\int_0^\infty dp_3\,H(p_1,p_2,p_3) \non[2ex]
&&\times\,[1-f(p_1-\mu)][1-f(p_2-\mu)]f(p_3-\mu_s)f(p_1+p_2-p_3-\mu)\, ,
\eea
where we have also explicitly evaluated the function $I(p_1,p_2,p_3,p_4)$, which, using Eqs.\ (\ref{Labcd}), (\ref{Jabcd}), and (\ref{I}), yields 
\bea \label{Hppp}
H(p_1,p_2,p_3) &\equiv& [\Theta(p_2-p_3)-\Theta(p_1-p_3)]^2[\Theta(p_1-p_2)h_1+\Theta(p_2-p_1)h_2]\non[2ex]
&&+[\Theta(p_1+p_2-p_3)-\Theta(p_2-p_3)-\Theta(p_1-p_3)]\Theta(2p_3-p_1-p_2)h_3\non[2ex]
&&+\{1-[\Theta(p_2-p_3)-\Theta(p_1-p_3)]^2\}\Theta(p_1+p_2-2p_3)h_4 \, , 
\eea
with 
\begin{subequations}
\bea
h_1&\equiv&  p_2^3(10p_1^2+5p_1p_2+p_2^2) \, , \\[2ex]
h_2&\equiv&  p_1^3(10p_2^2+5p_1p_2+p_1^2) \, , \\[2ex]
h_3&\equiv&  (p_1+p_2-p_3)^3[(p_1+p_2)^2+3(p_1+p_2)p_3+6p_3^2] \, , \\[2ex]
h_4&\equiv&  p_3^3[10(p_1+p_2)^2-15(p_1+p_2)p_3+6p_3^2] \, .
\eea
\end{subequations}
The integral in Eq.\ (\ref{normalphase1}) is still too complicated to be evaluated analytically for arbitrary $\mu$, $T$, and $\delta\mu$. We shall therefore discuss two limit cases where analytical results can be obtained. Firstly, let us discuss the case $T=0$. In this case, the Fermi functions in Eq.\ (\ref{normalphase1}) become step functions, $\lim_{T\to 0}f(x) = \Theta(-x)$. As a consequence, all step functions in Eq.\ (\ref{Hppp}) are 1 or 0 in the entire integration domain except for $\Theta(p_1+p_2-p_3)$. Due to this step function, one has to distinguish the cases $\mu_s<2\mu$ and $\mu_s>2\mu$. We are mostly interested in small deviations from chemical equilibrium. Therefore, we omit the case $\mu_s>2\mu$ (the result can be found in  Ref.\ \cite{PhysRevD.47.325}). For the other case  and setting $N_c=3$ we find 
\bea \label{normalzero}
\Gamma_0(T=0,\delta\mu<\mu)&=&\frac{6}{5\pi^5}G_F^2V^2_{ud}V^2_{us}\int_\mu^{\mu_s} dp_3\int_\mu^{p_3} dp_2
\int_\mu^{p_3-p_2+\mu} dp_1\, h_3  \non[2ex]
&=&  \frac{16}{5\pi^5}G_F^2 V_{us}^2V_{ud}^2\left(\mu^5\delta\mu^3+\frac{5}{16}\mu^4\delta\mu^4
-\frac{3}{16}\mu^3\delta\mu^5 +\frac{1}{32}\mu^2\delta\mu^6+\frac{5}{112}\mu\delta\mu^7-\frac{15}{896}\delta\mu^8\right) \, .
\eea
 Secondly, we compute the rate for  $\delta\mu\ll T\ll\mu$ (`subthermal regime'). In this case we observe that the prefactor of the integral in Eq.\ (\ref{normalphase1}) is linear in $\delta\mu/T$ to lowest order. Neglecting all higher order corrections, we can thus set $\delta\mu=0$ in the integral. We then introduce the dimensionless integration variables
\be
x = \frac{p_1-\mu}{T} \,  , \qquad y=\frac{p_2-\mu}{T} \, , \qquad z=\frac{p_3-\mu_s}{T} \, ,
\ee
and approximate the lower boundaries of the $x$, $y$ and $z$ integrals by $-\infty$ because of $\mu,\mu_s\gg T$. Moreover, we may expand the functions $h_i$ for large $\mu/T$, which yields to lowest order $h_1\simeq h_2\simeq h_3\simeq h_4\simeq 16\mu^5$. Then, all step functions in Eq.\ (\ref{Hppp}) combine to 1 such that we arrive at 
\be \label{Gam0subT}
\Gamma_0(\delta\mu\ll T\ll\mu) 
= \frac{64G_F^2V_{ud}^2V_{us}^2}{5\pi^3}T^2\mu^5\delta\mu \, , 
\ee
where we have used
\be
\int_{-\infty}^{\infty}dx\int_{-\infty}^{\infty}dy\int_{-\infty}^{\infty}dz\,
[1-f(x)][1-f(y)]f(z)f(x+y-z) = \frac{2\pi^2}{3} \, .
\ee
Here, in a slight abuse of notation, the Fermi-Dirac function is $f(x)=1/(e^x+1)$, not $f(x)=1/(e^{x/T}+1)$. The result (\ref{Gam0subT}) was first computed in Ref.\ \cite{Heiselberg:1992bd}.

\subsection{Comparing different color superconductors}

\begin{figure}[t]
   \begin{center}
     \includegraphics[width=0.6\textwidth]{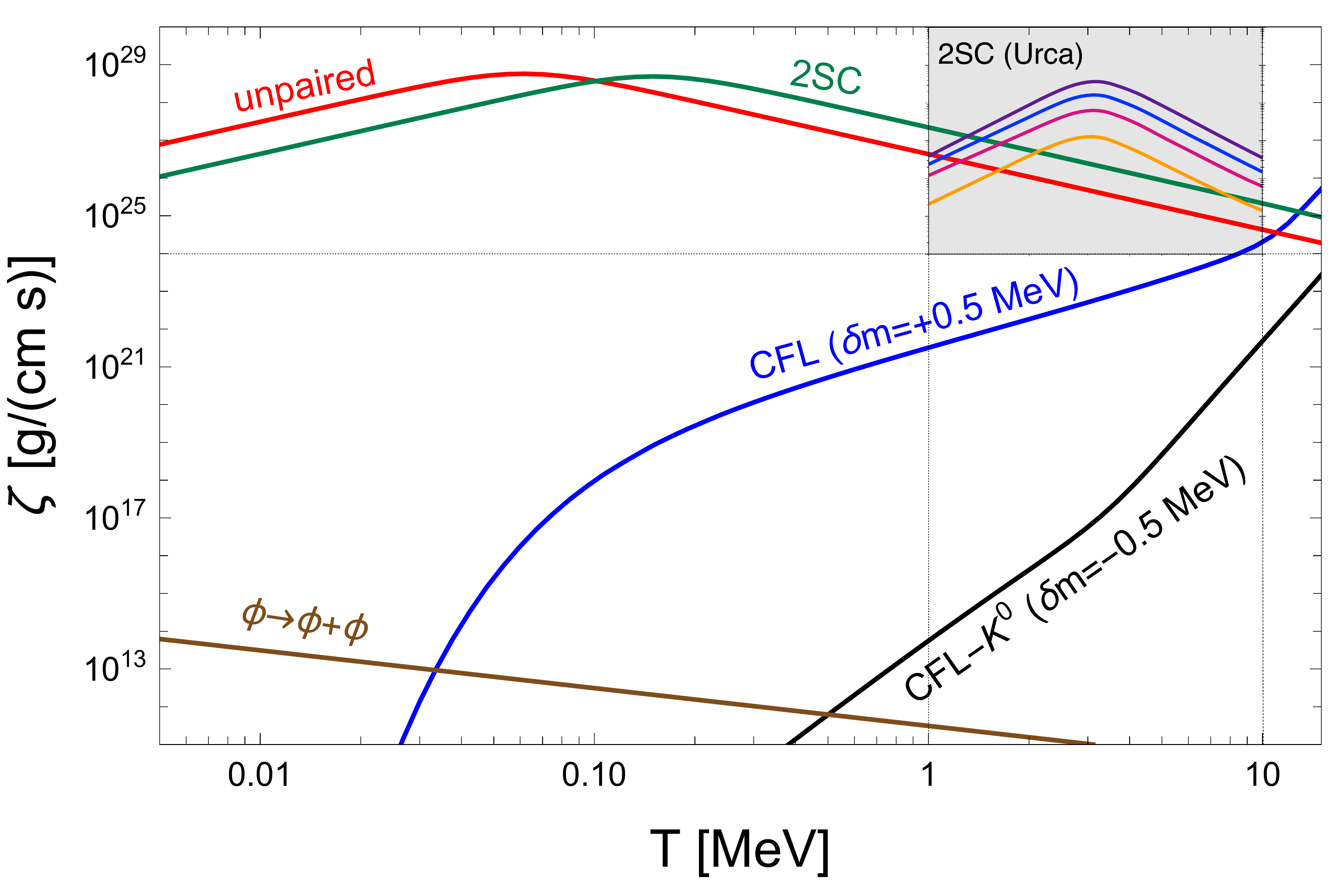}\end{center}
\caption{\label{fig:bulkvisc} Bulk viscosity $\zeta$ as a function of temperature for various color superconductors and unpaired quark matter. 
For all curves the external oscillation frequency is $\omega/(2\pi) = 1\, {\rm ms}^{-1}$, and for all curves except for the inset the quark chemical potential is $\mu=400\, {\rm MeV}$. The calculation presented in detail in the text gives the (red and green) curves for unpaired and 2SC results from the process $u+d\leftrightarrow u+s$ \cite{Alford:2006gy}.  The results for CFL are shown for thermal kaons \cite{Alford:2007rw},  $\delta m \equiv m_{K^0}-\mu_{K^0} > 0$, and in the presence of kaon condensation \cite{Alford:2008pb}, $\delta m <0$. In CFL there is also a contribution purely from the superfluid phonon $\phi$ \cite{Manuel:2007pz}. The inset shows the 2SC result from direct Urca processes \cite{Alford:2025tbp} for a baryon density of 7 times (nuclear) saturation density, employing an NJL model with different ratios for vector and scalar couplings, $G_V/G_S=0.6,0.8,1.0,1.2$ from top to bottom. }
\end{figure}

To summarize, we have computed the net $d$ quark production rate of the processes $u+d\leftrightarrow s+d$. We have derived the result for 2SC quark matter in the form of collision integrals which have to be evaluated numerically, and we have derived analytical limits for the case of unpaired quark matter. 
It remains to insert the result of the rate into the expression (\ref{zeta2}) for the frequency-dependent bulk viscosity. In this expression we already have assumed the rate to be linear in the deviation from equilibrium $\delta\mu$. Therefore, for unpaired quark matter, consistency requires us to work with the approximation (\ref{Gam0subT}).  As this analytical result shows explicitly, the reaction rate increases monotonically (and quadratically) with temperature. Therefore, as already anticipated  below Eq.\ (\ref{zeta2}), the bulk viscosity is expected to have a maximum for a certain temperature at given external frequency $\omega$. This behavior is borne out in Fig.\ \ref{fig:bulkvisc}, where the (red and green) curves for unpaired and 2SC quark matter are taken from Ref.\ \cite{Alford:2006gy}. For essentially the entire temperature range shown in this figure, the rate of $u+d\leftrightarrow s+d$ in the 2SC phase is well approximated by 1/9 of the value of the unpaired phase, due to the exponential suppression of all sub-processes that involve one or more gapped excitation branches (the green 2SC curve in Fig.\ \ref{fig:bulkvisc} assumes a critical temperature for the 2SC phase of $T_c=30\, {\rm MeV}$). 
While the electroweak rate of the 2SC phase is smaller than that of the unpaired phase, this is not true for the bulk viscosity. At a given temperature, it is not the largest reaction rate that gives the largest viscosity but the rate that is closest to the external frequency. For the slower rate of the 2SC phase a larger temperature is needed to be in resonance with the external oscillation. That is why the maximum of the bulk viscosity is shifted to a higher temperature due to  Cooper pairing.

The results of Fig.\ \ref{fig:bulkvisc} require an input for the thermodynamic functions $B$ and $C$, defined in Eqs.\ (\ref{BCabbr}). For the unpaired and 2SC curves from the process $u+d\leftrightarrow u+s$ shown here, the relations for non-interacting, zero-temperature quark matter have been used (augmented by the superconducting gap in the case of 2SC). As discussed above, see Eq.\ (\ref{Bms}), it is crucial to include a nonzero strange quark mass in the equilibrium quantities $B$ and $C$ because otherwise compression and expansion cannot induce an out-of-equilibrium situation (in the 2SC phase, the gap alone would suffice for this purpose because strange quarks are ungapped). Including the effect of interactions into the thermodynamic functions yields  corrections to the bulk viscosity. For unpaired quark matter, this question was addressed within perturbative QCD in Ref.\ \cite{CruzRojas:2024etx}, where also a comparison to strong-coupling results from the gauge-gravity duality was discussed. In the inset of Fig.\ \ref{fig:bulkvisc} we show 2SC results from direct Urca processes (\ref{leptonic}), where interactions computed in an NJL model are taken into account to evaluate the relevant susceptibilities. The curves shown in the inset, taken from Ref.\ \cite{Alford:2025tbp}, are valid for a fixed density that does not correspond exactly to the chemical potential of the other curves, and four different curves are shown for different values of the coupling strength in the model. Therefore, a direct quantitative comparison to the other curves has to be taken with care, the main point is rather as follows: If two reactions contribute to the {\it same} out-of-equilibrium chemical potential $\delta\mu$, the faster one dominates the bulk viscosity, even if the bulk viscosity was larger in the absence of the fast process. Here, however, Urca processes and non-leptonic processes re-equilibrate {\it different} $\delta\mu$'s. Therefore, the contribution from the Urca process, whose rate is slower than that of the non-leptonic one, {\it can} enhance the bulk viscosity at large temperatures and the inset suggests that this contribution is indeed relevant for instance for neutron star mergers. In unpaired quark matter, both classes of reactions were included within a single calculation in Ref.\ \cite{Sad:2007afd}. 

The other main point of Fig.\ \ref{fig:bulkvisc} is the comparison to the CFL bulk viscosity. In that case, the contribution of the quarks is negligible because all quarks acquire a gap in their dispersion. The main contribution comes from strangeness changing processes involving kaons. As discussed briefly in Sec.\ \ref{sec:cscstar}, it is expected that the CFL kaons condense. The figure shows the results for both cases, i.e., uncondensed thermal \cite{Alford:2007rw} and condensed kaons \cite{Alford:2008pb}. In both cases, the bulk viscosity arises from the interaction of the kaons with the Goldstone mode from superfluidity. The figure also shows the contribution from processes involving the Goldstone mode alone \cite{Manuel:2007pz}. We see that the bulk viscosity of CFL is, for most relevant  temperatures, negligibly small compared to that of 2SC and unpaired quark matter. Since the electroweak rate from kaons is smaller than the one from fermions, the bulk viscosity peaks at much larger temperatures. We see that a peak starts to develop at $T>10\, {\rm MeV}$, a regime particularly interesting for neutron star mergers. However, we need to keep in mind that in this regime we are approaching the critical temperature of kaon condensation \cite{Alford:2007qa}, and, in fact, the critical temperature of CFL itself. 

\section{Open questions and future directions}

Color superconductivity is well understood at ultra-high quark densities, where the QCD coupling is weak. Weak-coupling methods fail at moderate densities, and we have no first-principle understanding of color superconductivity in the regime relevant for neutron stars. Even if we were to ignore Cooper pairing or a strong coupling version thereof, the sign problem of lattice QCD prevents us from calculating by brute force the properties of cold and dense matter. Progress to overcome or circumvent the sign problem has been made from various different angles \cite{Aarts:2008rr,Cristoforetti:2014gsa,Langelage:2014vpa,Alexandru:2015sua,Marchis:2017oqi,Ratti:2022qgf,Basar:2023bwd}, but further breakthrough ideas are needed to compute the QCD phase diagram at finite density on the lattice. 

Besides the strong-coupling nature of the problem, color superconductivity at moderately high densities is difficult due to the effect of the strange quark mass, which disrupts the particularly symmetric paring pattern of CFL. A lot of progress has been made with the help of effective theories and phenomenological models, which give us an idea about the possible, less symmetric, candidate phases. It is still an open question, however, whether the CFL phase persists down to densities where the transition to nuclear matter occurs or whether there are non-CFL color superconductors in between. These phases are likely to contain unpaired fermionic modes, or at the very least fermionic modes with very small gaps,  and possibly break rotational and/or translational invariance. One possible route to better understand the breakdown density of CFL is to build on the recent progress in dense perturbative QCD with unpaired quark matter \cite{Kurkela:2009gj,Gorda:2021znl,Fernandez:2021jfr,Gorda:2023mkk}. Using these results and ideally combining them with the gap equation, progress could be made towards a more reliable (weak-coupling) calculation of the ratio $m_s^2/(\mu\Delta)$, which determines the mismatch in Fermi momenta that seeks to break CFL. 

The main `laboratory' to test the predictions and conjectured phases of color superconductivity are neutron stars. (It is conceivable that heavy-ion collisions may shed some light on color superconductivity in future experiments as well \cite{Kitazawa:2001ft,Schmitt:2016pre,Nishimura:2024kvz}.) Neutron stars are of course complicated objects, and many of their observable properties depend not only on the core but also on the outer low-density layers. It is therefore very difficult to find a `smoking gun' signature for the existence of deconfined quark matter (see Ref.\ \cite{Annala:2019puf} for an indication based on the equation of state) or color superconductivity. Nevertheless, more, and more precise, astrophysical data has become available and will become available in the near future. Most notably, new insights can be expected from the detection of gravitational waves from neutron star mergers \cite{LIGOScientific:2018hze} and possibly from isolated neutron stars \cite{2024arXiv240302066J}, with updated current \cite{Acernese_2015,Aasi_2015} and new \cite{Abbott_2017,Maggiore_2020} detectors, besides improved data to `measure' neutron star masses and radii, which also has seen impressive progress in recent years \cite{NANOGrav:2019jur,Riley:2019yda,Miller:2019cac,Fonseca:2021wxt,Miller:2021qha,Riley:2021pdl,Romani:2022jhd,Choudhury:2024xbk}. 

It is an interesting question whether color superconductivity plays a detectable role in a neutron star merger. Estimates of critical temperatures of most color superconductors indicate that Cooper pairing might survive even in the hot environment of a merger process. Therefore, it is worth investigating, for instance, if superfluid properties of CFL need to be taken into account in the hydrodynamic simulation, potentially within a two-fluid formalism that distinguishes between superfluid and normal components. Another aspect is the dynamical transition from hadronic to quark matter in a merger process, for instance if two neutron stars without quark core merge and the phase transition is triggered by the creation of hot and dense spatial regions. Even without color superconductivity this transition is treated non-dynamically in current simulations \cite{Most:2019onn,Tootle:2022pvd,Guo:2023som}. While adding color superconductivity to the equation of state is straightforward, further studies are needed to understand the effect of color superconductivity on the dynamical phase conversion under merger conditions. One may also ask whether color superconductivity can be linked to continuous gravitational waves from isolated neutron stars. Besides originating from potentially unstable $r$-modes, they can also be generated by `mountains', non-axisymmetric deformations of the star sustained by the internal structure. While the physics of the crust is the main candidate in the literature for the source of mountains, color superconductivity in the core can play a role if it is in a crystalline phase \cite{Knippel:2009st} or if it generates a lattice of magnetic flux tubes \cite{Glampedakis:2012qp,Haber:2017oqb}. 

Astrophysical constraints from mass and radius measurements and the tidal deformability extracted from neutron star mergers can be used for a model-independent constraint for the color-superconducting gap, as pointed out recently  \cite{Kurkela:2024xfh,Kurkela:2025acm}. The idea is to compute an upper bound on the condensation energy (\ref{OmOm0}) from  the astrophysical constraints, thermodynamic consistency, and the low-density nuclear matter equation of state \cite{Hebeler:2013nza}. This approach can be further improved in an obvious way by new, more constraining, data,  and by further improvements in our understanding of the low-density and ultra-high-density limits of the equation of state. The approach is agnostic to the mechanism that lowers the free energy of unpaired quark matter. One may ask whether other exotic non-perturbative phenomena compete with or coexist with color superconductivity at moderate densities. One candidate is the quarkyonic phase, the large-$N_c$ version of nuclear matter \cite{McLerran:2007qj}, modeled with the help of a layered structure of the Fermi sphere \cite{McLerran:2018hbz}. It is unknown whether a remnant or variant of this phase survives at $N_c=3$. Recently, Cooper pairing of quarks in the quarkyonic phase has been considered within a phenomenological approach \cite{Gartlein:2025zhd}, and it remains to be seen whether this phase can be confirmed in a more microscopic treatment. The quarkyonic phase is obtained fully dynamically within the gauge-gravity duality \cite{Kovensky:2020xif}. A natural question for future research is how realistic the holographic description of color superconductivity can be made. Two main lines of `holographic color superconductivity' exist, either focusing on the spontaneous breaking of color symmetry \cite{Chen:2009kx,Faedo:2018fjw,Preau:2025ubr} or on simply mimicking the Cooper pair condensate by a scalar field \cite{Basu:2011yg,BitaghsirFadafan:2018iqr,CruzRojas:2025fzs}. In either case, the theories under consideration are different from QCD and further studies towards a more realistic strong-coupling picture of color superconductivity are needed. 

Another unsolved fundamental question is the one about the quark-hadron continuity, a particular aspect of the QCD phase structure at high baryon densities. This question is often discussed without taking into account pairing, especially if the focus is on the equation of state and its astrophysical implications \cite{Baym:2017whm,Baym:2019iky}. In that case, since neither order parameter for deconfinement and chiral transitions corresponds to an exact symmetry, a continuous transition from nuclear matter to quark matter is allowed. This is more complicated in the presence of pairing, where the exact symmetry of baryon number conservation can be broken spontaneously, resulting in a superfluid. CFL is a superfluid, and although the 2SC in its usual definition is not, additional single flavor pairing can render this phase a superfluid too \cite{Fujimoto:2019sxg}. It was pointed out in Refs.\ \cite{Schafer:1998ef,Alford:1999pa} that a continuity between (hyper)nuclear matter and CFL is possible, and the role of the axial anomaly for the presence of a crossover and a high-density critical point was discussed  in Refs.\ \cite{Hatsuda:2006ps,Yamamoto:2007ah,Schmitt:2010pf}. The question of the quark-hadron continuity is not only of theoretical interest for the QCD phase diagram, but may be realized at a quark-hadron interface inside neutron stars. A particularly interesting aspect  of the interface between CFL and superfluid nuclear matter, yet to be fully understood, is the possible existence of continuous vortices 
\cite{Alford:2018mqj,Chatterjee:2018nxe}, also considered in two-flavor quark matter \cite{Fujimoto:2021bes}. Vortices are also, on the other hand,  argued to provide a qualitative difference between quark and hadronic matter despite the identical global symmetries \cite{Cherman:2018jir,Cherman:2020hbe}.

Astrophysical data (and possibly data from heavy-ion collisions) are expected to improve our understanding of color superconductivity, but the intricacies of different color-superconducting phases and the transition between them will always be difficult to extract from these highly complex systems. And, of course, we do not know for certain that the conditions in these systems are extreme enough to even exhibit color superconductivity. Therefore, one can ask if there is a more controlled (tabletop) experiment that can at least mimic color-superconducting matter. Experiments with ultracold fermionic atoms are a very good candidate. In the simplest situation of two fermionic species they show a BEC-BCS crossover 
\cite{2008NCimR..31..247K}. This crossover is conjectured in QCD at finite isospin chemical potential (and vanishing baryon chemical potential) \cite{Son:2000xc} and in $N_c=2$ QCD \cite{Iida:2019rah,Boz:2019enj}. However, in real-world dense QCD the situation is more complicated since we know that at low densities bound states are made of three quarks, not two. It is nevertheless desirable to reproduce certain aspects of color-superconducting quark matter in cold atomic systems, for instance pairing with mismatched Fermi surfaces
\cite{Zwierlein:2006zz,Partridge:2006zkb}. Also, atomic systems are routinely being prepared with more that two species and with `artificial' (non-abelian) gauge fields \cite{PhysRevLett.98.160405,PhysRevLett.101.203202,PhysRevLett.102.130401,PhysRevLett.109.240401,PhysRevA.97.023632,Wang_2020,2025arXiv250204714M} and it would be very interesting to create, for instance, an atomic analogue of color-flavor locking.

\bibliography{references}

@article{McLerran:2018hbz,
    author = "McLerran, Larry and Reddy, Sanjay",
    title = "{Quarkyonic Matter and Neutron Stars}",
    eprint = "1811.12503",
    archivePrefix = "arXiv",
    primaryClass = "nucl-th",
    reportNumber = "INT-PUB-18-060",
    doi = "10.1103/PhysRevLett.122.122701",
    journal = "Phys. Rev. Lett.",
    volume = "122",
    number = "12",
    pages = "122701",
    year = "2019"
}

@article{Schafer:2004zf,
    author = {Sch{\"a}fer, Thomas and Schwenzer, Kai},
    title = "{Non-Fermi liquid effects in QCD at high density}",
    eprint = "hep-ph/0405053",
    archivePrefix = "arXiv",
    doi = "10.1103/PhysRevD.70.054007",
    journal = "Phys. Rev. D",
    volume = "70",
    pages = "054007",
    year = "2004"
}

@article{Muller:2016fdr,
    author = {M{\"u}ller, Daniel and Buballa, Michael and Wambach, Jochen},
    title = "{Dyson-Schwinger Approach to Color-Superconductivity: Effects of Selfconsistent Gluon Dressing}",
    eprint = "1603.02865",
    archivePrefix = "arXiv",
    primaryClass = "hep-ph",
    month = "3",
    year = "2016"
}

@article{Muller:2013pya,
    author = {M{\"u}ller, D. and Buballa, M. and Wambach, J.},
    title = "{Dyson-Schwinger approach to color superconductivity at finite temperature and density}",
    eprint = "1303.2693",
    archivePrefix = "arXiv",
    primaryClass = "hep-ph",
    doi = "10.1140/epja/i2013-13096-5",
    journal = "Eur. Phys. J. A",
    volume = "49",
    pages = "96",
    year = "2013"
}

@article{Murgana:2025dfa,
    author = "Murgana, Fabrizio and Comitini, Giorgio and Ruggieri, Marco",
    title = "{Topological susceptibility in the superconductive phases of quantum chromodynamics: A Dyson-Schwinger perspective}",
    eprint = "2502.19327",
    archivePrefix = "arXiv",
    primaryClass = "hep-ph",
    doi = "10.1103/PhysRevD.111.096008",
    journal = "Phys. Rev. D",
    volume = "111",
    number = "9",
    pages = "096008",
    year = "2025"
}

@article{Alford:2009jm,
    author = "Alford, Mark G. and Braby, Matt and Mahmoodifar, Simin",
    title = "{Shear viscosity due to kaon condensation in color-flavor locked quark matter}",
    eprint = "0910.2180",
    archivePrefix = "arXiv",
    primaryClass = "nucl-th",
    doi = "10.1103/PhysRevC.81.025202",
    journal = "Phys. Rev. C",
    volume = "81",
    pages = "025202",
    year = "2010"
}

@article{Wang:2009if,
    author = "Wang, Xinyang and Malekzadeh, Hossein and Shovkovy, Igor A.",
    title = "{Non-leptonic weak processes in spin-one color superconducting quark matter}",
    eprint = "0912.3851",
    archivePrefix = "arXiv",
    primaryClass = "hep-ph",
    reportNumber = "TUM-EFT-4-09",
    doi = "10.1103/PhysRevD.81.045021",
    journal = "Phys. Rev. D",
    volume = "81",
    pages = "045021",
    year = "2010"
}

@article{Wang:2009xf,
    author = "Wang, Qun",
    title = "{Some aspects of color superconductivity: an introduction}",
    eprint = "0912.2485",
    archivePrefix = "arXiv",
    primaryClass = "nucl-th",
    journal = "Prog. Phys.",
    volume = "30",
    pages = "173",
    year = "2010"
}

@article{Pang:2011qh,
    author = "Pang, Jin-Yi and Wang, Jin-Cheng and Wang, Qun",
    title = "{Some recent progress on quark pairings in dense quark and nuclear matter}",
    eprint = "1109.3073",
    archivePrefix = "arXiv",
    primaryClass = "nucl-th",
    doi = "10.1088/0253-6102/57/2/16",
    journal = "Commun. Theor. Phys.",
    volume = "57",
    pages = "251--270",
    year = "2012"
}

@article{Berdermann:2016mwt,
    author = "Berdermann, J. and Blaschke, D. and Fischer, T. and Kachanovich, A.",
    title = "{Neutrino emissivities and bulk viscosity in neutral two-flavor quark matter}",
    eprint = "1609.05201",
    archivePrefix = "arXiv",
    primaryClass = "astro-ph.HE",
    doi = "10.1103/PhysRevD.94.123010",
    journal = "Phys. Rev. D",
    volume = "94",
    number = "12",
    pages = "123010",
    year = "2016"
}

@article{Aguilera:2005tg,
    author = "Aguilera, D. N. and Blaschke, D. and Buballa, M. and Yudichev, V. L.",
    title = "{Color-spin locking phase in two-flavor quark matter for compact star phenomenology}",
    eprint = "hep-ph/0503288",
    archivePrefix = "arXiv",
    doi = "10.1103/PhysRevD.72.034008",
    journal = "Phys. Rev. D",
    volume = "72",
    pages = "034008",
    year = "2005"
}

@article{Anglani:2011cw,
    author = "Anglani, Roberto and Mannarelli, Massimo and Ruggieri, Marco",
    title = "{Collective modes in the color flavor locked phase}",
    eprint = "1101.4277",
    archivePrefix = "arXiv",
    primaryClass = "hep-ph",
    reportNumber = "ICCUB-11-003, YITP-11-5, PHY-12921-TH-2011",
    doi = "10.1088/1367-2630/13/5/055002",
    journal = "New J. Phys.",
    volume = "13",
    pages = "055002",
    year = "2011"
}

@article{Bierkandt:2011zp,
    author = "Bierkandt, Robert and Manuel, Cristina",
    title = "{Bulk viscosity coefficients due to phonons and kaons in superfluid color-flavor locked quark matter}",
    eprint = "1104.5624",
    archivePrefix = "arXiv",
    primaryClass = "hep-ph",
    doi = "10.1103/PhysRevD.84.023004",
    journal = "Phys. Rev. D",
    volume = "84",
    pages = "023004",
    year = "2011"
}

@phdthesis{Schmitt:2004hg,
    author = "Schmitt, Andreas",
    title = "{Spin-one color superconductivity in cold and dense quark matter}",
    eprint = "nucl-th/0405076",
    archivePrefix = "arXiv",
    reportNumber = "GSI-2017-01245",
    month = "5",
    year = "2004"
}

@article{Fujimoto:2019sxg,
    author = "Fujimoto, Yuki and Fukushima, Kenji and Weise, Wolfram",
    title = "{Continuity from neutron matter to two-flavor quark matter with $^1 S_0$ and $^3 P_2$ superfluidity}",
    eprint = "1908.09360",
    archivePrefix = "arXiv",
    primaryClass = "hep-ph",
    doi = "10.1103/PhysRevD.101.094009",
    journal = "Phys. Rev. D",
    volume = "101",
    number = "9",
    pages = "094009",
    year = "2020"
}

@Article{LarkinOvchinnikov,
author = "Larkin, A. I. and Ovchinnikov, Yu. N.",
year = "1965",
journal = "Sov. Phys. JETP",
volume = "20",
pages = "762"
}

@Article{FuldeFerrell,
author = "Fulde, P and Ferrell, R. A.",
year = "1964",
journal = "Phys. Rev.",
volume = "135",
pages = "A550"
}

@article{Huang:2004bg,
    author = "Huang, Mei and Shovkovy, Igor A.",
    title = "{Chromomagnetic instability in dense quark matter}",
    eprint = "hep-ph/0407049",
    archivePrefix = "arXiv",
    doi = "10.1103/PhysRevD.70.051501",
    journal = "Phys. Rev. D",
    volume = "70",
    pages = "051501",
    year = "2004"
}

@article{Chabanov:2023abq,
    author = "Chabanov, Michail and Rezzolla, Luciano",
    title = "{Numerical modeling of bulk viscosity in neutron stars}",
    eprint = "2311.13027",
    archivePrefix = "arXiv",
    primaryClass = "gr-qc",
    doi = "10.1103/PhysRevD.111.044074",
    journal = "Phys. Rev. D",
    volume = "111",
    number = "4",
    pages = "044074",
    year = "2025"
}

@article{Fujimoto:2021bes,
    author = "Fujimoto, Yuki and Nitta, Muneto",
    title = "{Vortices penetrating two-flavor quark-hadron continuity}",
    eprint = "2102.12928",
    archivePrefix = "arXiv",
    primaryClass = "hep-ph",
    doi = "10.1103/PhysRevD.103.114003",
    journal = "Phys. Rev. D",
    volume = "103",
    number = "11",
    pages = "114003",
    year = "2021"
}

@article{Mannarelli:2014jsa,
    author = "Mannarelli, Massimo",
    editor = "Bombaci, I. and Covello, A. and Marcucci, L. E. and Rosati, S.",
    title = "{The amazing properties of crystalline color superconductors}",
    eprint = "1401.7551",
    archivePrefix = "arXiv",
    primaryClass = "hep-ph",
    doi = "10.1088/1742-6596/527/1/012020",
    journal = "J. Phys. Conf. Ser.",
    volume = "527",
    pages = "012020",
    year = "2014"
}

@article{Shovkovy:2004me,
    author = "Shovkovy, Igor A.",
    title = "{Two lectures on color superconductivity}",
    eprint = "nucl-th/0410091",
    archivePrefix = "arXiv",
    doi = "10.1007/s10701-005-6440-x",
    journal = "Found. Phys.",
    volume = "35",
    pages = "1309--1358",
    year = "2005"
}

@article{Buballa:2003qv,
    author = "Buballa, Michael",
    title = "{NJL model analysis of quark matter at large density}",
    eprint = "hep-ph/0402234",
    archivePrefix = "arXiv",
    doi = "10.1016/j.physrep.2004.11.004",
    journal = "Phys. Rept.",
    volume = "407",
    pages = "205--376",
    year = "2005"
}

@article{PhysRevLett.98.160405,
  title = {Color Superfluidity and ``Baryon'' Formation in Ultracold Fermions},
  author = {Rapp, \'Akos and Zar\'and, Gergely and Honerkamp, Carsten and Hofstetter, Walter},
  journal = {Phys. Rev. Lett.},
  volume = {98},
  issue = {16},
  pages = {160405},
  numpages = {4},
  year = {2007},
  month = {Apr},
  publisher = {American Physical Society},
  doi = {10.1103/PhysRevLett.98.160405}
}

@article{PhysRevLett.101.203202,
  title = {Collisional Stability of a Three-Component Degenerate Fermi Gas},
  author = {Ottenstein, T. B. and Lompe, T. and Kohnen, M. and Wenz, A. N. and Jochim, S.},
  journal = {Phys. Rev. Lett.},
  volume = {101},
  issue = {20},
  pages = {203202},
  numpages = {4},
  year = {2008},
  month = {Nov},
  publisher = {American Physical Society},
  doi = {10.1103/PhysRevLett.101.203202}
}

@article{PhysRevLett.109.240401,
  title = {New Type of Crossover Physics in Three-Component Fermi Gases},
  author = {Nishida, Yusuke},
  journal = {Phys. Rev. Lett.},
  volume = {109},
  issue = {24},
  pages = {240401},
  numpages = {5},
  year = {2012},
  month = {Dec},
  publisher = {American Physical Society},
  doi = {10.1103/PhysRevLett.109.240401}
}

@article{PhysRevLett.102.130401,
  title = {Bose-Einstein Condensate in a Uniform Light-Induced Vector Potential},
  author = {Lin, Y.-J. and Compton, R. L. and Perry, A. R. and Phillips, W. D. and Porto, J. V. and Spielman, I. B.},
  journal = {Phys. Rev. Lett.},
  volume = {102},
  issue = {13},
  pages = {130401},
  numpages = {4},
  year = {2009},
  month = {Mar},
  publisher = {American Physical Society},
  doi = {10.1103/PhysRevLett.102.130401}
}

@article{Wang_2020,
doi = {10.1088/1674-1056/abab72},
year = {2020},
month = {oct},
publisher = {Chinese Physical Society and IOP Publishing Ltd},
volume = {29},
number = {10},
pages = {100304},
author = {Wang, Ji-Guo and Li, Yue-Qing and Dong, Yu-Fei},
title = {Lattice configurations in spin-1 Bose-Einstein condensates with the SU(3) spin-orbit coupling},
journal = {Chinese Physics B}
}

@article{PhysRevA.97.023632,
  title = {Color superfluidity of neutral ultracold fermions in the presence of color-flip and color-orbit fields},
  author = {Kurkcuoglu, Doga Murat and S\'a de Melo, C. A. R.},
  journal = {Phys. Rev. A},
  volume = {97},
  issue = {2},
  pages = {023632},
  numpages = {21},
  year = {2018},
  month = {Feb},
  publisher = {American Physical Society},
  doi = {10.1103/PhysRevA.97.023632}
}

@ARTICLE{2025arXiv250204714M,
       author = {{Madasu}, Chetan S. and {Mitra}, Chirantan and {Gabardos}, Lucas and {Rathod}, Ketan D. and {Zanon-Willette}, Thomas and {Miniatura}, Christian and {Chevy}, Frederic and {Kwong}, Chi and {Wilkowski}, David},
        title = "{Experimental realization of a SU(3) color-orbit coupling in an ultracold gas}",
         year = 2025,
        month = feb,
          doi = {10.48550/arXiv.2502.04714},
archivePrefix = {arXiv},
       eprint = {2502.04714},
 primaryClass = {physics.atom-ph},
       adsurl = {https://ui.adsabs.harvard.edu/abs/2025arXiv250204714M}
}

@article{Boz:2019enj,
    author = "Boz, Tamer and Giudice, Pietro and Hands, Simon and Skullerud, Jon-Ivar",
    title = "{Dense two-color QCD towards continuum and chiral limits}",
    eprint = "1912.10975",
    archivePrefix = "arXiv",
    primaryClass = "hep-lat",
    doi = "10.1103/PhysRevD.101.074506",
    journal = "Phys. Rev. D",
    volume = "101",
    number = "7",
    pages = "074506",
    year = "2020"
}

@article{Iida:2019rah,
    author = "Iida, Kei and Itou, Etsuko and Lee, Tong-Gyu",
    title = "{Two-colour QCD phases and the topology at low temperature and high density}",
    eprint = "1910.07872",
    archivePrefix = "arXiv",
    primaryClass = "hep-lat",
    doi = "10.1007/JHEP01(2020)181",
    journal = "JHEP",
    volume = "01",
    pages = "181",
    year = "2020"
}

@article{Son:2000xc,
    author = "Son, D. T. and Stephanov, Misha A.",
    title = "{QCD at finite isospin density}",
    eprint = "hep-ph/0005225",
    archivePrefix = "arXiv",
    doi = "10.1103/PhysRevLett.86.592",
    journal = "Phys. Rev. Lett.",
    volume = "86",
    pages = "592--595",
    year = "2001"
}

@ARTICLE{2008NCimR..31..247K,
       author = {{Ketterle}, W. and {Zwierlein}, M.~W.},
        title = "{Making, probing and understanding ultracold Fermi gases}",
      journal = {Nuovo Cimento Rivista Serie},
         year = 2008,
        month = may,
       volume = {31},
       number = {5-6},
        pages = {247-422},
          doi = {10.1393/ncr/i2008-10033-1},
archivePrefix = {arXiv},
       eprint = {0801.2500},
 primaryClass = {cond-mat.other}
}

@article{Zwierlein:2006zz,
    author = "Zwierlein, Martin W. and Schirotzek, Andre and Schunck, Christian H. and Ketterle, Wolfgang",
    title = "{Fermionic Superfluidity with Imbalanced Spin Populations}",
    eprint = "cond-mat/0511197",
    archivePrefix = "arXiv",
    doi = "10.1126/science.1122318",
    journal = "Science",
    volume = "311",
    pages = "492--496",
    year = "2006"
}

@article{Partridge:2006zkb,
    author = "Partridge, Guthrie B. and Li, Wenhui and Kamar, Ramsey I. and Liao, Yean-an and Hulet, Randall G.",
    title = "{Pairing and Phase Separation in a Polarized Fermi Gas}",
    eprint = "cond-mat/0511752",
    archivePrefix = "arXiv",
    doi = "10.1126/science.1122876",
    journal = "Science",
    volume = "311",
    pages = "503--505",
    year = "2006"
}

@article{Baym:2019iky,
    author = "Baym, Gordon and Furusawa, Shun and Hatsuda, Tetsuo and Kojo, Toru and Togashi, Hajime",
    title = "{New Neutron Star Equation of State with Quark-Hadron Crossover}",
    eprint = "1903.08963",
    archivePrefix = "arXiv",
    primaryClass = "astro-ph.HE",
    reportNumber = "RIKEN-iTHEMS-Report-19",
    doi = "10.3847/1538-4357/ab441e",
    journal = "Astrophys. J.",
    volume = "885",
    pages = "42",
    year = "2019"
}

@article{Schmitt:2010pf,
    author = "Schmitt, Andreas and Stetina, Stephan and Tachibana, Motoi",
    title = "{Ginzburg-Landau phase diagram for dense matter with axial anomaly, strange quark mass, and meson condensation}",
    eprint = "1010.4243",
    archivePrefix = "arXiv",
    primaryClass = "hep-ph",
    doi = "10.1103/PhysRevD.83.045008",
    journal = "Phys. Rev. D",
    volume = "83",
    pages = "045008",
    year = "2011"
}

@article{Yamamoto:2007ah,
    author = "Yamamoto, Naoki and Tachibana, Motoi and Hatsuda, Tetsuo and Baym, Gordon",
    title = "{Phase structure, collective modes, and the axial anomaly in dense QCD}",
    eprint = "0704.2654",
    archivePrefix = "arXiv",
    primaryClass = "hep-ph",
    reportNumber = "TKYNT-07-10, SAGA-HE-233",
    doi = "10.1103/PhysRevD.76.074001",
    journal = "Phys. Rev. D",
    volume = "76",
    pages = "074001",
    year = "2007"
}

@article{Hatsuda:2006ps,
    author = "Hatsuda, Tetsuo and Tachibana, Motoi and Yamamoto, Naoki and Baym, Gordon",
    title = "{New critical point induced by the axial anomaly in dense QCD}",
    eprint = "hep-ph/0605018",
    archivePrefix = "arXiv",
    reportNumber = "TKYNT-06-8, SAGA-HE-228",
    doi = "10.1103/PhysRevLett.97.122001",
    journal = "Phys. Rev. Lett.",
    volume = "97",
    pages = "122001",
    year = "2006"
}

@article{Rajagopal:2005dg,
    author = "Rajagopal, Krishna and Schmitt, Andreas",
    title = "{Stressed pairing in conventional color superconductors is unavoidable}",
    eprint = "hep-ph/0512043",
    archivePrefix = "arXiv",
    reportNumber = "MIT-CTP-3712",
    doi = "10.1103/PhysRevD.73.045003",
    journal = "Phys. Rev. D",
    volume = "73",
    pages = "045003",
    year = "2006"
}

@article{Kovensky:2020xif,
    author = "Kovensky, Nicolas and Schmitt, Andreas",
    title = "{Holographic quarkyonic matter}",
    eprint = "2006.13739",
    archivePrefix = "arXiv",
    primaryClass = "hep-th",
    doi = "10.1007/JHEP09(2020)112",
    journal = "JHEP",
    volume = "09",
    pages = "112",
    year = "2020"
}

@article{McLerran:2007qj,
    author = "McLerran, Larry and Pisarski, Robert D.",
    title = "{Phases of cold, dense quarks at large N(c)}",
    eprint = "0706.2191",
    archivePrefix = "arXiv",
    primaryClass = "hep-ph",
    doi = "10.1016/j.nuclphysa.2007.08.013",
    journal = "Nucl. Phys. A",
    volume = "796",
    pages = "83--100",
    year = "2007"
}

@article{Hebeler:2013nza,
    author = "Hebeler, K. and Lattimer, J. M. and Pethick, C. J. and Schwenk, A.",
    title = "{Equation of state and neutron star properties constrained by nuclear physics and observation}",
    eprint = "1303.4662",
    archivePrefix = "arXiv",
    primaryClass = "astro-ph.SR",
    doi = "10.1088/0004-637X/773/1/11",
    journal = "Astrophys. J.",
    volume = "773",
    pages = "11",
    year = "2013"
}

@inproceedings{Kurkela:2025acm,
    author = "Kurkela, Aleksi and Rajagopal, Krishna and Steinhorst, Rachel",
    title = "{Updated Astrophysical Equation-of-State Constraints on the Color-Superconducting Gap}",
    booktitle = "{31st International Conference on Ultra-relativistic Nucleus-Nucleus Collisions}",
    eprint = "2508.20763",
    archivePrefix = "arXiv",
    primaryClass = "hep-ph",
    reportNumber = "MIT-CTP/5909",
    month = "8",
    year = "2025"
}

@article{Acernese_2015,
doi = {10.1088/0264-9381/32/2/024001},
year = {2014},
month = {dec},
publisher = {IOP Publishing},
volume = {32},
number = {2},
pages = {024001},
author = {Acernese, F and others},
title = {Advanced Virgo: a second-generation interferometric gravitational wave detector},
journal = {Classical and Quantum Gravity}
}

@article{Aasi_2015,
doi = {10.1088/0264-9381/32/7/074001},
year = {2015},
month = {mar},
publisher = {IOP Publishing},
volume = {32},
number = {7},
pages = {074001},
author = {Aasi, J and others},
title = {Advanced LIGO},
journal = {Classical and Quantum Gravity}
}

@article{Abbott_2017,
doi = {10.1088/1361-6382/aa51f4},
year = {2017},
month = {jan},
publisher = {IOP Publishing},
volume = {34},
number = {4},
pages = {044001},
author = {Abbott, B P and others},
title = {Exploring the sensitivity of next generation gravitational wave
					detectors},
journal = {Classical and Quantum Gravity}
}

@article{Maggiore_2020,
doi = {10.1088/1475-7516/2020/03/050},
year = {2020},
month = {mar},
publisher = {},
volume = {2020},
number = {03},
pages = {050},
author = {Maggiore, Michele and others},
title = {Science case for the Einstein telescope},
journal = {Journal of Cosmology and Astroparticle Physics}
}

@article{Guo:2023som,
    author = "Guo, Ling-Jun and Yang, Wen-Cong and Ma, Yong-Liang and Wu, Yue-Liang",
    title = "{Probing Hadron-quark Transition Through Binary Neutron Star Merger}",
    eprint = "2308.01770",
    archivePrefix = "arXiv",
    primaryClass = "astro-ph.HE",
    doi = "10.1088/1674-4527/adbc37",
    journal = "Res. Astron. Astrophys.",
    volume = "25",
    number = "3",
    pages = "035017",
    year = "2025"
}

@article{Most:2019onn,
    author = "Most, Elias R. and Jens Papenfort, L. and Dexheimer, Veronica and Hanauske, Matthias and Stoecker, Horst and Rezzolla, Luciano",
    title = "{On the deconfinement phase transition in neutron-star mergers}",
    eprint = "1910.13893",
    archivePrefix = "arXiv",
    primaryClass = "astro-ph.HE",
    doi = "10.1140/epja/s10050-020-00073-4",
    journal = "Eur. Phys. J. A",
    volume = "56",
    number = "2",
    pages = "59",
    year = "2020"
}

@article{Tootle:2022pvd,
    author = {Tootle, Samuel and Ecker, Christian and Topolski, Konrad and Demircik, Tuna and J{\"a}rvinen, Matti and Rezzolla, Luciano},
    title = "{Quark formation and phenomenology in binary neutron-star mergers using V-QCD}",
    eprint = "2205.05691",
    archivePrefix = "arXiv",
    primaryClass = "astro-ph.HE",
    reportNumber = "APCTP Pre2022 - 007",
    doi = "10.21468/SciPostPhys.13.5.109",
    journal = "SciPost Phys.",
    volume = "13",
    pages = "109",
    year = "2022"
}

@article{Aarts:2008rr,
    author = "Aarts, Gert and Stamatescu, Ion-Olimpiu",
    title = "{Stochastic quantization at finite chemical potential}",
    eprint = "0807.1597",
    archivePrefix = "arXiv",
    primaryClass = "hep-lat",
    doi = "10.1088/1126-6708/2008/09/018",
    journal = "JHEP",
    volume = "09",
    pages = "018",
    year = "2008"
}

@article{Marchis:2017oqi,
    author = "Marchis, Carlotta and Gattringer, Christof",
    title = "{Dual representation of lattice QCD with worldlines and worldsheets of abelian color fluxes}",
    eprint = "1712.07546",
    archivePrefix = "arXiv",
    primaryClass = "hep-lat",
    doi = "10.1103/PhysRevD.97.034508",
    journal = "Phys. Rev. D",
    volume = "97",
    number = "3",
    pages = "034508",
    year = "2018"
}

@article{Alexandru:2015sua,
    author = "Alexandru, Andrei and Basar, Gokce and Bedaque, Paulo F. and Ridgway, Gregory W. and Warrington, Neill C.",
    title = "{Sign problem and Monte Carlo calculations beyond Lefschetz thimbles}",
    eprint = "1512.08764",
    archivePrefix = "arXiv",
    primaryClass = "hep-lat",
    doi = "10.1007/JHEP05(2016)053",
    journal = "JHEP",
    volume = "05",
    pages = "053",
    year = "2016"
}

@article{Cristoforetti:2014gsa,
    author = "Cristoforetti, M. and Di Renzo, F. and Eruzzi, G. and Mukherjee, A. and Schmidt, C. and Scorzato, L. and Torrero, C.",
    title = "{An efficient method to compute the residual phase on a Lefschetz thimble}",
    eprint = "1403.5637",
    archivePrefix = "arXiv",
    primaryClass = "hep-lat",
    doi = "10.1103/PhysRevD.89.114505",
    journal = "Phys. Rev. D",
    volume = "89",
    number = "11",
    pages = "114505",
    year = "2014"
}

@article{Ratti:2022qgf,
    author = "Ratti, Claudia",
    title = "{Equation of state for QCD from lattice simulations}",
    doi = "10.1016/j.ppnp.2022.104007",
    journal = "Prog. Part. Nucl. Phys.",
    volume = "129",
    pages = "104007",
    year = "2023"
}

@article{Basar:2023bwd,
    author = "Basar, Gokce and Marincel, Joseph",
    title = "{Heavy-dense QCD, sign optimization, and Lefschetz thimbles}",
    eprint = "2311.06343",
    archivePrefix = "arXiv",
    primaryClass = "hep-th",
    doi = "10.1103/PhysRevC.109.045208",
    journal = "Phys. Rev. C",
    volume = "109",
    number = "4",
    pages = "045208",
    year = "2024"
}

@article{Langelage:2014vpa,
    author = "Langelage, Jens and Neuman, Mathias and Philipsen, Owe",
    title = "{Heavy dense QCD and nuclear matter from an effective lattice theory}",
    eprint = "1403.4162",
    archivePrefix = "arXiv",
    primaryClass = "hep-lat",
    doi = "10.1007/JHEP09(2014)131",
    journal = "JHEP",
    volume = "09",
    pages = "131",
    year = "2014"
}

@article{Nishimura:2024kvz,
    author = "Nishimura, Toru and Kitazawa, Masakiyo and Kunihiro, Teiji",
    title = "{Electromagnetic response of dense quark matter around color-superconducting phase transition and QCD critical point}",
    eprint = "2405.09240",
    archivePrefix = "arXiv",
    primaryClass = "hep-ph",
    reportNumber = "YITP-24-63, J-PARC-TH-0305",
    doi = "10.1016/j.aop.2024.169768",
    journal = "Annals Phys.",
    volume = "469",
    pages = "169768",
    year = "2024"
}

@article{Kitazawa:2001ft,
    author = "Kitazawa, M. and Koide, T. and Kunihiro, T. and Nemoto, Y.",
    title = "{Precursor of color superconductivity in hot quark matter}",
    eprint = "nucl-th/0111022",
    archivePrefix = "arXiv",
    doi = "10.1103/PhysRevD.65.091504",
    journal = "Phys. Rev. D",
    volume = "65",
    pages = "091504",
    year = "2002"
}

@article{Noronha:2007wg,
    author = "Noronha, Jorge L. and Shovkovy, Igor A.",
    title = "{Color-flavor locked superconductor in a magnetic field}",
    eprint = "0708.0307",
    archivePrefix = "arXiv",
    primaryClass = "hep-ph",
    doi = "10.1103/PhysRevD.76.105030",
    journal = "Phys. Rev. D",
    volume = "76",
    pages = "105030",
    year = "2007",
    note = "[Erratum: Phys.Rev.D 86, 049901 (2012)]"
}

@article{Ferrer:2005vd,
    author = "Ferrer, Efrain J. and de la Incera, Vivian and Manuel, Cristina",
    title = "{Magnetic color flavor locking phase in high density QCD}",
    eprint = "hep-ph/0503162",
    archivePrefix = "arXiv",
    doi = "10.1103/PhysRevLett.95.152002",
    journal = "Phys. Rev. Lett.",
    volume = "95",
    pages = "152002",
    year = "2005"
}

@article{Yu:2012jn,
    author = "Yu, Lang and Shovkovy, Igor A.",
    title = "{Directional dependence of color superconducting gap in two-flavor QCD in a magnetic field}",
    eprint = "1202.0872",
    archivePrefix = "arXiv",
    primaryClass = "hep-ph",
    doi = "10.1103/PhysRevD.85.085022",
    journal = "Phys. Rev. D",
    volume = "85",
    pages = "085022",
    year = "2012"
}

@article{Wang:2010ydb,
    author = "Wang, Xinyang and Shovkovy, Igor A.",
    title = "{Bulk viscosity of spin-one color superconducting strange quark matter}",
    eprint = "1006.1293",
    archivePrefix = "arXiv",
    primaryClass = "hep-ph",
    doi = "10.1103/PhysRevD.82.085007",
    journal = "Phys. Rev. D",
    volume = "82",
    pages = "085007",
    year = "2010"
}

@ARTICLE{2024arXiv240302066J,
       author = {{Jones}, D.~I. and {Riles}, K.},
        title = "{Multimessenger observations and the science enabled: Continuous waves and their progenitors, equation of state of dense matter}",
         year = 2024,
        month = mar,
           doi = {10.48550/arXiv.2403.02066},
archivePrefix = {arXiv},
       eprint = {2403.02066},
 primaryClass = {astro-ph.HE},
       adsurl = {https://ui.adsabs.harvard.edu/abs/2024arXiv240302066J}
}

@article{Choudhury:2024xbk,
    author = "Choudhury, Devarshi and others",
    title = "{A NICER View of the Nearest and Brightest Millisecond Pulsar: PSR J0437{\textendash}4715}",
    eprint = "2407.06789",
    archivePrefix = "arXiv",
    primaryClass = "astro-ph.HE",
    doi = "10.3847/2041-8213/ad5a6f",
    journal = "Astrophys. J. Lett.",
    volume = "971",
    number = "1",
    pages = "L20",
    year = "2024"
}

@article{Romani:2022jhd,
    author = "Romani, Roger W. and Kandel, D. and Filippenko, Alexei V. and Brink, Thomas G. and Zheng, WeiKang",
    title = "{PSR J0952\ensuremath{-}0607: The Fastest and Heaviest Known Galactic Neutron Star}",
    eprint = "2207.05124",
    archivePrefix = "arXiv",
    primaryClass = "astro-ph.HE",
    doi = "10.3847/2041-8213/ac8007",
    journal = "Astrophys. J. Lett.",
    volume = "934",
    number = "2",
    pages = "L17",
    year = "2022"
}

@article{Fonseca:2021wxt,
    author = "Fonseca, E. and others",
    title = "{Refined Mass and Geometric Measurements of the High-mass PSR J0740+6620}",
    eprint = "2104.00880",
    archivePrefix = "arXiv",
    primaryClass = "astro-ph.HE",
    doi = "10.3847/2041-8213/ac03b8",
    journal = "Astrophys. J. Lett.",
    volume = "915",
    number = "1",
    pages = "L12",
    year = "2021"
}

@article{NANOGrav:2019jur,
    author = "Cromartie, H. T. and others",
    collaboration = "NANOGrav",
    title = "{Relativistic Shapiro delay measurements of an extremely massive millisecond pulsar}",
    eprint = "1904.06759",
    archivePrefix = "arXiv",
    primaryClass = "astro-ph.HE",
    doi = "10.1038/s41550-019-0880-2",
    journal = "Nature Astron.",
    volume = "4",
    number = "1",
    pages = "72--76",
    year = "2019"
}

@article{Miller:2021qha,
    author = "Miller, M. C. and others",
    title = "{The Radius of PSR J0740+6620 from NICER and XMM-Newton Data}",
    eprint = "2105.06979",
    archivePrefix = "arXiv",
    primaryClass = "astro-ph.HE",
    doi = "10.3847/2041-8213/ac089b",
    journal = "Astrophys. J. Lett.",
    volume = "918",
    number = "2",
    pages = "L28",
    year = "2021"
}

@article{Riley:2021pdl,
    author = "Riley, Thomas E. and others",
    title = "{A NICER View of the Massive Pulsar PSR J0740+6620 Informed by Radio Timing and XMM-Newton Spectroscopy}",
    eprint = "2105.06980",
    archivePrefix = "arXiv",
    primaryClass = "astro-ph.HE",
    doi = "10.3847/2041-8213/ac0a81",
    journal = "Astrophys. J. Lett.",
    volume = "918",
    number = "2",
    pages = "L27",
    year = "2021"
}

@article{Miller:2019cac,
    author = "Miller, M. C. and others",
    title = "{PSR J0030+0451 Mass and Radius from $NICER$ Data and Implications for the Properties of Neutron Star Matter}",
    eprint = "1912.05705",
    archivePrefix = "arXiv",
    primaryClass = "astro-ph.HE",
    doi = "10.3847/2041-8213/ab50c5",
    journal = "Astrophys. J. Lett.",
    volume = "887",
    number = "1",
    pages = "L24",
    year = "2019"
}

@article{Riley:2019yda,
    author = "Riley, Thomas E. and others",
    title = "{A $NICER$ View of PSR J0030+0451: Millisecond Pulsar Parameter Estimation}",
    eprint = "1912.05702",
    archivePrefix = "arXiv",
    primaryClass = "astro-ph.HE",
    doi = "10.3847/2041-8213/ab481c",
    journal = "Astrophys. J. Lett.",
    volume = "887",
    number = "1",
    pages = "L21",
    year = "2019"
}

@article{LIGOScientific:2018hze,
    author = "Abbott, B. P. and others",
    collaboration = "LIGO Scientific, Virgo",
    title = "{Properties of the binary neutron star merger GW170817}",
    eprint = "1805.11579",
    archivePrefix = "arXiv",
    primaryClass = "gr-qc",
    doi = "10.1103/PhysRevX.9.011001",
    journal = "Phys. Rev. X",
    volume = "9",
    number = "1",
    pages = "011001",
    year = "2019"
}

@article{Annala:2019puf,
    author = {Annala, Eemeli and Gorda, Tyler and Kurkela, Aleksi and N{\"a}ttil{\"a}, Joonas and Vuorinen, Aleksi},
    title = "{Evidence for quark-matter cores in massive neutron stars}",
    eprint = "1903.09121",
    archivePrefix = "arXiv",
    primaryClass = "astro-ph.HE",
    reportNumber = "CERN-TH-2019-031, HIP-2019-7/TH",
    doi = "10.1038/s41567-020-0914-9",
    journal = "Nature Phys.",
    volume = "16",
    number = "9",
    pages = "907--910",
    year = "2020"
}

@article{Schmitt:2016pre,
    author = "Schmitt, Andreas",
    title = "{From ultra-dense QCD towards NICA densities: Color-flavor locking and other color superconductors}",
    doi = "10.1140/epja/i2016-16226-7",
    journal = "Eur. Phys. J. A",
    volume = "52",
    number = "8",
    pages = "226",
    year = "2016"
}

@article{Gartlein:2025zhd,
    author = {G{\"a}rtlein, Christoph and Ivanytskyi, Oleksii and Sagun, Violetta and Lopes, Il{\'\i}dio},
    title = "{Color-superconducting quarkyonic matter}",
    eprint = "2509.03517",
    archivePrefix = "arXiv",
    primaryClass = "nucl-th",
    month = "9",
    year = "2025"
}

@article{CruzRojas:2024etx,
    author = {Cruz Rojas, Jes{\'u}s and Gorda, Tyler and Hoyos, Carlos and Jokela, Niko and J{\"a}rvinen, Matti and Kurkela, Aleksi and Paatelainen, Risto and S{\"a}ppi, Saga and Vuorinen, Aleksi},
    title = "{Estimate for the Bulk Viscosity of Strongly Coupled Quark Matter Using Perturbative QCD and Holography}",
    eprint = "2402.00621",
    archivePrefix = "arXiv",
    primaryClass = "hep-ph",
    reportNumber = "APCTP Pre2024 - 003, HIP-2024-1/TH, TUM-EFT 187/24",
    doi = "10.1103/PhysRevLett.133.071901",
    journal = "Phys. Rev. Lett.",
    volume = "133",
    number = "7",
    pages = "071901",
    year = "2024"
}

@article{Sedrakian:2013pva,
    author = "Sedrakian, Armen",
    title = "{Rapid cooling of the compact star in Cassiopea A as a phase transition in dense QCD}",
    eprint = "1303.5380",
    archivePrefix = "arXiv",
    primaryClass = "astro-ph.HE",
    doi = "10.1051/0004-6361/201321541",
    journal = "Astron. Astrophys.",
    volume = "555",
    pages = "L10",
    year = "2013"
}

@article{Jaikumar:2005hy,
    author = "Jaikumar, Prashanth and Roberts, Craig D. and Sedrakian, Armen",
    title = "{Direct Urca neutrino rate in colour superconducting quark matter}",
    eprint = "nucl-th/0509093",
    archivePrefix = "arXiv",
    reportNumber = "ANL-PHY-11346-TH-2005, MPG-VT-UR-262-05, UNITU-THEP-04-2005",
    doi = "10.1103/PhysRevC.73.042801",
    journal = "Phys. Rev. C",
    volume = "73",
    pages = "042801",
    year = "2006"
}

@article{Anglani:2006br,
    author = "Anglani, Roberto and Nardulli, Giuseppe and Ruggieri, Marco and Mannarelli, Massimo",
    title = "{Neutrino emission from compact stars and inhomogeneous color superconductivity}",
    eprint = "hep-ph/0607341",
    archivePrefix = "arXiv",
    reportNumber = "BARI-TH-06-539, MIT-CTP-3758",
    doi = "10.1103/PhysRevD.74.074005",
    journal = "Phys. Rev. D",
    volume = "74",
    pages = "074005",
    year = "2006"
}

@article{Sad:2006egl,
    author = "Sa'd, Basil A. and Shovkovy, Igor A. and Rischke, Dirk H.",
    title = "{Bulk viscosity of spin-one color superconductors with two quark flavors}",
    eprint = "astro-ph/0607643",
    archivePrefix = "arXiv",
    doi = "10.1103/PhysRevD.75.065016",
    journal = "Phys. Rev. D",
    volume = "75",
    pages = "065016",
    year = "2007"
}

@article{Sarkar:2016gib,
    author = "Sarkar, Sreemoyee and Sharma, Rishi",
    title = "{The shear viscosity of two-flavor crystalline color superconducting quark matter}",
    eprint = "1701.00010",
    archivePrefix = "arXiv",
    primaryClass = "hep-ph",
    reportNumber = "TIFR-TH-16-48",
    doi = "10.1103/PhysRevD.96.094025",
    journal = "Phys. Rev. D",
    volume = "96",
    pages = "094025",
    year = "2017"
}

@article{Alford:2025jtm,
    author = "Alford, Mark G. and Brodie, Liam and Buballa, Michael and Gholami, Hosein and Haber, Alexander and Hofmann, Marco",
    title = "{Neutrino absorption in two-flavor color-superconducting quark matter}",
    eprint = "2509.04240",
    archivePrefix = "arXiv",
    primaryClass = "nucl-th",
    month = "9",
    year = "2025"
}

@article{Alford:2014doa,
    author = "Alford, Mark G. and Nishimura, Hiromichi and Sedrakian, Armen",
    title = "{Transport coefficients of two-flavor superconducting quark matter}",
    eprint = "1408.4999",
    archivePrefix = "arXiv",
    primaryClass = "hep-ph",
    doi = "10.1103/PhysRevC.90.055205",
    journal = "Phys. Rev. C",
    volume = "90",
    number = "5",
    pages = "055205",
    year = "2014"
}

@article{Jaikumar:2002vg,
    author = {Jaikumar, Prashanth and Prakash, Madappa and Sch{\"a}fer, Thomas},
    title = "{Neutrino emission from Goldstone modes in dense quark matter}",
    eprint = "astro-ph/0203088",
    archivePrefix = "arXiv",
    doi = "10.1103/PhysRevD.66.063003",
    journal = "Phys. Rev. D",
    volume = "66",
    pages = "063003",
    year = "2002"
}

@article{Shovkovy:2002sg,
    author = "Shovkovy, Igor A. and Ellis, Paul J.",
    title = "{Optically opaque color flavor locked phase inside compact stars}",
    eprint = "hep-ph/0211049",
    archivePrefix = "arXiv",
    reportNumber = "NUC-MINN-02-8-T",
    doi = "10.1103/PhysRevC.67.048801",
    journal = "Phys. Rev. C",
    volume = "67",
    pages = "048801",
    year = "2003"
}

@article{Andersson:1997xt,
    author = "Andersson, Nils",
    title = "{A New class of unstable modes of rotating relativistic stars}",
    eprint = "gr-qc/9706075",
    archivePrefix = "arXiv",
    doi = "10.1086/305919",
    journal = "Astrophys. J.",
    volume = "502",
    pages = "708--713",
    year = "1998"
}

@article{Kraav:2024cus,
    author = "Kraav, Kirill Y. and Gusakov, Mikhail E. and Kantor, Elena M.",
    title = "{Instability windows of relativistic r-modes}",
    eprint = "2401.06200",
    archivePrefix = "arXiv",
    primaryClass = "astro-ph.HE",
    doi = "10.1103/PhysRevD.109.043012",
    journal = "Phys. Rev. D",
    volume = "109",
    number = "4",
    pages = "043012",
    year = "2024"
}

@article{Antonopoulou:2022rpq,
    author = "Antonopoulou, Danai and Haskell, Brynmor and Espinoza, Crist{\'o}bal M.",
    title = "{Pulsar glitches: observations and physical interpretation}",
    doi = "10.1088/1361-6633/ac9ced",
    journal = "Rept. Prog. Phys.",
    volume = "85",
    number = "12",
    pages = "126901",
    year = "2022"
}

@article{Potekhin:2015qsa,
    author = "Potekhin, A. Y. and Pons, J. A. and Page, Dany",
    title = "{Neutron stars - cooling and transport}",
    eprint = "1507.06186",
    archivePrefix = "arXiv",
    primaryClass = "astro-ph.HE",
    doi = "10.1007/s11214-015-0180-9",
    journal = "Space Sci. Rev.",
    volume = "191",
    number = "1-4",
    pages = "239--291",
    year = "2015"
}

@article{Marino:2024gpm,
    author = "Marino, Alessio and Dehman, Clara and Kovlakas, Konstantinos and Rea, Nanda and Pons, Jose A. and Vigan{\`o}, D.",
    title = "{Constraints on the dense matter equation of state from young and cold isolated neutron stars}",
    eprint = "2404.05371",
    archivePrefix = "arXiv",
    primaryClass = "astro-ph.HE",
    doi = "10.1038/s41550-024-02291-y",
    journal = "Nature Astron.",
    volume = "8",
    number = "8",
    pages = "1020--1030",
    year = "2024"
}

@article{Alford:2017rxf,
    author = "Alford, Mark G. and Bovard, Luke and Hanauske, Matthias and Rezzolla, Luciano and Schwenzer, Kai",
    title = "{Viscous Dissipation and Heat Conduction in Binary Neutron-Star Mergers}",
    eprint = "1707.09475",
    archivePrefix = "arXiv",
    primaryClass = "gr-qc",
    doi = "10.1103/PhysRevLett.120.041101",
    journal = "Phys. Rev. Lett.",
    volume = "120",
    number = "4",
    pages = "041101",
    year = "2018"
}

@article{Pons:2019zyc,
    author = "Pons, Jos{\'e} A. and Vigan{\`o}, Daniele",
    title = "{Magnetic, thermal and rotational evolution of isolated neutron stars}",
    eprint = "1911.03095",
    archivePrefix = "arXiv",
    primaryClass = "astro-ph.HE",
    doi = "10.1007/s41115-019-0006-7",
    journal = "Liv. Rev. Comput. Astrophys.",
    volume = "5",
    number = "1",
    pages = "3",
    year = "2019"
}

@article{Alford:2025tbp,
    author = "Alford, Mark and Harutyunyan, Arus and Sedrakian, Armen and Tsiopelas, Stefanos",
    title = "{Bulk viscosity of two-flavor color superconducting quark matter in neutron star mergers}",
    eprint = "2506.08144",
    archivePrefix = "arXiv",
    primaryClass = "nucl-th",
    doi = "10.3389/fspas.2025.1648066",
    journal = "Front. Astron. Space Sci.",
    volume = "12",
    pages = "1648066",
    year = "2025"
}

@article{Cherman:2020hbe,
    author = "Cherman, Aleksey and Jacobson, Theodore and Sen, Srimoyee and Yaffe, Laurence G.",
    title = "{Higgs-confinement phase transitions with fundamental representation matter}",
    eprint = "2007.08539",
    archivePrefix = "arXiv",
    primaryClass = "hep-th",
    doi = "10.1103/PhysRevD.102.105021",
    journal = "Phys. Rev. D",
    volume = "102",
    number = "10",
    pages = "105021",
    year = "2020"
}

@article{Cherman:2018jir,
    author = "Cherman, Aleksey and Sen, Srimoyee and Yaffe, Laurence G.",
    title = "{Anyonic particle-vortex statistics and the nature of dense quark matter}",
    eprint = "1808.04827",
    archivePrefix = "arXiv",
    primaryClass = "hep-th",
    reportNumber = "INT-PUB-18-043",
    doi = "10.1103/PhysRevD.100.034015",
    journal = "Phys. Rev. D",
    volume = "100",
    number = "3",
    pages = "034015",
    year = "2019"
}

@article{Chatterjee:2018nxe,
    author = "Chatterjee, Chandrasekhar and Nitta, Muneto and Yasui, Shigehiro",
    title = "{Quark-hadron continuity under rotation: Vortex continuity or boojum?}",
    eprint = "1806.09291",
    archivePrefix = "arXiv",
    primaryClass = "hep-ph",
    doi = "10.1103/PhysRevD.99.034001",
    journal = "Phys. Rev. D",
    volume = "99",
    number = "3",
    pages = "034001",
    year = "2019"
}

@article{Alford:2018mqj,
    author = "Alford, Mark G. and Baym, Gordon and Fukushima, Kenji and Hatsuda, Tetsuo and Tachibana, Motoi",
    title = "{Continuity of vortices from the hadronic to the color-flavor locked phase in dense matter}",
    eprint = "1803.05115",
    archivePrefix = "arXiv",
    primaryClass = "hep-ph",
    reportNumber = "RIKEN-QHP-367, RIKEN-iTHEMS-Report-18, RIKEN-ITHEMS-REPORT-18",
    doi = "10.1103/PhysRevD.99.036004",
    journal = "Phys. Rev. D",
    volume = "99",
    number = "3",
    pages = "036004",
    year = "2019"
}

@article{Alford:1999pa,
    author = "Alford, Mark G. and Berges, Juergen and Rajagopal, Krishna",
    title = "{Unlocking color and flavor in superconducting strange quark matter}",
    eprint = "hep-ph/9903502",
    archivePrefix = "arXiv",
    reportNumber = "MIT-CTP-2844",
    doi = "10.1016/S0550-3213(99)00410-1",
    journal = "Nucl. Phys. B",
    volume = "558",
    pages = "219--242",
    year = "1999"
}

@article{Schafer:1998ef,
    author = {Sch{\"a}fer, Thomas and Wilczek, Frank},
    title = "{Continuity of quark and hadron matter}",
    eprint = "hep-ph/9811473",
    archivePrefix = "arXiv",
    reportNumber = "IASSNS-HEP-98-100",
    doi = "10.1103/PhysRevLett.82.3956",
    journal = "Phys. Rev. Lett.",
    volume = "82",
    pages = "3956--3959",
    year = "1999"
}

@article{Sogabe:2024yfl,
    author = "Sogabe, Noriyuki and Yin, Yi",
    title = "{Berry Curvature and Spin-One Color Superconductivity}",
    eprint = "2411.08005",
    archivePrefix = "arXiv",
    primaryClass = "nucl-th",
    doi = "10.1103/PhysRevLett.134.171903",
    journal = "Phys. Rev. Lett.",
    volume = "134",
    number = "17",
    pages = "171903",
    year = "2025"
}

@article{Schmitt:2005wg,
    author = "Schmitt, Andreas and Shovkovy, Igor A. and Wang, Qun",
    title = "{Neutrino emission and cooling rates of spin-one color superconductors}",
    eprint = "hep-ph/0510347",
    archivePrefix = "arXiv",
    doi = "10.1103/PhysRevD.73.034012",
    journal = "Phys. Rev. D",
    volume = "73",
    pages = "034012",
    year = "2006"
}

@article{Feng:2007bg,
    author = "Feng, Bo and Hou, De-fu and Ren, Hai-cang",
    title = "{Angular momentum mixing in a non-spherical color superconductor}",
    eprint = "0711.0496",
    archivePrefix = "arXiv",
    primaryClass = "hep-ph",
    doi = "10.1016/j.nuclphysb.2007.12.005",
    journal = "Nucl. Phys. B",
    volume = "796",
    pages = "500--520",
    year = "2008"
}

@article{Schmitt:2003xq,
    author = "Schmitt, Andreas and Wang, Qun and Rischke, Dirk H.",
    title = "{Electromagnetic Meissner effect in spin one color superconductors}",
    eprint = "nucl-th/0301090",
    archivePrefix = "arXiv",
    doi = "10.1103/PhysRevLett.91.242301",
    journal = "Phys. Rev. Lett.",
    volume = "91",
    pages = "242301",
    year = "2003"
}

@article{Pang:2010wk,
    author = "Pang, Jin-yi and Brauner, Tomas and Wang, Qun",
    title = "{Spin-one color superconductors: Collective modes and effective Lagrangian}",
    eprint = "1010.1986",
    archivePrefix = "arXiv",
    primaryClass = "nucl-th",
    doi = "10.1016/j.nuclphysa.2011.01.014",
    journal = "Nucl. Phys. A",
    volume = "852",
    pages = "175--196",
    year = "2011"
}

@article{Brauner:2008ma,
    author = "Brauner, Tomas",
    title = "{Helical ordering in the ground state of spin-one color superconductors as a consequence of parity violation}",
    eprint = "0810.3481",
    archivePrefix = "arXiv",
    primaryClass = "hep-ph",
    doi = "10.1103/PhysRevD.78.125027",
    journal = "Phys. Rev. D",
    volume = "78",
    pages = "125027",
    year = "2008"
}

@article{Alford:2005yy,
    author = "Alford, Mark G. and Cowan, Greig A.",
    title = "{Single-flavor and two-flavor pairing in three-flavor quark matter}",
    eprint = "hep-ph/0512104",
    archivePrefix = "arXiv",
    doi = "10.1088/0954-3899/32/4/009",
    journal = "J. Phys. G",
    volume = "32",
    pages = "511--528",
    year = "2006"
}

@article{Lau:2017qtz,
    author = "Lau, S. Y. and Leung, P. T. and Lin, L. -M.",
    title = "{Tidal deformations of compact stars with crystalline quark matter}",
    eprint = "1705.01710",
    archivePrefix = "arXiv",
    primaryClass = "astro-ph.HE",
    doi = "10.1103/PhysRevD.95.101302",
    journal = "Phys. Rev. D",
    volume = "95",
    number = "10",
    pages = "101302",
    year = "2017"
}

@article{Carignano:2017meb,
    author = "Carignano, Stefano and Mannarelli, Massimo and Anzuini, Filippo and Benhar, Omar",
    title = "{Crystalline phases by an improved gradient expansion technique}",
    eprint = "1711.08607",
    archivePrefix = "arXiv",
    primaryClass = "hep-ph",
    doi = "10.1103/PhysRevD.97.036009",
    journal = "Phys. Rev. D",
    volume = "97",
    number = "3",
    pages = "036009",
    year = "2018"
}

@article{Giannakis:2002jh,
    author = "Giannakis, Ioannis and Liu, James T. and Ren, Hai-cang",
    title = "{Angular momentum mixing in crystalline color superconductivity}",
    eprint = "hep-ph/0202138",
    archivePrefix = "arXiv",
    reportNumber = "RU02-2-B, MCTP-02-06",
    doi = "10.1103/PhysRevD.66.031501",
    journal = "Phys. Rev. D",
    volume = "66",
    pages = "031501",
    year = "2002"
}

@article{Casalbuoni:2001gt,
    author = "Casalbuoni, R. and Gatto, Raoul and Mannarelli, M. and Nardulli, G.",
    title = "{Effective field theory for the crystalline color superconductive phase of QCD}",
    eprint = "hep-ph/0101326",
    archivePrefix = "arXiv",
    reportNumber = "BARI-TH-410-01, UGVA-DPT-2001-01-1093",
    doi = "10.1016/S0370-2693(01)00645-1",
    journal = "Phys. Lett. B",
    volume = "511",
    pages = "218--228",
    year = "2001"
}

@article{Casalbuoni:2005zp,
    author = "Casalbuoni, R. and Gatto, R. and Ippolito, N. and Nardulli, G. and Ruggieri, M.",
    title = "{Ginzburg-Landau approach to the three flavor LOFF phase of QCD}",
    eprint = "hep-ph/0507247",
    archivePrefix = "arXiv",
    reportNumber = "BARI-TH-515-05",
    doi = "10.1016/j.physletb.2005.08.123",
    journal = "Phys. Lett. B",
    volume = "627",
    pages = "89--96",
    year = "2005",
    note = "[Erratum: Phys.Lett.B 634, 565--566 (2006)]"
}

@article{Lau:2018mae,
    author = "Lau, S. Y. and Leung, P. T. and Lin, L. -M.",
    title = "{Two-layer compact stars with crystalline quark matter: Screening effect on the tidal deformability}",
    eprint = "1808.08107",
    archivePrefix = "arXiv",
    primaryClass = "astro-ph.HE",
    doi = "10.1103/PhysRevD.99.023018",
    journal = "Phys. Rev. D",
    volume = "99",
    number = "2",
    pages = "023018",
    year = "2019"
}

@article{Forbes:2004ww,
    author = "Forbes, Michael McNeil",
    title = "{Kaon condensation in an NJL model at high density}",
    eprint = "hep-ph/0411001",
    archivePrefix = "arXiv",
    reportNumber = "MIT-CTP-3554",
    doi = "10.1103/PhysRevD.72.094032",
    journal = "Phys. Rev. D",
    volume = "72",
    pages = "094032",
    year = "2005"
}

@article{Buballa:2004sx,
    author = "Buballa, Michael",
    title = "{NJL-model description of Goldstone boson condensation in the color-flavor locked phase}",
    eprint = "hep-ph/0410397",
    archivePrefix = "arXiv",
    doi = "10.1016/j.physletb.2005.01.027",
    journal = "Phys. Lett. B",
    volume = "609",
    pages = "57--67",
    year = "2005"
}

@article{Alford:2002kj,
    author = "Alford, Mark and Rajagopal, Krishna",
    title = "{Absence of two flavor color superconductivity in compact stars}",
    eprint = "hep-ph/0204001",
    archivePrefix = "arXiv",
    reportNumber = "GUTPA-02-03-02, MIT-CTP-3258",
    doi = "10.1088/1126-6708/2002/06/031",
    journal = "JHEP",
    volume = "06",
    pages = "031",
    year = "2002"
}

@article{Heiselberg:1992dx,
    author = "Heiselberg, H. and Pethick, C. J. and Staubo, E. F.",
    title = "{Quark matter droplets in neutron stars}",
    reportNumber = "NORDITA-92-39-A",
    doi = "10.1103/PhysRevLett.70.1355",
    journal = "Phys. Rev. Lett.",
    volume = "70",
    pages = "1355--1359",
    year = "1993"
}

@article{Glendenning:1992vb,
    author = "Glendenning, Norman K.",
    title = "{First order phase transitions with more than one conserved charge: Consequences for neutron stars}",
    reportNumber = "LBL-30295-REV, LBL-30295",
    doi = "10.1103/PhysRevD.46.1274",
    journal = "Phys. Rev. D",
    volume = "46",
    pages = "1274--1287",
    year = "1992"
}

@article{Schmitt:2020tac,
    author = "Schmitt, Andreas",
    title = "{Chiral pasta: Mixed phases at the chiral phase transition}",
    eprint = "2002.01451",
    archivePrefix = "arXiv",
    primaryClass = "hep-ph",
    doi = "10.1103/PhysRevD.101.074007",
    journal = "Phys. Rev. D",
    volume = "101",
    number = "7",
    pages = "074007",
    year = "2020"
}

@article{Glampedakis:2012qp,
    author = "Glampedakis, K. and Jones, D. I. and Samuelsson, L.",
    title = "{Gravitational waves from color-magnetic `mountains' in neutron stars}",
    eprint = "1204.3781",
    archivePrefix = "arXiv",
    primaryClass = "astro-ph.SR",
    doi = "10.1103/PhysRevLett.109.081103",
    journal = "Phys. Rev. Lett.",
    volume = "109",
    pages = "081103",
    year = "2012"
}

@article{Evans:2020uui,
    author = "Evans, Geraint W. and Schmitt, Andreas",
    title = "{Strange quark mass turns magnetic domain walls into multi-winding flux tubes}",
    eprint = "2009.01141",
    archivePrefix = "arXiv",
    primaryClass = "hep-ph",
    doi = "10.1088/1361-6471/abcb9d",
    journal = "J. Phys. G",
    volume = "48",
    number = "3",
    pages = "035002",
    year = "2021"
}

@article{Alford:2010qf,
    author = "Alford, Mark G. and Sedrakian, Armen",
    title = "{Color-magnetic flux tubes in quark matter cores of neutron stars}",
    eprint = "1001.3346",
    archivePrefix = "arXiv",
    primaryClass = "astro-ph.SR",
    doi = "10.1088/0954-3899/37/7/075202",
    journal = "J. Phys. G",
    volume = "37",
    pages = "075202",
    year = "2010"
}

@article{Iida:2004if,
    author = "Iida, Kei",
    title = "{Magnetic vortex in color-flavor locked quark matter}",
    eprint = "hep-ph/0412426",
    archivePrefix = "arXiv",
    doi = "10.1103/PhysRevD.71.054011",
    journal = "Phys. Rev. D",
    volume = "71",
    pages = "054011",
    year = "2005"
}

@article{Iida:2002ev,
    author = "Iida, Kei and Baym, Gordon",
    title = "{Superfluid phases of quark matter. 3. Supercurrents and vortices}",
    eprint = "hep-ph/0204124",
    archivePrefix = "arXiv",
    doi = "10.1103/PhysRevD.66.014015",
    journal = "Phys. Rev. D",
    volume = "66",
    pages = "014015",
    year = "2002"
}

@article{Haber:2018tqw,
    author = "Haber, Alexander and Schmitt, Andreas",
    title = "{Multi-winding flux tubes in CFL quark matter}",
    eprint = "1811.12302",
    archivePrefix = "arXiv",
    primaryClass = "hep-ph",
    doi = "10.22323/1.336.0213",
    journal = "PoS Confinement2018",
    pages = "213",
    year = "2018"
}

@article{Alford:2016dco,
    author = "Alford, Mark G. and Mallavarapu, S. Kumar and Vachaspati, Tanmay and Windisch, Andreas",
    title = "{Stability of superfluid vortices in dense quark matter}",
    eprint = "1601.04656",
    archivePrefix = "arXiv",
    primaryClass = "nucl-th",
    doi = "10.1103/PhysRevC.93.045801",
    journal = "Phys. Rev. C",
    volume = "93",
    number = "4",
    pages = "045801",
    year = "2016"
}

@article{Vinci:2012mc,
    author = "Vinci, Walter and Cipriani, Mattia and Nitta, Muneto",
    title = "{Spontaneous Magnetization through Non-Abelian Vortex Formation in Rotating Dense Quark Matter}",
    eprint = "1206.3535",
    archivePrefix = "arXiv",
    primaryClass = "hep-ph",
    doi = "10.1103/PhysRevD.86.085018",
    journal = "Phys. Rev. D",
    volume = "86",
    pages = "085018",
    year = "2012"
}

@article{Eto:2009kg,
    author = "Eto, Minoru and Nitta, Muneto",
    title = "{Color Magnetic Flux Tubes in Dense QCD}",
    eprint = "0907.1278",
    archivePrefix = "arXiv",
    primaryClass = "hep-ph",
    reportNumber = "RIKEN-TH-160",
    doi = "10.1103/PhysRevD.80.125007",
    journal = "Phys. Rev. D",
    volume = "80",
    pages = "125007",
    year = "2009"
}

@article{Eto:2013hoa,
    author = "Eto, Minoru and Hirono, Yuji and Nitta, Muneto and Yasui, Shigehiro",
    title = "{Vortices and Other Topological Solitons in Dense Quark Matter}",
    eprint = "1308.1535",
    archivePrefix = "arXiv",
    primaryClass = "hep-ph",
    doi = "10.1093/ptep/ptt095",
    journal = "PTEP",
    volume = "2014",
    number = "1",
    pages = "012D01",
    year = "2014"
}

@article{Rischke:2002rz,
    author = "Rischke, Dirk H. and Shovkovy, Igor A.",
    title = "{Longitudinal gluons and Nambu-Goldstone bosons in a two flavor color superconductor}",
    eprint = "nucl-th/0205080",
    archivePrefix = "arXiv",
    doi = "10.1103/PhysRevD.66.054019",
    journal = "Phys. Rev. D",
    volume = "66",
    pages = "054019",
    year = "2002"
}

@article{Malekzadeh:2006ud,
    author = "Malekzadeh, H. and Rischke, Dirk H.",
    title = "{Gluon self-energy in the color-flavor-locked phase}",
    eprint = "hep-ph/0602082",
    archivePrefix = "arXiv",
    doi = "10.1103/PhysRevD.73.114006",
    journal = "Phys. Rev. D",
    volume = "73",
    pages = "114006",
    year = "2006"
}

@article{Geissel:2025vnp,
    author = "Gei{\ss}el, Andreas and Gorda, Tyler and Braun, Jens",
    title = "{Color superconductivity under neutron-star conditions at next-to-leading order}",
    eprint = "2504.03834",
    archivePrefix = "arXiv",
    primaryClass = "hep-ph",
    month = "4",
    year = "2025"
}

@article{Geissel:2024nmx,
    author = "Gei{\ss}el, Andreas and Gorda, Tyler and Braun, Jens",
    title = "{Pressure and speed of sound in two-flavor color-superconducting quark matter at next-to-leading order}",
    eprint = "2403.18010",
    archivePrefix = "arXiv",
    primaryClass = "hep-ph",
    doi = "10.1103/PhysRevD.110.014034",
    journal = "Phys. Rev. D",
    volume = "110",
    number = "1",
    pages = "014034",
    year = "2024"
}

@article{Kurkela:2009gj,
    author = "Kurkela, Aleksi and Romatschke, Paul and Vuorinen, Aleksi",
    title = "{Cold Quark Matter}",
    eprint = "0912.1856",
    archivePrefix = "arXiv",
    primaryClass = "hep-ph",
    reportNumber = "BI-TP-2009-30, CERN-PH-TH-2009-229, INT-PUB-09-060, TUW-09-19",
    doi = "10.1103/PhysRevD.81.105021",
    journal = "Phys. Rev. D",
    volume = "81",
    pages = "105021",
    year = "2010"
}

@article{Gorda:2023mkk,
    author = {Gorda, Tyler and Paatelainen, Risto and S{\"a}ppi, Saga and Sepp{\"a}nen, Kaapo},
    title = "{Equation of State of Cold Quark Matter to $O(\alpha_s^3 \ln \alpha_s)$}",
    eprint = "2307.08734",
    archivePrefix = "arXiv",
    primaryClass = "hep-ph",
    reportNumber = "HIP-2023-10/TH, TUM-EFT 181/23",
    doi = "10.1103/PhysRevLett.131.181902",
    journal = "Phys. Rev. Lett.",
    volume = "131",
    number = "18",
    pages = "181902",
    year = "2023"
}

@article{Fernandez:2021jfr,
    author = {Fernandez, Lo{\"\i}c and Kneur, Jean-Lo{\"\i}c},
    title = "{All Order Resummed Leading and Next-to-Leading Soft Modes of Dense QCD Pressure}",
    eprint = "2109.02410",
    archivePrefix = "arXiv",
    primaryClass = "hep-ph",
    doi = "10.1103/PhysRevLett.129.212001",
    journal = "Phys. Rev. Lett.",
    volume = "129",
    number = "21",
    pages = "212001",
    year = "2022"
}

@article{Gorda:2021znl,
    author = {Gorda, Tyler and Kurkela, Aleksi and Paatelainen, Risto and S{\"a}ppi, Saga and Vuorinen, Aleksi},
    title = "{Soft Interactions in Cold Quark Matter}",
    eprint = "2103.05658",
    archivePrefix = "arXiv",
    primaryClass = "hep-ph",
    reportNumber = "HIP-2021-9/TH",
    doi = "10.1103/PhysRevLett.127.162003",
    journal = "Phys. Rev. Lett.",
    volume = "127",
    number = "16",
    pages = "162003",
    year = "2021"
}

@Article{Freedman:1976ub,
      author    = "Freedman, Barry A. and McLerran, Larry D.",
      title     = "Fermions and Gauge Vector Mesons at Finite Temperature and
                   Density. 3. The Ground State Energy of a Relativistic Quark
                   Gas",
      journal   = "Phys. Rev.",
      volume    = "D16",
      year      = "1977",
      pages     = "1169",
      SLACcitation  = "%%CITATION = PHRVA,D16,1169;%%"
}

@Article{Deryagin:1992rw,
     author    = "Deryagin, D. V. and Grigoriev, Dmitri Yu. and Rubakov, V.
                  A.",
     title     = "Standing wave ground state in high density, zero
                  temperature QCD at large N(c)",
     journal   = "Int. J. Mod. Phys.",
     volume    = "A7",
     year      = "1992",
     pages     = "659-681",
     SLACcitation  = "%%CITATION = IMPAE,A7,659;%%"
}

@article{Braun:2021uua,
    author = "Braun, Jens and Schallmo, Benedikt",
    title = "{From quarks and gluons to color superconductivity at supranuclear densities}",
    eprint = "2106.04198",
    archivePrefix = "arXiv",
    primaryClass = "hep-ph",
    doi = "10.1103/PhysRevD.105.036003",
    journal = "Phys. Rev. D",
    volume = "105",
    number = "3",
    pages = "036003",
    year = "2022"
}

@article{Nickel:2006vf,
    author = "Nickel, Dominik and Wambach, Jochen and Alkofer, Reinhard",
    title = "{Color-superconductivity in the strong-coupling regime of Landau gauge QCD}",
    eprint = "hep-ph/0603163",
    archivePrefix = "arXiv",
    doi = "10.1103/PhysRevD.73.114028",
    journal = "Phys. Rev. D",
    volume = "73",
    pages = "114028",
    year = "2006"
}

@article{Marhauser:2006hy,
    author = "Marhauser, Florian and Nickel, Dominik and Buballa, Michael and Wambach, Jochen",
    title = "{Color-spin locking in a selfconsistent Dyson-Schwinger approach}",
    eprint = "hep-ph/0612027",
    archivePrefix = "arXiv",
    doi = "10.1103/PhysRevD.75.054022",
    journal = "Phys. Rev. D",
    volume = "75",
    pages = "054022",
    year = "2007"
}

@article{Gholami:2025afm,
    author = "Gholami, Hosein and Kurth, Lennart and Mire, Ugo and Buballa, Michael and Schaefer, Bernd-Jochen",
    title = "{Renormalizing the Quark-Meson-Diquark Model}",
    eprint = "2505.22542",
    archivePrefix = "arXiv",
    primaryClass = "hep-ph",
    month = "5",
    year = "2025"
}

@article{Basu:2011yg,
    author = "Basu, Pallab and Nogueira, Fernando and Rozali, Moshe and Stang, Jared B. and Van Raamsdonk, Mark",
    title = "{Towards A Holographic Model of Color Superconductivity}",
    eprint = "1101.4042",
    archivePrefix = "arXiv",
    primaryClass = "hep-th",
    doi = "10.1088/1367-2630/13/5/055001",
    journal = "New J. Phys.",
    volume = "13",
    pages = "055001",
    year = "2011"
}

@article{BitaghsirFadafan:2018iqr,
    author = "Bitaghsir Fadafan, Kazem and Cruz Rojas, Jesus and Evans, Nick",
    title = "{Holographic description of color superconductivity}",
    eprint = "1803.03107",
    archivePrefix = "arXiv",
    primaryClass = "hep-ph",
    doi = "10.1103/PhysRevD.98.066010",
    journal = "Phys. Rev. D",
    volume = "98",
    number = "6",
    pages = "066010",
    year = "2018"
}

@article{Faedo:2018fjw,
    author = "Faedo, Ant{\'o}n F. and Mateos, David and Pantelidou, Christiana and Tarr{\'\i}o, Javier",
    title = "{A Supersymmetric Color Superconductor from Holography}",
    eprint = "1807.09712",
    archivePrefix = "arXiv",
    primaryClass = "hep-th",
    reportNumber = "ICCUB-18-016",
    doi = "10.1007/JHEP05(2019)106",
    journal = "JHEP",
    volume = "05",
    pages = "106",
    year = "2019"
}

@article{Chen:2009kx,
    author = "Chen, Heng-Yu and Hashimoto, Koji and Matsuura, Shunji",
    title = "{Towards a Holographic Model of Color-Flavor Locking Phase}",
    eprint = "0909.1296",
    archivePrefix = "arXiv",
    primaryClass = "hep-th",
    reportNumber = "MAD-TH-09-07, NSF-KITP-09-172, RIKEN-TH-166",
    doi = "10.1007/JHEP02(2010)104",
    journal = "JHEP",
    volume = "02",
    pages = "104",
    year = "2010"
}

@article{CruzRojas:2025fzs,
    author = {Cruz Rojas, Jes{\'u}s and Demircik, Tuna and Ecker, Christian and J{\"a}rvinen, Matti},
    title = "{Towards holographic color superconductivity in QCD}",
    eprint = "2505.06338",
    archivePrefix = "arXiv",
    primaryClass = "hep-th",
    month = "5",
    year = "2025"
}

@article{Preau:2025ubr,
    author = "Pr{\'e}au, Edwan",
    title = "{Revisiting color superconductors in bottom-up holography}",
    eprint = "2501.16472",
    archivePrefix = "arXiv",
    primaryClass = "hep-th",
    doi = "10.1007/JHEP05(2025)204",
    journal = "JHEP",
    volume = "05",
    pages = "204",
    year = "2025"
}

@article{Prakash:2000jr,
    author = "Prakash, Madappa and Lattimer, James M. and Pons, Jose A. and Steiner, Andrew W. and Reddy, Sanjay",
    editor = "Blaschke, D. and Glendenning, N. K. and Sedrakian, A.",
    title = "{Evolution of a neutron star from its birth to old age}",
    eprint = "astro-ph/0012136",
    archivePrefix = "arXiv",
    journal = "Lect. Notes Phys.",
    volume = "578",
    pages = "364--423",
    year = "2001"
}

@article{Hammond:2021vtv,
    author = "Hammond, Peter and Hawke, Ian and Andersson, Nils",
    title = "{Thermal aspects of neutron star mergers}",
    eprint = "2108.08649",
    archivePrefix = "arXiv",
    primaryClass = "astro-ph.HE",
    doi = "10.1103/PhysRevD.104.103006",
    journal = "Phys. Rev. D",
    volume = "104",
    number = "10",
    pages = "103006",
    year = "2021"
}

@article{Fabrocini:2005lvb,
    author = "Fabrocini, Adelchi and Fantoni, Stefano and Illarionov, Alexey Yu. and Schmidt, Kevin E.",
    title = "{S-1(0) superfluid phase-transition in neutron matter with realistic nuclear potentials and modern many-body theories}",
    eprint = "nucl-th/0607034",
    archivePrefix = "arXiv",
    doi = "10.1103/PhysRevLett.95.192501",
    journal = "Phys. Rev. Lett.",
    volume = "95",
    pages = "192501",
    year = "2005"
}

@article{Schulze:1996zz,
    author = "Schulze, H. J. and Cugnon, J. and Lejeune, A. and Baldo, M. and Lombardo, U.",
    title = "{Medium polarization effects on neutron matter superfluidity}",
    doi = "10.1016/0370-2693(96)00213-4",
    journal = "Phys. Lett. B",
    volume = "375",
    pages = "1--8",
    year = "1996"
}

@article{Cao:2006gq,
    author = "Cao, L. G. and Lombardo, U. and Schuck, P.",
    title = "{Screening effects in superfluid nuclear and neutron matter within Brueckner theory}",
    eprint = "nucl-th/0608005",
    archivePrefix = "arXiv",
    doi = "10.1103/PhysRevC.74.064301",
    journal = "Phys. Rev. C",
    volume = "74",
    pages = "064301",
    year = "2006"
}

@article{Wambach:1992ik,
    author = "Wambach, J. and Ainsworth, T. L. and Pines, D.",
    title = "{Quasiparticle interactions in neutron matter for applications in neutron stars}",
    reportNumber = "KFA-IKP-TH-1992-6",
    doi = "10.1016/0375-9474(93)90317-Q",
    journal = "Nucl. Phys. A",
    volume = "555",
    pages = "128--150",
    year = "1993"
}

@article{Gholami:2024diy,
    author = "Gholami, Hosein and Hofmann, Marco and Buballa, Michael",
    title = "{Renormalization-group consistent treatment of color superconductivity in the NJL model}",
    eprint = "2408.06704",
    archivePrefix = "arXiv",
    primaryClass = "hep-ph",
    doi = "10.1103/PhysRevD.111.014006",
    journal = "Phys. Rev. D",
    volume = "111",
    number = "1",
    pages = "014006",
    year = "2025"
}

@article{Warringa:2006dk,
    author = "Warringa, Harmen J.",
    title = "{The Phase diagram of neutral quark matter with pseudoscalar condensates in the color-flavor locked phase}",
    eprint = "hep-ph/0606063",
    archivePrefix = "arXiv",
    month = "6",
    year = "2006"
}

@Article{Ruster:2005jc,
     author    = "R{\"u}ster, Stefan B. and Werth, Verena and Buballa, Michael
                  and Shovkovy, Igor A. and Rischke, Dirk H.",
     title     = "The phase diagram of neutral quark matter: Self-consistent
                  treatment of  quark masses",
     journal   = "Phys. Rev.",
     volume    = "D72",
     year      = "2005",
     pages     = "034004",
     eprint    = "hep-ph/0503184",
     SLACcitation  = "%%CITATION = HEP-PH/0503184;%%"
}

@Article{Steiner:2002gx,
     author    = "Steiner, Andrew W. and Reddy, Sanjay and Prakash, Madappa",
     title     = "Color-neutral superconducting quark matter",
     journal   = "Phys. Rev.",
     volume    = "D66",
     year      = "2002",
     pages     = "094007",
     eprint    = "hep-ph/0205201",
     SLACcitation  = "%%CITATION = HEP-PH/0205201;%%"
}

@Article{Blaschke:2005uj,
     author    = "Blaschke, D. and Fredriksson, S. and Grigorian, H. and
                  Oztas, A. M. and Sandin, F.",
     title     = "The phase diagram of three-flavor quark matter under
                  compact star  constraints",
     journal   = "Phys. Rev.",
     volume    = "D72",
     year      = "2005",
     pages     = "065020",
     eprint    = "hep-ph/0503194",
     SLACcitation  = "%%CITATION = HEP-PH/0503194;%%"
}

@Article{Abuki:2004zk,
     author    = "Abuki, Hiroaki and Kitazawa, Masakiyo and Kunihiro, Teiji",
     title     = "How do chiral condensates affect color superconducting
                  quark matter under charge neutrality constraints?",
     journal   = "Phys. Lett.",
     volume    = "B615",
     year      = "2005",
     pages     = "102-110",
     eprint    = "hep-ph/0412382",
     SLACcitation  = "%%CITATION = HEP-PH/0412382;%%"
}

@Article{Buballa:2001gj,
     author    = "Buballa, M. and Oertel, M.",
     title     = "Color-flavor unlocking and phase diagram with self-
                  consistently  determined strange quark masses",
     journal   = "Nucl. Phys.",
     volume    = "A703",
     year      = "2002",
     pages     = "770-784",
     eprint    = "hep-ph/0109095",
     SLACcitation  = "%%CITATION = HEP-PH/0109095;%%"
}

@article{Nambu:1961tp,
    author = "Nambu, Yoichiro and Jona-Lasinio, G.",
    editor = "Eguchi, T.",
    title = "{Dynamical Model of Elementary Particles Based on an Analogy with Superconductivity. 1.}",
    doi = "10.1103/PhysRev.122.345",
    journal = "Phys. Rev.",
    volume = "122",
    pages = "345--358",
    year = "1961"
}

@article{Nambu:1961fr,
    author = "Nambu, Yoichiro and Jona-Lasinio, G.",
    editor = "Eguchi, T.",
    title = "{Dynamical model of elementary particles based on an analogy with superconductivity. II.}",
    doi = "10.1103/PhysRev.124.246",
    journal = "Phys. Rev.",
    volume = "124",
    pages = "246--254",
    year = "1961"
}

@article{Kurkela:2024xfh,
    author = "Kurkela, Aleksi and Rajagopal, Krishna and Steinhorst, Rachel",
    title = "{Astrophysical Equation-of-State Constraints on the Color-Superconducting Gap}",
    eprint = "2401.16253",
    archivePrefix = "arXiv",
    primaryClass = "astro-ph.HE",
    reportNumber = "MIT-CTP/5671",
    doi = "10.1103/PhysRevLett.132.262701",
    journal = "Phys. Rev. Lett.",
    volume = "132",
    number = "26",
    pages = "262701",
    year = "2024"
}

@article{Anglani:2013gfu,
    author = "Anglani, Roberto and Casalbuoni, Roberto and Ciminale, Marco and Ippolito, Nicola and Gatto, Raoul and Mannarelli, Massimo and Ruggieri, Marco",
    title = "{Crystalline color superconductors}",
    eprint = "1302.4264",
    archivePrefix = "arXiv",
    primaryClass = "hep-ph",
    doi = "10.1103/RevModPhys.86.509",
    journal = "Rev. Mod. Phys.",
    volume = "86",
    pages = "509--561",
    year = "2014"
}

@article{Fukushima:2010bq,
    author = "Fukushima, Kenji and Hatsuda, Tetsuo",
    title = "{The phase diagram of dense QCD}",
    eprint = "1005.4814",
    archivePrefix = "arXiv",
    primaryClass = "hep-ph",
    reportNumber = "YITP-10-28, TKYNT-10-06",
    doi = "10.1088/0034-4885/74/1/014001",
    journal = "Rept. Prog. Phys.",
    volume = "74",
    pages = "014001",
    year = "2011"
}

@Article{Rajagopal:2000wf,
     author    = "Rajagopal, Krishna and Wilczek, Frank",
     title     = "The condensed matter physics of QCD",
     year      = "2000",
     eprint    = "hep-ph/0011333",
     SLACcitation  = "%%CITATION = HEP-PH/0011333;%%"
}

@inproceedings{Shovkovy:2005fy,
    author = "Shovkovy, Igor A.",
    title = "{Color superconductivity in quark matter}",
    booktitle = "{Workshop on Extreme QCD}",
    eprint = "nucl-th/0511014",
    archivePrefix = "arXiv",
    pages = "37--46",
    month = "11",
    year = "2005"
}

@Article{Balachandran:2005ev,
     author    = "Balachandran, A. P. and Digal, S. and Matsuura, T.",
     title     = "Semi-superfluid strings in high density QCD",
     journal   = "Phys. Rev.",
     volume    = "D73",
     year      = "2006",
     pages     = "074009",
     eprint    = "hep-ph/0509276",
     SLACcitation  = "%%CITATION = HEP-PH/0509276;%%"
}

@Article{Forbes:2001gj,
     author    = "Forbes, Michael McNeil and Zhitnitsky, Ariel R.",
     title     = "Global strings in high density QCD",
     journal   = "Phys. Rev.",
     volume    = "D65",
     year      = "2002",
     pages     = "085009",
     eprint    = "hep-ph/0109173",
     SLACcitation  = "%%CITATION = HEP-PH/0109173;%%"
}

@Article{Barrois:1977xd,
     author    = "Barrois, Bertrand C.",
     title     = "SUPERCONDUCTING QUARK MATTER",
     journal   = "Nucl. Phys.",
     volume    = "B129",
     year      = "1977",
     pages     = "390",
     SLACcitation  = "%%CITATION = NUPHA,B129,390;%%"
}

@Article{Frautschi:1978rz,
     author    = "Frautschi, Steven C.",
     title     = "ASYMPTOTIC FREEDOM AND COLOR SUPERCONDUCTIVITY IN DENSE
                  QUARK MATTER",
     note      = "Presented at Workshop on Hadronic Matter at Extreme Energy
                  Density, Erice, Italy, Oct 13-21, 1978",
     year      = "1978"
}

@article{Alford:1997zt,
    author = "Alford, Mark G. and Rajagopal, Krishna and Wilczek, Frank",
    title = "{QCD at finite baryon density: Nucleon droplets and color superconductivity}",
    eprint = "hep-ph/9711395",
    archivePrefix = "arXiv",
    reportNumber = "IASSNS-HEP-97-119, MIT-CTP-2695",
    doi = "10.1016/S0370-2693(98)00051-3",
    journal = "Phys. Lett. B",
    volume = "422",
    pages = "247--256",
    year = "1998"
}

@article{Rischke:2000ra,
    author = "Rischke, Dirk H.",
    title = "{Debye screening and Meissner effect in a three flavor color superconductor}",
    eprint = "nucl-th/0003063",
    archivePrefix = "arXiv",
    doi = "10.1103/PhysRevD.62.054017",
    journal = "Phys. Rev. D",
    volume = "62",
    pages = "054017",
    year = "2000"
}

@book{lebellac, 
place={Cambridge}, 
series={Cambridge Monographs on Mathematical Physics}, 
title={Thermal Field Theory}, 
DOI={10.1017/CBO9780511721700}, 
publisher={Cambridge University Press}, 
author={Le Bellac, Michel}, 
year={1996}, 
collection={Cambridge Monographs on Mathematical Physics}
}

@book{Kadanoff,
  address = {New York},
  author = {Kadanoff, L.P. and Baym, G.},
  publisher = {W.A. Benjamin Inc.},
  title = {Quantum Statistical Mechanics},
  year = 1962
}

@article{Sad:2007afd,
    author = "Sa'd, Basil A. and Shovkovy, Igor A. and Rischke, Dirk H.",
    archivePrefix = "arXiv",
    doi = "10.1103/PhysRevD.75.125004",
    eprint = "astro-ph/0703016",
    journal = "Phys.\ Rev.\ D",
    pages = "125004",
    title = "{Bulk viscosity of strange quark matter: Urca versus non-leptonic processes}",
    volume = "75",
    year = "2007"
}

@article{Iwamoto:1982zz,
    author = "Iwamoto, Naoki",
    doi = "10.1016/0003-4916(82)90271-8",
    journal = "Annals Phys.",
    pages = "1--49",
    title = "{Neutrino emissivities and mean free paths of degenerate quark matter}",
    volume = "141",
    year = "1982"
}

@article{Iwamoto:1980eb,
    author = "Iwamoto, N.",
    doi = "10.1103/PhysRevLett.44.1637",
    journal = "Phys.\ Rev.\ Lett.",
    pages = "1637--1640",
    title = "{QUARK BETA DECAY AND THE COOLING OF NEUTRON STARS}",
    volume = "44",
    year = "1980"
}

@article{Mannarelli:2009ia,
    author = "Mannarelli, Massimo and Manuel, Cristina",
    archivePrefix = "arXiv",
    doi = "10.1103/PhysRevD.81.043002",
    eprint = "0909.4486",
    journal = "Phys.\ Rev.\ D",
    pages = "043002",
    primaryClass = "hep-ph",
    title = "{Bulk viscosities of a cold relativistic superfluid: Color-flavor locked quark matter}",
    volume = "81",
    year = "2010"
}

@article{Alford:2007qa,
      author         = "Alford, Mark G. and Braby, Matt and Schmitt, Andreas",
      title          = "{Critical temperature for kaon condensation in
                        color-flavor locked quark matter}",
      journal        = "J. Phys.",
      volume         = "G35",
      year           = "2008",
      pages          = "025002",
      doi            = "10.1088/0954-3899/35/2/025002",
      eprint         = "0707.2389",
      archivePrefix  = "arXiv",
      primaryClass   = "nucl-th",
      SLACcitation   = "%%CITATION = ARXIV:0707.2389;%%"
}

@article{Schwenzer:2012ga,
      author         = "Schwenzer, Kai",
      title          = "{How long-range interactions tune the damping in compact
                        stars}",
      year           = "2012",
      eprint         = "1212.5242",
      archivePrefix  = "arXiv",
      primaryClass   = "nucl-th",
      SLACcitation   = "%%CITATION = ARXIV:1212.5242;%%"
}

@BOOK{Vollhardt1990,
author = {{Vollhardt}, D. and {W\"{o}lfle}, P.},
title = "The superfluid phases of Helium 3",
year = 1990,
publisher={Taylor \& Francis},
address={Bristol}
}

@article{Alford:2010gw,
      author         = "Alford, Mark G. and Mahmoodifar, Simin and Schwenzer,
                        Kai",
      title          = "{Large amplitude behavior of the bulk viscosity of dense
                        matter}",
      journal        = "J. Phys.",
      volume         = "G37",
      year           = "2010",
      pages          = "125202",
      doi            = "10.1088/0954-3899/37/12/125202",
      eprint         = "1005.3769",
      archivePrefix  = "arXiv",
      primaryClass   = "nucl-th",
      SLACcitation   = "%%CITATION = ARXIV:1005.3769;%%"
}

@article{PhysRevD.46.3290,
  title = {Bulk viscosity of strange quark matter, damping of quark star vibration, and the maximum rotation rate of pulsars},
  author = {Madsen, Jes},
  journal = {Phys. Rev. D},
  volume = {46},
  issue = {8},
  pages = {3290--3295},
  numpages = {0},
  year = {1992},
  month = {Oct},
  publisher = {American Physical Society},
  doi = {10.1103/PhysRevD.46.3290}
}

@article{Heiselberg:1992bd,
      author         = "Heiselberg, H.",
      title          = "{The Weak conversion rate in quark matter}",
      journal        = "Phys. Scripta",
      volume         = "46",
      year           = "1992",
      pages          = "485-488",
      doi            = "10.1088/0031-8949/46/6/002",
      SLACcitation   = "%%CITATION = PHSTB,46,485;%%"
}

@article{PhysRevD.47.325,
  title = {Rate of the weak reaction $s+u\ensuremath{\rightarrow}u+d$ in quark matter},
  author = {Madsen, Jes},
  journal = {Phys. Rev. D},
  volume = {47},
  issue = {1},
  pages = {325--330},
  numpages = {0},
  year = {1993},
  month = {Jan},
  publisher = {American Physical Society},
  doi = {10.1103/PhysRevD.47.325}
}

@article{Manuel:2007pz,
      author         = "Manuel, Cristina and Llanes-Estrada, Felipe J.",
      title          = "{Bulk viscosity in a cold CFL superfluid}",
      journal        = "JCAP",
      volume         = "0708",
      year           = "2007",
      pages          = "001",
      doi            = "10.1088/1475-7516/2007/08/001",
      eprint         = "0705.3909",
      archivePrefix  = "arXiv",
      primaryClass   = "hep-ph",
      SLACcitation   = "%%CITATION = ARXIV:0705.3909;%%"
}

@article{Alford:2007rw,
      author         = "Alford, Mark G. and Braby, Matt and Reddy, Sanjay and
                        Schäfer, Thomas",
      title          = "{Bulk viscosity due to kaons in color-flavor-locked quark
                        matter}",
      journal        = "Phys. Rev.",
      volume         = "C75",
      year           = "2007",
      pages          = "055209",
      doi            = "10.1103/PhysRevC.75.055209",
      eprint         = "nucl-th/0701067",
      archivePrefix  = "arXiv",
      primaryClass   = "nucl-th",
      reportNumber   = "LA-UR-07-0429",
      SLACcitation   = "%%CITATION = NUCL-TH/0701067;%%"
}

@article{Alford:2008pb,
      author         = "Alford, Mark G. and Braby, Matt and Schmitt, Andreas",
      title          = "{Bulk viscosity in kaon-condensed color-flavor locked
                        quark matter}",
      journal        = "J. Phys.",
      volume         = "G35",
      year           = "2008",
      pages          = "115007",
      doi            = "10.1088/0954-3899/35/11/115007",
      eprint         = "0806.0285",
      archivePrefix  = "arXiv",
      primaryClass   = "nucl-th",
      SLACcitation   = "%%CITATION = ARXIV:0806.0285;%%"
}

@article{Alford:2006gy,
      author         = "Alford, Mark G. and Schmitt, Andreas",
      title          = "{Bulk viscosity in 2SC quark matter}",
      journal        = "J. Phys.",
      volume         = "G34",
      year           = "2007",
      pages          = "67-102",
      doi            = "10.1088/0954-3899/34/1/005",
      eprint         = "nucl-th/0608019",
      archivePrefix  = "arXiv",
      primaryClass   = "nucl-th",
      SLACcitation   = "%%CITATION = NUCL-TH/0608019;%%"
}

@Article{Son:1998uk,
     author    = "Son, D. T.",
     title     = "Superconductivity by long-range color magnetic interaction
                  in  high-density quark matter",
     journal   = "Phys. Rev.",
     volume    = "D59",
     year      = "1999",
     pages     = "094019",
     eprint    = "hep-ph/9812287",
     SLACcitation  = "%%CITATION = HEP-PH/9812287;%%"
}

@phdthesis{Barrois:1979pv,
     author    = "Barrois, Bertrand C.",
     title     = "Non-perturbative effects in dense quark matter",
     school    = "California Institute of Technology, Pasadena, California",     
     year      = "1979",
     note      = "{UMI} 79-04847"
}

@book{Bellac:2011kqa,
      author         = "Bellac, Michel Le",
      title          = "{Thermal Field Theory}",
      publisher      = "Cambridge University Press",
      year           = "2011",
      series         = "Cambridge Monographs on Mathematical Physics",
      doi            = "10.1017/CBO9780511721700",
      SLACcitation   = "%%CITATION = INSPIRE-1384874;%%"
}

@article{Pisarski:1999av,
      author         = "Pisarski, Robert D. and Rischke, Dirk H.",
      title          = "{Superfluidity in a model of massless fermions coupled to
                        scalar bosons}",
      journal        = "Phys. Rev.",
      volume         = "D60",
      year           = "1999",
      pages          = "094013",
      doi            = "10.1103/PhysRevD.60.094013",
      eprint         = "nucl-th/9903023",
      archivePrefix  = "arXiv",
      primaryClass   = "nucl-th",
      SLACcitation   = "%%CITATION = NUCL-TH/9903023;%%"
}

@article{Bailin:1983bm,
      author         = "Bailin, D. and Love, A.",
      title          = "{Superfluidity and Superconductivity in Relativistic
                        Fermion Systems}",
      journal        = "Phys. Rept.",
      volume         = "107",
      year           = "1984",
      pages          = "325",
      doi            = "10.1016/0370-1573(84)90145-5",
      reportNumber   = "PRINT-83-1015 (SUSSEX)",
      SLACcitation   = "%%CITATION = PRPLC,107,325;%%"
}

@article{Gerhold:2005uu,
      author         = "Gerhold, Andreas and Rebhan, Anton",
      title          = "{Fermionic dispersion relations in ultradegenerate
                        relativistic plasmas beyond leading logarithmic order}",
      journal        = "Phys. Rev.",
      volume         = "D71",
      year           = "2005",
      pages          = "085010",
      doi            = "10.1103/PhysRevD.71.085010",
      eprint         = "hep-ph/0501089",
      archivePrefix  = "arXiv",
      primaryClass   = "hep-ph",
      reportNumber   = "TUW-05-01",
      SLACcitation   = "%%CITATION = HEP-PH/0501089;%%"
}

@article{Schafer:2000tw,
      author         = "Sch{\"a}fer, Thomas",
      title          = "{Quark hadron continuity in QCD with one flavor}",
      journal        = "Phys. Rev.",
      volume         = "D62",
      year           = "2000",
      pages          = "094007",
      doi            = "10.1103/PhysRevD.62.094007",
      eprint         = "hep-ph/0006034",
      archivePrefix  = "arXiv",
      primaryClass   = "hep-ph",
      reportNumber   = "SUNY-NTG-00-13",
      SLACcitation   = "%%CITATION = HEP-PH/0006034;%%"
}

@article{Ruester:2005jc,
      author         = "Ruester, Stefan B. and Werth, Verena and Buballa, Michael
                        and Shovkovy, Igor A. and Rischke, Dirk H.",
      title          = "{The Phase diagram of neutral quark matter:
                        Self-consistent treatment of quark masses}",
      journal        = "Phys. Rev.",
      volume         = "D72",
      year           = "2005",
      pages          = "034004",
      doi            = "10.1103/PhysRevD.72.034004",
      eprint         = "hep-ph/0503184",
      archivePrefix  = "arXiv",
      primaryClass   = "hep-ph",
      SLACcitation   = "%%CITATION = HEP-PH/0503184;%%"
}

@article{Mannarelli:2007bs,
      author         = "Mannarelli, Massimo and Rajagopal, Krishna and Sharma,
                        Rishi",
      title          = "{The Rigidity of crystalline color superconducting quark
                        matter}",
      journal        = "Phys. Rev.",
      volume         = "D76",
      year           = "2007",
      pages          = "074026",
      doi            = "10.1103/PhysRevD.76.074026",
      eprint         = "hep-ph/0702021",
      archivePrefix  = "arXiv",
      primaryClass   = "hep-ph",
      reportNumber   = "MIT-CTP-3807",
      SLACcitation   = "%%CITATION = HEP-PH/0702021;%%"
}

@article{Knippel:2009st,
      author         = "Knippel, Bettina and Sedrakian, Armen",
      title          = "{Gravitational radiation from crystalline
                        color-superconducting hybrid stars}",
      journal        = "Phys. Rev.",
      volume         = "D79",
      year           = "2009",
      pages          = "083007",
      doi            = "10.1103/PhysRevD.79.083007",
      eprint         = "0901.4637",
      archivePrefix  = "arXiv",
      primaryClass   = "astro-ph.SR",
      SLACcitation   = "%%CITATION = ARXIV:0901.4637;%%"
}

@article{Rajagopal:2006ig,
      author         = "Rajagopal, Krishna and Sharma, Rishi",
      title          = "{The Crystallography of Three-Flavor Quark Matter}",
      journal        = "Phys. Rev.",
      volume         = "D74",
      year           = "2006",
      pages          = "094019",
      doi            = "10.1103/PhysRevD.74.094019",
      eprint         = "hep-ph/0605316",
      archivePrefix  = "arXiv",
      primaryClass   = "hep-ph",
      reportNumber   = "MIT-CTP-3749",
      SLACcitation   = "%%CITATION = HEP-PH/0605316;%%"
}

@article{Alford:2000ze,
      author         = "Alford, Mark G. and Bowers, Jeffrey A. and Rajagopal,
                        Krishna",
      title          = "{Crystalline color superconductivity}",
      journal        = "Phys. Rev.",
      volume         = "D63",
      year           = "2001",
      pages          = "074016",
      doi            = "10.1103/PhysRevD.63.074016",
      eprint         = "hep-ph/0008208",
      archivePrefix  = "arXiv",
      primaryClass   = "hep-ph",
      reportNumber   = "MIT-CTP-3012",
      SLACcitation   = "%%CITATION = HEP-PH/0008208;%%"
}

@article{Schafer:2005ym,
      author         = "Sch{\"a}fer, Thomas",
      title          = "{Meson supercurrent state in high density QCD}",
      journal        = "Phys. Rev. Lett.",
      volume         = "96",
      year           = "2006",
      pages          = "012305",
      doi            = "10.1103/PhysRevLett.96.012305",
      eprint         = "hep-ph/0508190",
      archivePrefix  = "arXiv",
      primaryClass   = "hep-ph",
      SLACcitation   = "%%CITATION = HEP-PH/0508190;%%"
}

@article{Kryjevski:2008zz,
      author         = "Kryjevski, Andrei",
      title          = "{Spontaneous superfluid current generation in the kaon
                        condensed color flavor locked phase at nonzero strange
                        quark mass}",
      journal        = "Phys. Rev.",
      volume         = "D77",
      year           = "2008",
      pages          = "014018",
      doi            = "10.1103/PhysRevD.77.014018",
      eprint         = "hep-ph/0508180",
      archivePrefix  = "arXiv",
      primaryClass   = "hep-ph",
      SLACcitation   = "%%CITATION = HEP-PH/0508180;%%"
}

@article{Bedaque:2001je,
      author         = "Bedaque, Paulo F. and Sch{\"a}fer, Thomas",
      title          = "{High density quark matter under stress}",
      journal        = "Nucl. Phys.",
      volume         = "A697",
      year           = "2002",
      pages          = "802-822",
      doi            = "10.1016/S0375-9474(01)01272-6",
      eprint         = "hep-ph/0105150",
      archivePrefix  = "arXiv",
      primaryClass   = "hep-ph",
      reportNumber   = "SUNY-NTG-01-07",
      SLACcitation   = "%%CITATION = HEP-PH/0105150;%%"
}

@article{Son:1999cm,
      author         = "Son, D. T. and Stephanov, Misha A.",
      title          = "{Inverse meson mass ordering in color flavor locking
                        phase of high density QCD}",
      journal        = "Phys. Rev.",
      volume         = "D61",
      year           = "2000",
      pages          = "074012",
      doi            = "10.1103/PhysRevD.61.074012",
      eprint         = "hep-ph/9910491",
      archivePrefix  = "arXiv",
      primaryClass   = "hep-ph",
      SLACcitation   = "%%CITATION = HEP-PH/9910491;%%"
}

@article{Fukushima:2005cm,
      author         = "Fukushima, Kenji",
      title          = "{Analytical and numerical evaluation of the Debye and
                        Meissner masses in dense neutral three-flavor quark
                        matter}",
      journal        = "Phys. Rev.",
      volume         = "D72",
      year           = "2005",
      pages          = "074002",
      doi            = "10.1103/PhysRevD.72.074002",
      eprint         = "hep-ph/0506080",
      archivePrefix  = "arXiv",
      primaryClass   = "hep-ph",
      reportNumber   = "RBRC-524",
      SLACcitation   = "%%CITATION = HEP-PH/0506080;%%"
}

@article{Casalbuoni:2004tb,
      author         = "Casalbuoni, R. and Gatto, Raoul and Mannarelli, M. and
                        Nardulli, G. and Ruggieri, M.",
      title          = "{Meissner masses in the gCFL phase of QCD}",
      journal        = "Phys. Lett.",
      volume         = "B605",
      year           = "2005",
      pages          = "362-368",
      doi            = "10.1016/j.physletb.2004.11.045,
                        10.1016/j.physletb.2005.04.025",
      note           = "[Erratum: Phys. Lett.B615,297(2005)]",
      eprint         = "hep-ph/0410401",
      archivePrefix  = "arXiv",
      primaryClass   = "hep-ph",
      reportNumber   = "BARI-TH-498-04",
      SLACcitation   = "%%CITATION = HEP-PH/0410401;%%"
}

@Article{Clogston:1962,
     author    = "Clogston, A. M. ",
     title     = "Upper Limit for the Critical Field in Hard Superconductors",
     journal    = "Phys. Rev. Lett.",
     volume    = "9",
     pages     = "266",
     year      = "1962"
}

@Article{Chandrasekhar:1962,
     author    = "Chandrasekhar, B. S.",
     title     = "Upper Limit for the Critical Field in Hard Superconductors",
     journal    = "Appl. Phys. Lett.",
     volume    = "1",
     pages     = "7",
     year      = "1962"
}

@article{Rischke:2001py,
      author         = "Rischke, Dirk H.",
      title          = "{Gluon selfenergy in a two flavor color superconductor}",
      journal        = "Phys. Rev.",
      volume         = "D64",
      year           = "2001",
      pages          = "094003",
      doi            = "10.1103/PhysRevD.64.094003",
      eprint         = "nucl-th/0103050",
      archivePrefix  = "arXiv",
      primaryClass   = "nucl-th",
      SLACcitation   = "%%CITATION = NUCL-TH/0103050;%%"
}

@article{PhysRev.108.1175,
  title = {Theory of Superconductivity},
  author = {Bardeen, J. and Cooper, L. N. and Schrieffer, J. R.},
  journal = {Phys. Rev.},
  volume = {108},
  issue = {5},
  pages = {1175--1204},
  numpages = {0},
  year = {1957},
  month = {Dec},
  publisher = {American Physical Society},
  doi = {10.1103/PhysRev.108.1175}
}

@article{Alford:1998mk,
      author         = "Alford, Mark G. and Rajagopal, Krishna and Wilczek,
                        Frank",
      title          = "{Color flavor locking and chiral symmetry breaking in
                        high density QCD}",
      journal        = "Nucl. Phys.",
      volume         = "B537",
      year           = "1999",
      pages          = "443-458",
      doi            = "10.1016/S0550-3213(98)00668-3",
      eprint         = "hep-ph/9804403",
      archivePrefix  = "arXiv",
      primaryClass   = "hep-ph",
      reportNumber   = "IASSNS-HEP-98-29, MIT-CTP-2731",
      SLACcitation   = "%%CITATION = HEP-PH/9804403;%%"
}

@article{Shuster:1999tn,
      author         = "Shuster, E. and Son, D. T.",
      title          = "{On finite density QCD at large N(c)}",
      journal        = "Nucl. Phys.",
      volume         = "B573",
      year           = "2000",
      pages          = "434-446",
      doi            = "10.1016/S0550-3213(99)00615-X",
      eprint         = "hep-ph/9905448",
      archivePrefix  = "arXiv",
      primaryClass   = "hep-ph",
      reportNumber   = "MIT-CTP-2865",
      SLACcitation   = "%%CITATION = HEP-PH/9905448;%%"
}

@article{Giannakis:2004xt,
      author         = "Giannakis, Ioannis and Hou, De-fu and Ren, Hai-cang and
                        Rischke, Dirk H.",
      title          = "{Gauge field fluctuations and first-order phase
                        transition in color superconductivity}",
      journal        = "Phys. Rev. Lett.",
      volume         = "93",
      year           = "2004",
      pages          = "232301",
      doi            = "10.1103/PhysRevLett.93.232301",
      eprint         = "hep-ph/0406031",
      archivePrefix  = "arXiv",
      primaryClass   = "hep-ph",
      SLACcitation   = "%%CITATION = HEP-PH/0406031;%%"
}

@article{Schmitt:2002sc,
      author         = "Schmitt, Andreas and Wang, Qun and Rischke, Dirk H.",
      title          = "{When the transition temperature in color superconductors
                        is not like in BCS theory}",
      journal        = "Phys. Rev.",
      volume         = "D66",
      year           = "2002",
      pages          = "114010",
      doi            = "10.1103/PhysRevD.66.114010",
      eprint         = "nucl-th/0209050",
      archivePrefix  = "arXiv",
      primaryClass   = "nucl-th",
      SLACcitation   = "%%CITATION = NUCL-TH/0209050;%%"
}

@article{Schmitt:2003aa,
      author         = "Schmitt, Andreas and Wang, Qun and Rischke, Dirk H.",
      title          = "{Mixing and screening of photons and gluons in a color
                        superconductor}",
      journal        = "Phys. Rev.",
      volume         = "D69",
      year           = "2004",
      pages          = "094017",
      doi            = "10.1103/PhysRevD.69.094017",
      eprint         = "nucl-th/0311006",
      archivePrefix  = "arXiv",
      primaryClass   = "nucl-th",
      SLACcitation   = "%%CITATION = NUCL-TH/0311006;%%"
}

@article{Rajagopal:2000rs,
      author         = "Rajagopal, Krishna and Shuster, Eugene",
      title          = "{On the applicability of weak coupling results in high
                        density QCD}",
      journal        = "Phys. Rev.",
      volume         = "D62",
      year           = "2000",
      pages          = "085007",
      doi            = "10.1103/PhysRevD.62.085007",
      eprint         = "hep-ph/0004074",
      archivePrefix  = "arXiv",
      primaryClass   = "hep-ph",
      reportNumber   = "MIT-CTP-2969",
      SLACcitation   = "%%CITATION = HEP-PH/0004074;%%"
}

@article{Schmitt:2004et,
      author         = "Schmitt, Andreas",
      title          = "{The Ground state in a spin-one color superconductor}",
      journal        = "Phys. Rev.",
      volume         = "D71",
      year           = "2005",
      pages          = "054016",
      doi            = "10.1103/PhysRevD.71.054016",
      eprint         = "nucl-th/0412033",
      archivePrefix  = "arXiv",
      primaryClass   = "nucl-th",
      SLACcitation   = "%%CITATION = NUCL-TH/0412033;%%"
}

@article{Wang:2001aq,
      author         = "Wang, Qun and Rischke, Dirk H.",
      title          = "{How the quark selfenergy affects the color
                        superconducting gap}",
      journal        = "Phys. Rev.",
      volume         = "D65",
      year           = "2002",
      pages          = "054005",
      doi            = "10.1103/PhysRevD.65.054005",
      eprint         = "nucl-th/0110016",
      archivePrefix  = "arXiv",
      primaryClass   = "nucl-th",
      SLACcitation   = "%%CITATION = NUCL-TH/0110016;%%"
}

@article{Pisarski:1999tv,
      author         = "Pisarski, Robert D. and Rischke, Dirk H.",
      title          = "{Color superconductivity in weak coupling}",
      journal        = "Phys. Rev.",
      volume         = "D61",
      year           = "2000",
      pages          = "074017",
      doi            = "10.1103/PhysRevD.61.074017",
      eprint         = "nucl-th/9910056",
      archivePrefix  = "arXiv",
      primaryClass   = "nucl-th",
      SLACcitation   = "%%CITATION = NUCL-TH/9910056;%%"
}

@article{Haber:2017oqb,
      author         = "Haber, Alexander and Schmitt, Andreas",
      title          = "{New color-magnetic defects in dense quark matter}",
      journal        = "J. Phys.",
      volume         = "G45",
      year           = "2018",
      number         = "6",
      pages          = "065001",
      doi            = "10.1088/1361-6471/aabc1a",
      eprint         = "1712.08587",
      archivePrefix  = "arXiv",
      primaryClass   = "hep-ph",
      SLACcitation   = "%%CITATION = ARXIV:1712.08587;%%"
}

@Article{Cornwall:1974vz,
     author    = "Cornwall, John M. and Jackiw, R. and Tomboulis, E.",
     title     = "Effective Action for Composite Operators",
     journal   = "Phys. Rev.",
     volume    = "D10",
     year      = "1974",
     pages     = "2428-2445",
     SLACcitation  = "%%CITATION = PHRVA,D10,2428;%%"
}

@Article{Baym:1962sx,
     author    = "Baym, Gordon",
     title     = "Selfconsistent approximation in many body systems",
     journal   = "Phys. Rev.",
     volume    = "127",
     year      = "1962",
     pages     = "1391-1401",
     SLACcitation  = "%%CITATION = PHRVA,127,1391;%%"
}

@Article{Luttinger:1960ua,
     author    = "Luttinger, J. M. and Ward, John Clive",
     title     = "Ground state energy of a many fermion system. 2",
     journal   = "Phys. Rev.",
     volume    = "118",
     year      = "1960",
     pages     = "1417-1427",
     SLACcitation  = "%%CITATION = PHRVA,118,1417;%%"
}

@article{Schmitt:2017efp,
      author         = "Schmitt, Andreas and Shternin, Peter",
      title          = "{Reaction rates and transport in neutron stars}",
      journal        = "Astrophys. Space Sci. Libr.",
      volume         = "457",
      year           = "2018",
      pages          = "455-574",
      doi            = "10.1007/978-3-319-97616-7_9",
      eprint         = "1711.06520",
      archivePrefix  = "arXiv",
      primaryClass   = "astro-ph.HE",
      SLACcitation   = "%%CITATION = ARXIV:1711.06520;%%"
}

@article{Schmitt:2014eka,
      author         = "Schmitt, Andreas",
      title          = "{Introduction to Superfluidity}",
      journal        = "Lect. Notes Phys.",
      volume         = "888",
      year           = "2015",
      pages          = "pp.1-155",
      doi            = "10.1007/978-3-319-07947-9",
      eprint         = "1404.1284",
      archivePrefix  = "arXiv",
      primaryClass   = "hep-ph",
      SLACcitation   = "%%CITATION = ARXIV:1404.1284;%%"
}

@article{Alford:2007xm,
      author         = "Alford, Mark G. and Schmitt, Andreas and Rajagopal,
                        Krishna and Schäfer, Thomas",
      title          = "{Color superconductivity in dense quark matter}",
      journal        = "Rev. Mod. Phys.",
      volume         = "80",
      year           = "2008",
      pages          = "1455-1515",
      doi            = "10.1103/RevModPhys.80.1455",
      eprint         = "0709.4635",
      archivePrefix  = "arXiv",
      primaryClass   = "hep-ph",
      reportNumber   = "MIT-CTP-3861",
      SLACcitation   = "%%CITATION = ARXIV:0709.4635;%%"
}

@article{Baym:2017whm,
      author         = "Baym, Gordon and Hatsuda, Tetsuo and Kojo, Toru and
                        Powell, Philip D. and Song, Yifan and Takatsuka,
                        Tatsuyuki",
      title          = "{From hadrons to quarks in neutron stars: a review}",
      journal        = "Rept. Prog. Phys.",
      volume         = "81",
      year           = "2018",
      number         = "5",
      pages          = "056902",
      doi            = "10.1088/1361-6633/aaae14",
      eprint         = "1707.04966",
      archivePrefix  = "arXiv",
      primaryClass   = "astro-ph.HE",
      reportNumber   = "RIKEN-ITHEMS-REPORT-17, RIKEN-QHP-316,
                        RIKEN-iTHEMS-Report-17",
      SLACcitation   = "%%CITATION = ARXIV:1707.04966;%%"
}

@article{Schmitt:2010pn,
      author         = "Schmitt, Andreas",
      title          = "{Dense matter in compact stars: A pedagogical
                        introduction}",
      journal        = "Lect. Notes Phys.",
      volume         = "811",
      year           = "2010",
      pages          = "1-111",
      doi            = "10.1007/978-3-642-12866-0",
      eprint         = "1001.3294",
      archivePrefix  = "arXiv",
      primaryClass   = "astro-ph.SR",
      SLACcitation   = "%%CITATION = ARXIV:1001.3294;%%"
}

\end{document}